\documentclass[a4paper, amsfonts, amssymb, amsmath, reprint, showkeys, nofootinbib, twoside, balancelastpage, longbibliography, nobibnotes]{revtex4-1}
\usepackage[english]{babel}
\usepackage[utf8]{inputenc}
\usepackage[colorinlistoftodos, color=green!40, prependcaption]{todonotes}
\usepackage{tikz}
\usetikzlibrary{decorations.pathreplacing,decorations.pathmorphing,calligraphy,calc,fadings,patterns}
\usetikzlibrary{hobby}
\usetikzlibrary{fadings}
\tikzfading[name=fade out, inner color=transparent!0,
         outer color=transparent!100]

\usepackage[colorlinks=true
,urlcolor=blue
,anchorcolor=blue
,citecolor=blue
,filecolor=magenta
,linkcolor=red
,menucolor=blue
,linktocpage=true
,pdfproducer=medialab
,pdfa=true
]{hyperref}

\usepackage{bm}
\usepackage{bbold}
\renewcommand{\d}{\mathrm{d}}
\renewcommand{\v}[1]{\bm{#1}}

\newcommand{\E}{\mathbb{E}}

\newcommand{\bin}{B}

\usepackage{algorithm}
\usepackage{algpseudocode}

\usepackage{caption}
\usepackage{subcaption}

\begin{document}

\title{A Compound Poisson Generator approach to Point-Source Inference in Astrophysics}
\author{Gabriel H. Collin}
\affiliation{Department of Physics, Massachusetts Institute of Technology, Cambridge, MA 02139, USA}
\affiliation{Institute for Data, Systems, and Society, Massachusetts Institute of Technology, Cambridge, MA 02139, USA}
\author{Nicholas L. Rodd}
\affiliation{Berkeley Center for Theoretical Physics, University of California, Berkeley, CA 94720, USA}
\affiliation{Theoretical Physics Group, Lawrence Berkeley National Laboratory, Berkeley, CA 94720, USA}
\author{Tyler Erjavec}
\affiliation{Department of Physics, Massachusetts Institute of Technology, Cambridge, MA 02139, USA}
\author{Kerstin Perez}
\affiliation{Department of Physics, Massachusetts Institute of Technology, Cambridge, MA 02139, USA}
\affiliation{The NSF AI Institute for Artificial Intelligence and Fundamental Interactions}

\begin{abstract}
    The identification and description of point sources is one of the oldest problems in astronomy; yet, even today the correct statistical treatment for point sources remains one of the field's hardest problems. For dim or crowded sources, likelihood based inference methods are required to estimate the uncertainty on the characteristics of the source population. In this work, a new parametric likelihood is constructed for this problem using Compound Poisson Generator (CPG) functionals which incorporate instrumental effects from first principles. We demonstrate that the CPG approach exhibits a number of advantages over Non-Poissonian Template Fitting (NPTF) - an existing method - in a series of test scenarios in the context of X-ray astronomy. These demonstrations show that the effect of the point-spread function, effective area, and choice of point-source spatial distribution cannot, generally, be factorised as they are in NPTF, while the new CPG construction is validated in these scenarios. Separately, an examination of the diffuse-flux emission limit is used to show that most simple choices of priors on the standard parameterisation of the population model can result in unexpected biases: when a model comprising both a point-source population and diffuse component is applied to this limit, nearly all observed flux will be assigned to either the population or to the diffuse component. A new parametrisation is presented for these priors which properly estimates the uncertainties in this limit. In this choice of priors, CPG correctly identifies that the fraction of flux assigned to the population model cannot be constrained by the data. 
\end{abstract}

\maketitle


\begin{figure*}
    \centering
    \begin{subfigure}{\textwidth}
         \centering
            \includegraphics{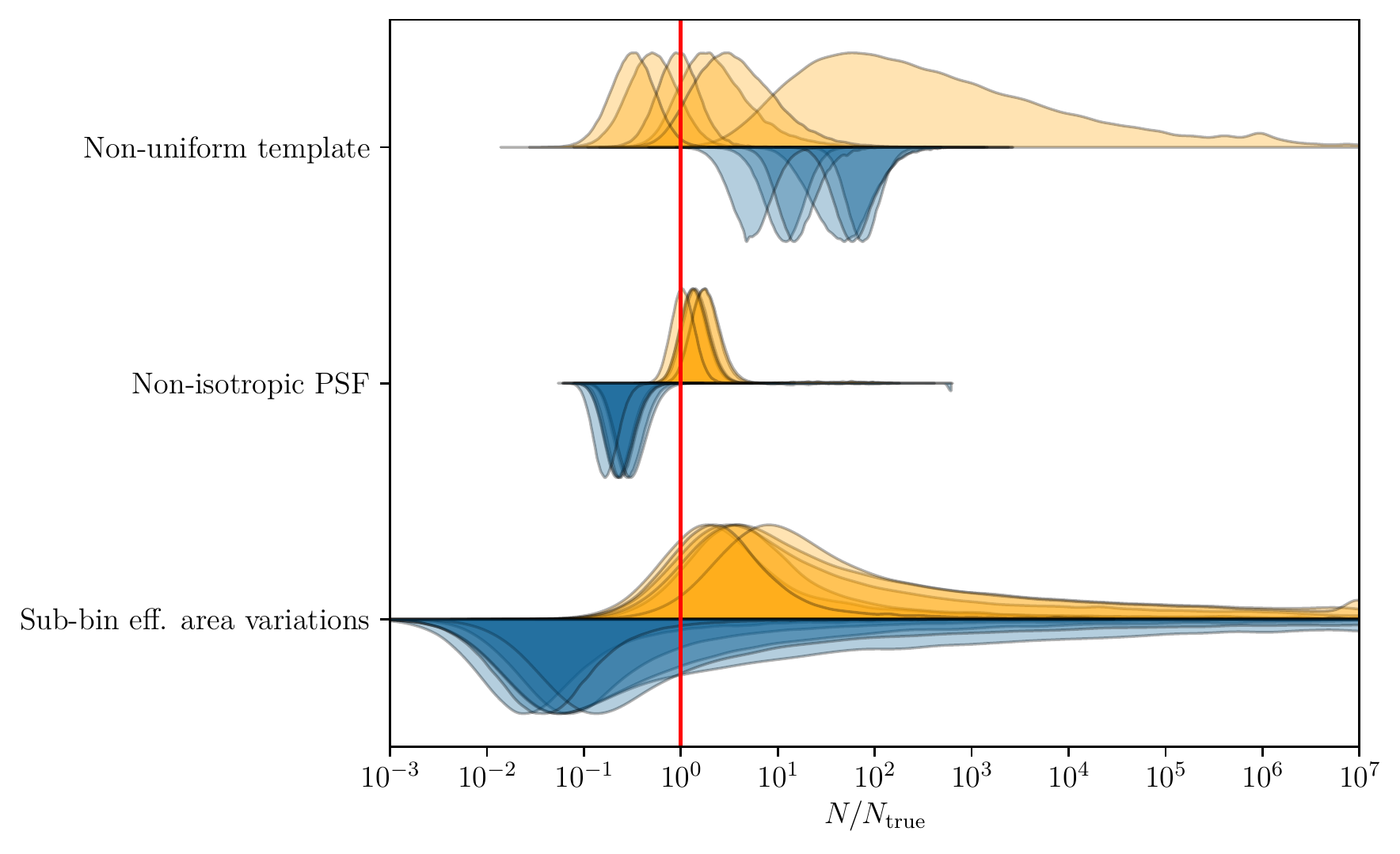}
     \end{subfigure}
    \caption{The recovered posterior in the number of point sources in the population, $N$, for both the novel CPG construction and the existing NPTF approach, for the three test scenarios considered in this work.
    Upright distributions are the posteriors for individual trials from the CPG construction.
    Inverted distributions are the posteriors recovered by NPTF.
    A clear bias away from the true number of sources, $N_{\mathrm{true}}$ (red line), is observed from NPTF.
    We will describe the results shown here in detail in Sec.~\ref{sec:nptf}.
    For example, the extension of many posteriors to $N \gg N_\textrm{true}$ is explained in Sec.~\ref{sub:nptf:effarea}.}
    \label{fig:N_summary}
\end{figure*}
\begin{figure*}
    \begin{subfigure}{\textwidth}
         \centering
            \includegraphics{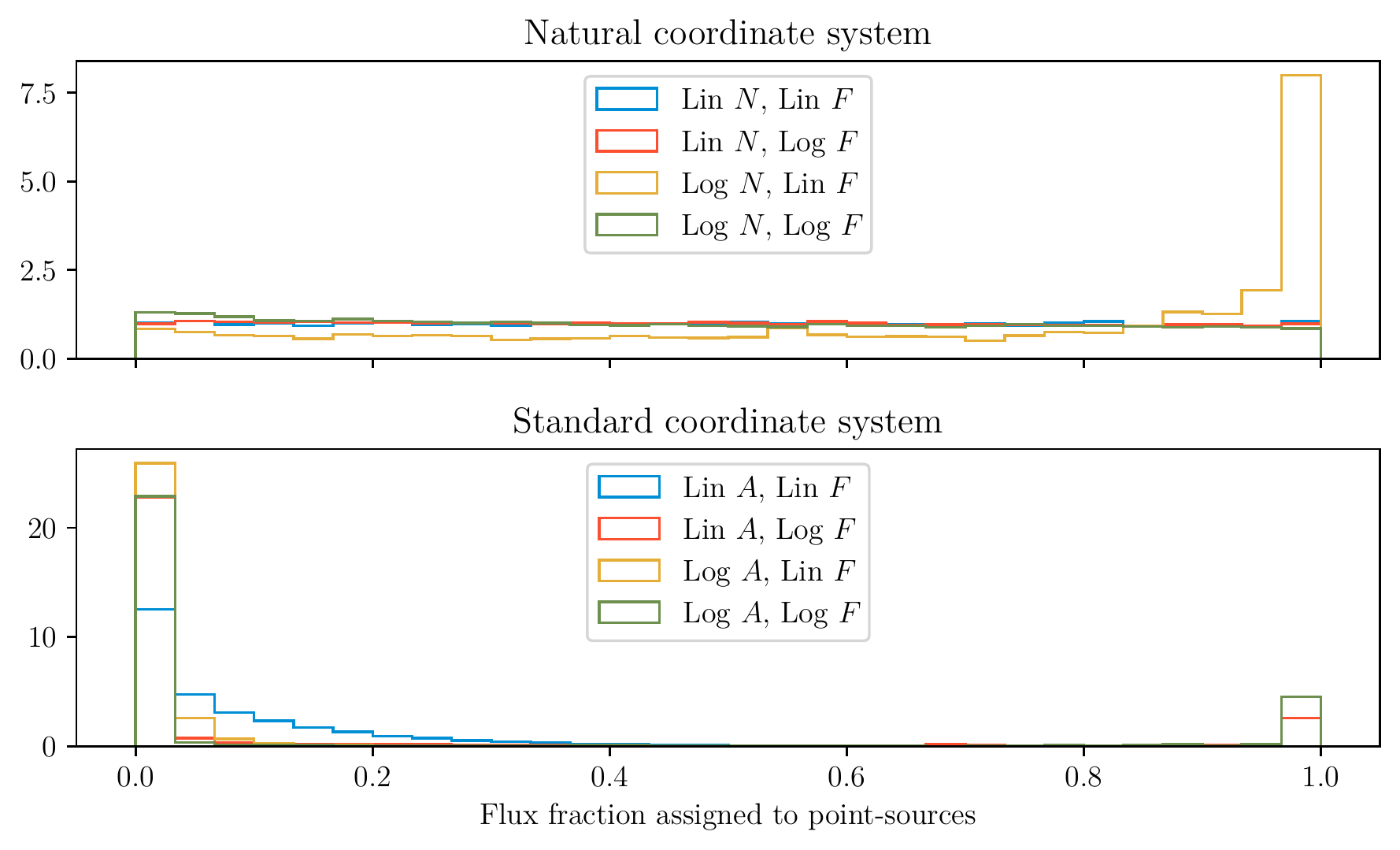}
     \end{subfigure}
    \caption{The median recovered posterior for the novel natural coordinate system (above) and standard coordinate system (below), for multiple choices of priors on the number of sources ($N$, natural only), flux ($F$), and differential source-count function normalisation ($A$, standard only).
    The posterior is shown for the fraction of flux assigned to the population model, as compared to a diffuse background model.
    In this scenario, only diffuse flux is present, and so these posteriors should be uniform, as diffuse emission is formally indistinguishable from a population of dim point sources.
    This is observed only for the natural coordinate system, with one exception discussed further in Sec.~\ref{sec:priors}.}
    \label{fig:coords_summary}
\end{figure*}

A practical and unbiased method of point-source inference is central to the task of recovering the physical characteristics of point-source populations.
When point sources are bright and well separated in the sky, their identification and cataloguing is straightforward and requires little statistics beyond quantification of uncertainties on location and brightness.
The primary difficulty arises when point sources are dim, where one must distinguish a putative source from a background fluctuation, and account for the possibility of multiple closely overlapping sources --- called a crowded field.
The calculation of the uncertainty on the number of sources is particularly demanding, as the number of sources is a discrete parameter.
Parametric point-source inference methods, such as the Non-Poissonian Template Fitting (NPTF) method \cite{Lee:2015fea}, side-step this issue by specifying a likelihood conditioned on the characteristics of a population.
In particular, the likelihood can be expressed in terms of the mean number of sources --- a continuous parameter for which uncertainty calculation is considerably easier.

NPTF is currently the most widely applied parametric point-source inference method in gamma-ray and neutrino astronomy, especially in the analysis of gamma-rays from the Galactic Center, however, there is growing concern that results obtained with NPTF must be interpreted cautiously to avoid potential biases, see the discussion in Refs.~\cite{Leane:2019xiy,Chang:2019ars,Buschmann:2020adf,Leane:2020nmi,Leane:2020pfc}.
In the present work, we will identify several clear biases of the NPTF framework which become most obvious in the following three specific test scenarios motivated by X-ray astronomy:
\hyperref[sub:nptf:nonuniform]{(i)} a spatially non-uniform point-source population model; \hyperref[sub:nptf:anisoPSF]{(ii)} a non-isotropic instrumental point-spread function (PSF); and \hyperref[sub:nptf:effarea]{(iii)} variation in the instrumental detector response, namely the effective area, on scales smaller than the choice of binning --- that is typically, in turn, on the scale of the PSF.
In all cases, we observe a bias in the recovered number of point-sources as summarised in Fig.~\ref{fig:N_summary}.

To resolve these biases, we develop a new parametric point-source inference likelihood called the Compound Poisson Generator (CPG) likelihood.
This new likelihood addresses the biases in the NPTF apparent in the X-ray domain, and further provides a rigorous treatment of several approximations made in the conventional NPTF approach.
In addition, we demonstrate the importance of a carefully chosen prior parameterisation in Bayesian analyses.
Priors chosen directly on the standard parameterisation of the differential source-count function can, in the limit where these models are formally indistinguishable, lead to posteriors where the observed flux is assigned to either the point-source population model or an associated diffuse emission model.
We define a new coordinate system for the specification of priors that removes this unwanted effect as summarised in Fig.~\ref{fig:coords_summary}.

This work does not address biases that may result from incorrect modelling of the spatial distributions of populations (see in particular Refs.~\cite{Leane:2020nmi,Leane:2020pfc}), effective area, exposure area, or the PSF.
Such biases are not limitations of the point-source inference method; instead, they are concerns for individual applications of point-source inference to specific analyses --- although they are no less important for obtaining correct results.
In this work, we assume that these effects are modelled correctly, and investigate the statistical methodology of point-source inference.

We organise the remainder of the discussion as follows.
Section~\ref{sec:review} reviews the current state of point-source inference, with a more detailed discussion on the concerns with NPTF, where the CPG enters, and how it resolves some of these issues.
Section~\ref{sec:ppi} outlines the point-source population model and the various instrumental effects introduced by standard X-ray instruments such as the NuSTAR satellite telescope, which we use as an example.
Having outlined the background and challenges, in Sec.~\ref{sec:cpg}, the CPG likelihood is constructed from first principles.
Setion~\ref{sec:priors} introduces the priors on the population model parameters and demonstrates how the new coordinate system correctly handles the combination of point source and diffuse models.
The NPTF method is introduced in more detail in Sec.~\ref{sec:nptf}, where we consider several scenarios motivated by instrumental effects, results from CPG and NPTF on these cases are directly compared, and the general efficacy of the new likelihood is demonstrated.
An implementation of the CPG is made publicly available \href{https://github.com/ghcollin/cpg_likelihood}{here}.

\section{Review of point-source inference\label{sec:review}}

To begin with, we will review the history and current state of point-source inference as it applies to parametric inference in the regime of dim sources and crowded fields.
In this regime automated source extraction is most commonly employed.
This involves such methods as thresholding, peak searches, and wavelet decomposition among others~\cite{masiasReviewSourceDetection2012}.
Source extraction algorithms can be broadly categorised as point estimation methods --- they produce a single best-fit or most-significant solution, and this solution can be represented as a point in the parameter space of point-sources.
For example, in thresholding and peak searches, a pixel is considered part of a source if the measured pixel intensity exceeds some defined threshold; which may be manually selected, or defined by a statistical significance estimated from the image.
The set of pixels that are classified as belonging to a source is a single point in the parameter space of all such sets of pixels, and may be further transformed into point-source locations by taking the pixels in each set, and, for example, computing the centroid of those pixels.
The resultant point in parameter space is the algorithm's best guess as to the true parameters.
However, this guess is just that: an estimate of the true locations of the point-sources in an image.
As such, the calculation of uncertainties is required in order to quantify the quality of this single best guess.

In this problem, there are two classes of parameters: the individual point-source parameters, $\v{\vartheta}$, such as location and brightness, which are continuous quantities; and the number of sources in the image, $\tilde{N}$, which is a discrete quantity.
The measured data, such as the observed image, will be distributed according to the likelihood $\mathcal{L}(\v{d} | \tilde{N}, \{\v{\vartheta}_i\})$, where $\{\v{\vartheta}_i\}$ is a set of size $\tilde{N}$, one for each of the $\tilde{N}$ sources.
The pair $(\tilde{N}, \{\v{\vartheta}_i\})$ form a point in the parameter space of the likelihood, and is often referred to as a catalogue.
When it comes to the calculation of uncertainties, discrete and continuous parameters must be handled in different ways.

Deriving uncertainties on the continuous individual source parameters is the easier of the two.
Assuming the likelihood distribution is available, uncertainties can be derived using variational methods, or confidence regions can be constructed should Wilks' theorem apply.
Deriving uncertainties on the discrete number of sources, $\tilde{N}$, is more fraught.
The calculation of uncertainty on a discrete parameter is equivalent to the problem of model selection.
Each choice for the number of sources should be considered as a separate model.
Two choices for the number of sources may then be compared using a likelihood ratio between the likelihoods for each of the associated models.
This likelihood ratio is a test statistic, and the probability distribution for the ratio must be known if it is to be converted to a $p$-value.
The $p$-values may then be used to construct a confidence interval on the number of sources.
Most often, Wilks' theorem is used to assume that the likelihood ratio is $\chi^2$ distributed; however, this is only guaranteed in the asymptotic limit and any deviation of the true distribution from this assumption will result in incorrect confidence intervals.

An alternative approach to model selection is the calculation of marginal likelihoods:\footnote{Point-source inference can be approached in a frequentist or Bayesian framework, and we will use the language of both throughout.
Nevertheless, when adopting a Bayesian approach -- as has commonly been done for the NPTF -- priors must be chosen carefully, as we discuss in Sec.~\ref{sec:priors}.
}
\begin{equation}
    \mathcal{L}(\v{d} | \tilde{N}) = \int \mathcal{L}(\v{d} | \tilde{N}, \{\v{\vartheta}_i\}) \left[\prod_{i=1}^{\tilde{N}} p(\v{\vartheta}_i ) \d\v{\vartheta}_i \right],
\end{equation}
where $p(\v{\vartheta}_i)$ is a prior on the individual source parameters.
From this expression, the posterior $p(\tilde{N} | \v{d})$ can be found through Bayes' theorem, and so a credible region can be constructed on the number of sources --- avoiding the need for a test statistic as in the likelihood ratio case.
Variational methods provide only a lower bound on the marginal likelihood of the model, and thus provide no more than an estimate for the posterior distribution on the number of sources.

A more common approach is to sample the posterior $p(\{\v{\vartheta}_i\} | \v{d}, \tilde{N})$ directly using nested sampling \cite{skillingNestedSampling2004} or a Markov Chain Monte-Carlo (MCMC). 
Nested sampling provides an estimate of the marginal likelihood directly, while various techniques exist for deriving the marginal likelihood from MCMC --- for details consult the review \textcite{gelmanSimulatingNormalizingConstants1998}.
Although this provides the most principled approach to the problem for small numbers of sources, in practice it exhibits a critical flaw.
When the number of sources is potentially large, a marginal likelihood must be computed for each possible choice of $\tilde{N}$ -- a flaw also shared by the variational and likelihood ratio methods.
For hundreds of sources, this can rapidly become computationally infeasible.

Ultimately, this flaw can be understood to be a manifestation of a one-dimensional grid search over $\tilde{N}$.
In fact, MCMC is designed to avoid grid searches by concentrating sampling on regions of parameter space that are most likely; however, the MCMC algorithm is designed exclusively for continuous parameters, while $\tilde{N}$ is a discrete parameter which when varied also modifies the number of $\v{\vartheta}$ parameters.
The Reversible Jump Markov Chain Monte-Carlo (RJMCMC) algorithm is an extension of MCMC to discrete parameters that control the number of parameters in the distribution~\cite{greenReversibleJumpMarkov1995}.
Probabilistic cataloguing \cite{Daylan:2016tia,2017AJ....154..132P} is the application of RJMCMC to the problem of point-source inference.
While the point estimation methods produce a single best guess catalogue, probabilistic cataloguing produces posterior samples in the space of all possible catalogues.
From this catalogue posterior, a posterior on the number of sources can be calculated.
A catalogue posterior provides -- essentially by definition -- the most general solution to the problem of point-source inference.
This generality comes at a high cost: the trans-dimensional parameter transformations required by RJMCMC has to be carefully selected, as they must be tuned to the specific application to ensure that trans-dimensional moves are efficient.
In addition, the dimensionality of the likelihood is at least two or three times larger than the number of sources --- depending on the number of parameters per source.
For thousands of sources, it can be unrealistic to assume that the RJMCMC will properly equilibrate for even a given number of sources, let alone across the number of sources; the application of RJMCMC will require careful and extensive diagnostics, none of which are foolproof.

The solution to this problem requires a revisitation to the stated inference goal.
If the goal truly necessitates the location and intensity of each individual source, then probabilistic cataloguing is necessary.
But in this problem area, where the number of sources is large, they also likely form a crowded field.
In this case, locations of individual sources will be poorly defined, as they form a near-uniform density within which sources are easily interchangeable.
Instead of attempting to track each individual source, a model can be defined for the entire population of sources.
The most common choice of model is a differential source-count function: $\d N/\d \v{\vartheta} (\v{\theta})$.
This function then has its own parameters, $\v{\theta}$, which are characteristics of the population as a whole, such as the spatial distribution of sources, or the average flux of the population.
The differential source-count function gives the density of sources as a function of $\v{\vartheta}$ given $\v{\theta}$, and thus implicitly defines a probability distribution:
\begin{equation}
    \frac{\d N}{\d \v{\vartheta}}(\v{\theta}) = N p(\v{\vartheta} | \v{\theta}),    
\end{equation}
where $N$ is the mean number of sources in the population.
This allows marginalisation of the likelihood over both the set of $\v{\theta}$ and further the possible number of sources,
\begin{equation}
    \mathcal{L}(\v{d} | \v{\theta}) = \sum_{\tilde{N}} \int \mathcal{L}(\v{d} | \tilde{N}, \{\v{\vartheta}_i\}) \left[\prod_{i=1}^n p(\v{\vartheta}_i | \v{\theta}) \d\v{\vartheta}_i \right] p(\tilde{N} | N),
    \label{eq:non_parametric_marginal}
\end{equation}
where $p(\tilde{N} | N)$ is a prior on the number of sources given the mean number of sources --- usually assumed to be a Poisson distribution.
Thus, sampling directly from the posterior of $\mathcal{L}(\v{d} | \v{\theta})$ reduces the high dimensional non-parametric sampling problem in the space of all possible catalogues to a low dimensional parametric sampling problem in $\v{\theta}$.
However, to achieve this we have marginalised out the $\v{\vartheta}$ parameters, and so identification of the location or brightness of individual sources is no longer possible --- the inference problem is now on the population itself.

To achieve this vast simplification, the likelihood $\mathcal{L}(\v{d} | \v{\theta})$ must be constructed directly so that it may be evaluated without computing the multi-dimensional integrals in Eq.~\ref{eq:non_parametric_marginal}.
An early attempt for this construction, called the $P(D)$ method \cite{1957PCPS...53..764S,Miyaji:2001dp,1992ApJ...396..460B}, used instrument simulations to numerically estimate the likelihood as a histogram over the number of detected photons.
This procedure estimates $\mathcal{L}(\v{d} | \v{\theta})$ by sampling $\v{d}$ from the right-hand side of Eq.~\ref{eq:non_parametric_marginal}.
As this is effectively equivalent to an importance weighted integration of those multi-dimensional integrals, this method ultimately carries the same computational burden of probabilistic cataloguing.

In \textcite{Malyshev:2011zi}, the authors noted that the number of detected photons is a sum of the number of detected photons produced by each source --- itself distributed by a Poisson distribution.
This allows the likelihood to be constructed using probability generating functions, and provides an analytic formula for $\mathcal{L}(\v{d} | \v{\theta})$ assuming a perfect instrument.
To incorporate instrumental effects, a heuristic argument is used to justify a semi-analytic expression for the likelihood in terms of a detector effect correction function, $\rho(f)$, that is generated through Monte-Carlo simulation.

This method was further extended by \textcite{Lee:2015fea} to images of multiple pixels by parameterising the source-count function in terms of a spatial template.
Known as Non-Poissonian Template Fitting (NPTF), this is currently a leading method for parametric point-source inference in gamma-ray astronomy, and the method has also been applied to the search for astrophysical neutrino point-sources in data from the IceCube telescope \cite{Aartsen:2019mbc}.
The NPTF was primarily developed to analyse the excess of Galactic Center (GCE) gamma-rays observed by the \emph{Fermi} telescope~\cite{Goodenough:2009gk,Hooper:2010mq,Hooper:2011ti,Abazajian:2012pn,Hooper:2013rwa,Gordon:2013vta,Abazajian:2014fta,Daylan:2014rsa,Calore:2014xka,Abazajian:2014hsa,TheFermi-LAT:2015kwa,Linden:2016rcf,Macias:2016nev,Clark:2016mbb} and concluded that the observed excess was better described by a population of point sources, in comparison to a dark-matter annihilation origin \cite{Lee:2014mza,Lee:2015fea} (see also \cite{Bartels:2015aea}).\footnote{The NPTF has also been applied to the problem of determining the point-source contribution to the extragalactic gamma-ray background~\cite{Lisanti:2016jub}.}
A similar approach to the NPTF that has also been widely used is the one-point fluctuation analysis or one-point PDF method, see for example~\cite{Lee:2008fm,Feyereisen:2015cea,Zechlin:2015wdz,Zechlin:2016pme,Feyereisen:2016fzb,Zechlin:2017uzo,Manconi:2019ynl,Calore:2021bty}.
In the present work we will focus on comparisons between the CPG and the NPTF, but many of our conclusions would be similar if we compared to these alternative approaches.

Recent investigations have suggested that the NPTF results must be interpreted carefully, raising the possibility that the nature of the GCE has not yet been conclusively resolved.
Firstly, \textcite{Leane:2019xiy} demonstrated that the NPTF could attribute an injected dark-matter signal to the point-source model, although it appears this concern can be addressed.
In particular, an improved treatment of the background models largely resolves the issue \cite{Buschmann:2020adf}, and further, as was emphasised in \textcite{Chang:2019ars}, a degree of confusion is unavoidable given the inherent degeneracy between purely Poisson emission and a population of point sources that produce at most one photon each in the data set.
Nevertheless, when performing a Bayesian analysis, confusion between point source and pure Poisson emission can be exacerbated by a poor choice of priors, a point we will return to in Sec.~\ref{sec:priors}.
An additional concern regarding the NPTF that has not yet been addressed, is that it appears the presence of an unmodelled asymmetry in the data can significantly bias the method, in that the asymmetry leads the NPTF to return strong evidence for a point-source population, even when there is none \cite{Leane:2020pfc,Leane:2020nmi}.
Taken together, the above results emphasise that the output of the NPTF must be interpreted cautiously --- indeed, whether the GCE contains the first hints of the particle nature of dark matter remains an open problem, that even more recent wavelet~\cite{Zhong:2019ycb} and Machine Learning~\cite{List:2020mzd} approaches have not conclusively resolved.

Nevertheless, as yet, the modelling of the detector effect correction function of NPTF, $\rho(f)$, has not been questioned, despite the heuristic justification for its original inclusion in the method.
In this work we will show that there are instances where the NPTF construction explicitly breaks down as a result of the mathematical construction of $\rho(f)$, and that this represents an obstacle to extending the method to X-ray data sets.
In detail, we present a first principles construction of the parametric likelihood in Eq.~\ref{eq:non_parametric_marginal}, which incorporates the correction of detector effects in a statistically justified manner.
The result, which we call the Compound Poisson Generator, includes a new detector effect correction function that is an analytic expression of the basic components of instrumental effects: the PSF, effective area, and photon detection probability.
The new construction demonstrates that the spatial distribution of the point-source population cannot be disentangled from the detector effect correction function, nor can the PSF be disentangled from the effective area.
This shows that the NPTF -- which factorises the spatial distribution, PSF, and effective area -- cannot describe point-source statistics in general. 
To show this lack of generality, the NuSTAR X-ray telescope is used as a test case for both the CPG and NPTF.
The substantial differences between the detector response in NuSTAR and \emph{Fermi} will reveal the stated deficiencies in NPTF.

As for the effect of unmodelled asymmetries, these occur when the spatial distribution for an emitter is specified incorrectly.
As NPTF and CPG have additional explanatory power above that of a simple Poisson model, they will produce a better fit to the data even if the emission is truly diffuse.
This same effect would be observed when using probabilistic cataloguing, as the location of the sources will migrate to explain the deviation from the specified spatial distribution. 
The CPG likelihood does not address this issue; as it is, in fact, not an issue with the point-source model at all.
To see this, consider the following analogy with statistical mechanics.
We would be surprised to observe a box where all gas molecules are located in one corner, as this situation is a very unlikely macrostate for a gas.
On the other hand, from the perspective of the microstate description of the gas, the gas is in as likely a configuration as any other configuration including those where the system is more thoroughly mixed.
Probabilistic cataloguing, NPTF and CPG all specify the likelihood of the data, which is a microstate description of the population.
Thus, a configuration where all sources are on one side of the image is as likely as any other, and these methods make no distinction.

The problem here lies in the diffuse model that these population models are compared to.
The introduction of an unmodelled asymmetry will have only a small effect on the population likelihood; in comparison, the likelihood for the diffuse model will suffer greatly due to this mismodelling.
When the two models are compared, it appears that the population model is erroneously describing sources, but it is the diffuse model that is erroneously rejecting the diffuse hypothesis.
To resolve the issue, the spatial distribution must be given additional degrees of freedom so that the diffuse model can account for deviations from the expected spatial distribution.
Much like in statistical mechanics, care can be taken by examining the macrostate of the fitted population.
The analogous quantities to macrostates in point-source inference are the population parameters.
The differential source-count function can be adjusted to include, for example, an $N_{\text{top}}$ and $N_{\text{bottom}}$ for the top and bottom of the image.
The ratio of these means can then be computed, and an unlikely value for this ratio will form a diagnostic signal for a mismodelling issue.

\section{Point-source population and Instrument Models\label{sec:ppi}}

This section describes the point-source population model that will be used in this investigation, as well as the instrument model that is necessary to generate simulated observations.
The instrument model is also needed in Sec.~\ref{sec:cpg} to incorporate a detector correction into the likelihood.

\subsection{Population Model}

In the parametric approach, an assumption must be made to select a model that describes the population of sources. 
Here, we describe the population using a differential source-count function: $\d N/\d F$.
This function describes the number of sources, $N$, as a differential over the individual point-source flux, $F$.

The point-source flux $F$ is defined as a normalisation factor in a power-law flux energy spectrum:
\begin{equation}
    \frac{\d \Phi}{\d E} = F \left(\frac{E}{E_0}\right)^{-\gamma}, \label{eq:energy_spectrum}
\end{equation}
where $\Phi$ is the number of photons per unit area and time, $E$ is the photon energy, $E_0$ is the power-law scale, and $\gamma$ is the power-law index (also known as the photon index).
As a result, the dimensions of $F$ are photons per unit area, time and energy. 
Power law spectra are common for the high-energy tails of X-ray sources \cite{Fleishman:2007mc,Mukai:2017qww,Hong:2016qjq} -- due to various processes such as synchrotron emission and inverse Compton scattering -- and thus make a natural choice for this investigation.
A power law index of $\gamma = 1.5$ is used for all scenarios, as spectra with this index are common near the Galactic Center \cite{Hong:2016qjq}, and the power-law scale is set to $E_0 = 1$ keV.

The contribution of each source to the image is determined by converting the flux of the individual source to an expected number of photons based on the response of the detector.
For a telescope, this response is commonly specified in terms of the exposure time, $t_{\text{exp}}$, which is the time the instrument spent collecting the flux of interest, and further, the effective area, $\mathcal{A}_{\text{eff}}$, which is the collecting area of an equivalent idealised telescope that detects all incident photons, so that the effective area is strictly less than the actual size of the real instrument.
The mean number of detected photons (also called the mean number of counts), $S$, is then
\begin{equation}
    S = t_{\text{exp}} \int_{E_i}^{E_f}\d E\,  \mathcal{A}_{\text{eff}}(E) \frac{\d \Phi}{\d E}, \label{eq:counts_def}
\end{equation}
where the number of counts is defined within an energy band of interest, $E \in [E_i, E_f]$.
As the only model parameter in this equation is $F$ -- through $\d \Phi/\d E$ -- this expression can be simplified to $S = F \kappa$, where $\kappa$ is called the detector response.
Generally speaking, the exposure time and effective area are position dependent, so this is a function of position in the image, $\v{x}$:
\begin{equation}
    \kappa(\v{x}) = t_{\text{exp}}(\v{x}) \int_{E_i}^{E_f}\d E\, \mathcal{A}_{\text{eff}}(E, \v{x}) \left(\frac{E}{E_0}\right)^{-\gamma}.
\end{equation}

Equations~\ref{eq:energy_spectrum} and \ref{eq:counts_def} are not necessary choices for the methods employed in this investigations.
All that is required, is for the number of counts, $S_{\v{x}}$, received at location $\v{x}$ to be able to be specified in terms of the individual source model parameter $F$, and another quantity $\kappa(\v{x})$.
The value of $\kappa(\v{x})$ need not even be based on an assumed energy spectrum, it must only be known and calculable for any location $\v{x}$ and satisfy the relation $S_{\v{x}} = \kappa(\v{x}) F$.

As reviewed above, for a single source the key parameter dictating how many photons we expect to observe is the flux, $F$.
When studying a population of sources, these fluxes can vary between the individual sources.
The distribution of fluxes is described with a differential source-count function, $\d N/\d F$, which encodes the number of sources with flux between $F$ and $F + dF$.
For the investigations in this article, the differential source-count function is assumed to be a broken power-law.
A singly-broken power-law has the following form,
\begin{equation}
    \frac{\d N}{\d F} = A \begin{cases} 
        \left(\frac{F}{F_{b(2)}}\right)^{-n_2} & F \leq F_{b(2)} \\ 
        \left(\frac{F}{F_{b(2)}}\right)^{-n_1} & F_{b(2)} < F
        \end{cases},
\label{eq:singlebreak-PL}
\end{equation}
where $A$ is a normalisation factor, $F_{b(2)}$ is the location of the break in flux, $n_1$ is the power index after the break, and $n_2$ is the index before the break.
The generalisation of this distribution to multiple breaks is discussed in Sec.~\ref{sec:priors}.
The ultimate goal is then to infer the $\d N/\d F$ model parameters, denoted in aggregate by the vector $\v{\theta}$, from the statistics of the number of counted photons received by the telescope or detector.
To be explicit, for the parameterisation given in Eq.~\ref{eq:singlebreak-PL}, $\v{\theta} = \{A, F_{b(2)},n_1,n_2\}$.

Generally, the differential source-count function can be posed in terms of a distribution, $p(F)$, which gives the probability density of an individual source having flux, $F$, through the relation
\begin{equation}
    \frac{\d N}{\d F} = N p(F),
\label{eq:dndf_def}
\end{equation}
where $N$ is the mean number of sources.
This representation is more useful when describing the statistical process underlying point sources.

Another common representation is the cumulative source-count function (also called simply the source-count function):
\begin{equation}
    N(F' > F) = \int_F^\infty \frac{\d N}{\d F'} \d F', \label{eq:source_count_def}
\end{equation}
which specifies the number of sources in the population with a flux greater than $F$.
If the population is large in spatial extent, this may be written as the areal source-density function, $n(F' > F)$, often in numbers of sources per steradian or arc-minute squared.

Power-law type source-count functions are common for astrophysical populations generally, and for X-ray emitter populations specifically \cite{Mukai:2017qww,Hong:2016qjq}. 
This is not unexpected, as the inverse square reduction in apparent brightness with distance will give most populations a power-law like distribution in the observed brightness.
More generally, power-law distributions are common in nature (see, for example, the Pareto distribution) and are the maximum entropy distributions for a logarithmic parameter with a specified mean.

\subsection{X-ray Instrumentation}

\begin{figure*}
    \centering
    \includegraphics[page=8]{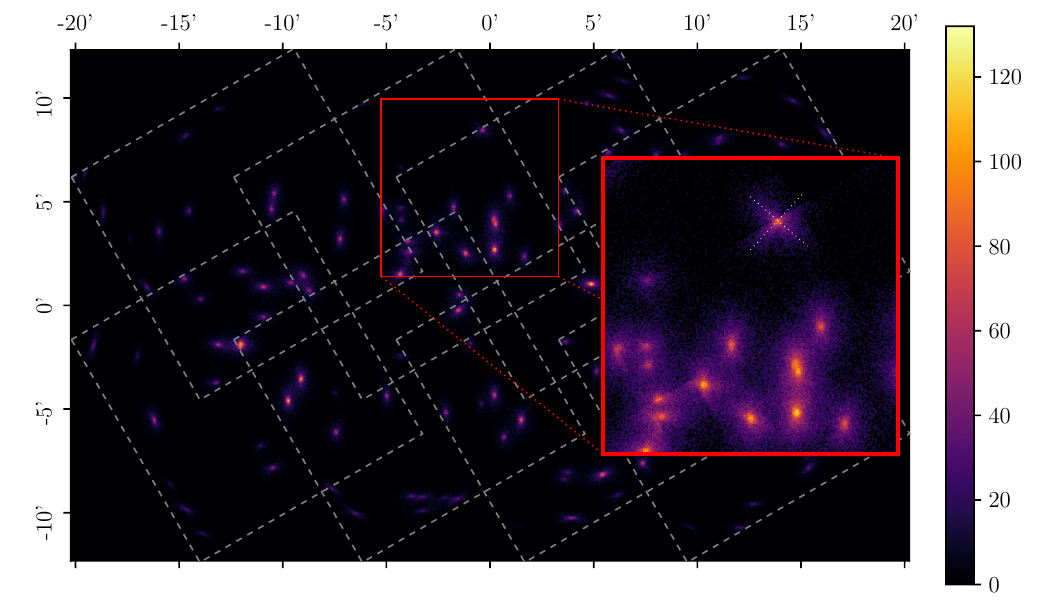}
    \caption{Multiple simulated observations are formed into a mosaic for the final image. The colour scale shows the total number of detected photons per bin. The grey dashed lines show the individual observations that comprise the mosaic. The axes are centered around an arbitrary sky location in units of arcminutes. The red inset shows the detail of a few sources in this image. The colour scale of the inset has been scaled by a square root to improve contrast. For one particular source within the inset, the white dotted lines are each perpendicular to the optical axes of the nearby observations, and show the direction along which the PSF lies for that observation. The simulation details are the same as Sec.~\ref{sub:nptf:nustar}, except that the density of sources is $5$ times less, the average flux per source $600$ times greater, and the bin size is $5$ times smaller than the NuSTAR detector pixelisation.}
    \label{fig:sim_obs}
\end{figure*}

Although the $P(D)$ method has been used extensively to study X-ray sources \cite{Miyaji:2001dp}, the current leading parametric inference method, NPTF, has yet to be applied to this regime.
X-ray astronomy poses a number of novel complications -- not present in gamma-ray and neutrino astronomy -- to the application of parametric point-source inference.
Overcoming these challenges is one of the central aims of this work.

The NuSTAR telescope \cite{Harrison:2013md,Wik:2014boa,Madsen:2015jea,madsenObservationalArtifactsNuSTAR2017}, in particular, possesses most of these complications; for this reason, NuSTAR is used as detector model to explore the effect that these complications have on parametric point-source inference.
As compared to gamma-ray and neutrino astronomy, the NuSTAR X-ray data set presents the following unique challenges:
\begin{itemize}
    \item NuSTAR, like most X-ray telescopes, has a far narrower field of view (FOV) of $12$ arcmin per side, with hard edges due to the use of a detector plane with focusing optics. \emph{Fermi} and IceCube have a FOV of approximately $40$ degrees and the full sky respectively;
    \item Although NuSTAR has a much narrower angular resolution, the NuSTAR PSF is a larger fraction of the FOV compared to \emph{Fermi} and IceCube;
    \item The NuSTAR PSF varies significantly as a function of position on the detector plane within a given observation; and
    \item To compensate for the narrow FOV, multiple observations are often compiled together into a  mosaic. This creates a complex and discontinuous detector response, with multiple overlapping PSFs for each observation. While the instrument response of both \emph{Fermi} and particularly IceCube do vary across the sky, the variation is significantly smoother than common for X-ray data sets.
\end{itemize}

In order to study all of these effects, we have developed a detector simulation suite for NuSTAR.
This simulation injects point sources into an binned map that forms a image.
In this investigation, the bin sizes are much larger than the pixelisation\footnote{For the remainder of this article, pixel will be used to refer to the physical detector pixelisation of the NuSTAR detector, while bin will refer the aggregation of pixels according to a scheme chosen by the analyser. For the investigations presented here, the bins are twenty or more times larger than the physical pixelisation.} of the NuSTAR X-ray detectors, and so the effect of this intrinsic pixelisation on the image binning is not considered here. 
The NuSTAR effective area -- including vignetting effects -- and PSF are incorporated into the simulation, and can be individually altered or simplified to assess the effect of each on the performance of the point-source methods investigated here.
The simulation also requires a spatial distribution for the point-source population, as well a $\d N/\d F$ function as defined by Eq.~\ref{eq:dndf_def}. Further details on NuSTAR and the simulation procedure are given in App.~\ref{app:nustar_sim}.

An example image generated by the simulation is shown in Fig.~\ref{fig:sim_obs}.
This image demonstrates many of the complications arising from the NuSTAR telescope.
Multiple observations are stacked into a mosaic, and the boundaries between the observations, shown by grey dashed lines, cause discontinuities in the images of nearby point-sources.
In addition, the anisotropic PSF of NuSTAR causes complicated PSF shapes for sources near these boundaries, as shown in the red inset.
The white dotted lines within the inset highlight one particular source that lies on the boundary of an observation.
Two observations contribute to the image of this source, and so two PSF shapes (along the dotted lines) combine to create a highly irregular PSF shape.

The X-ray data will contain emission from contributions other than point sources. In particular, we expect both detector backgrounds to populate the collected data, as well as smooth astrophysical emission associated with, for instance, the cosmic X-ray background.
We collectively refer to these non-point source flux contributions as diffuse emission.
The total diffuse emission will have an associated spatial map which specifies the mean expected flux at each location, which we can then combine with the detector response -- as we did for the point-source flux $F$ -- to determine the mean predicted diffuse counts at each location.
Simulation of these contributions is then performed by generating a map that is a draw from a Poisson distribution that is associated with the mean value of the diffuse map.

\section{Derivation of the CPG likelihood\label{sec:cpg}}

We now turn to towards the central goal of this paper, the construction of the CPG likelihood.
The construction of the likelihood will heavily involve the use of probability generating functions and functionals, as we will exploit their useful properties for constructing compound distributions.
Generating functions are an alternative representation of a discrete probability distribution over non-negative integers.
Suppose we have a distribution $P(k)$, which determines the probability of observing an integer $k$, the generating function is then defined by the Z-transform of the distribution:
\begin{equation}
    G(z) = \E[z^k] = \sum_{k=0}^\infty P(k) z^k. 
\end{equation}
For example, the generating function for the Poisson distribution with mean $\lambda$ is
\begin{equation}
    \sum_{k=0}^\infty \frac{\lambda^k}{k!} e^{-\lambda} z^k = e^{\lambda(z-1)}.
\end{equation}

The generating function approach will be convenient for a number of reasons, however, a central property that we will exploit is as follows. Consider forming a sum of $M$ independent and identically distributed random variates, each of which has a generating function $G_1(z)$, however with $M$ itself an independent random variate, with its own generating function $G_2$. Then the generating function for the sum is given by $G_2(G_1(z))$, which follows directly from the expectation value definition of the generating function and the law of total expectation.
For the problem at hand, this will arise in the context of counting the total number of counts detected from a population of point sources, for which the number of sources is itself a random variable.

We will build up the full CPG likelihood over the course of several steps, where each part describes the distributions and generating functions for a major concept in the construction. Specifically, we divide the discussion as follows.
\begin{itemize}
    \item[\ref{sub:cpg:single}]\hspace{-0.1cm}: Generating function for a single point-source.
    \item[\ref{sub:cpg:multi}]\hspace{-0.1cm}: Generating function for multiple sources.
    \item[\ref{sub:cpg:detector}]\hspace{-0.1cm}: Definition of the detector effect correction.
    \item[\ref{sub:cpg:likelihood}]\hspace{-0.1cm}: Calculation of the single bin likelihood from the full generating function.
    \item[\ref{sub:cpg:whole_lh}]\hspace{-0.1cm}: Extending the single bin likelihood to an image.
    \item[\ref{sub:cpg:multimodel}]\hspace{-0.1cm}: Incorporating multiple population and diffuse emission models.
\end{itemize}

The goal of this section is to build an intuition for the moving parts of this construction. A more direct, but abstract, construction is provided in App.~\ref{app:gen}, along with further discussion of the properties of generating functions and functionals. In that appendix we also show the unbinned likelihood, as well as the likelihood that accounts for the correlations point sources induce between neighboring bins, as well as a discussion on why these correlations must be ignored for computational reasons in these demonstrations.

\begin{figure*}
    \centering
     \begin{subfigure}[t]{0.495\textwidth}
         \centering
            \includegraphics[page=1]{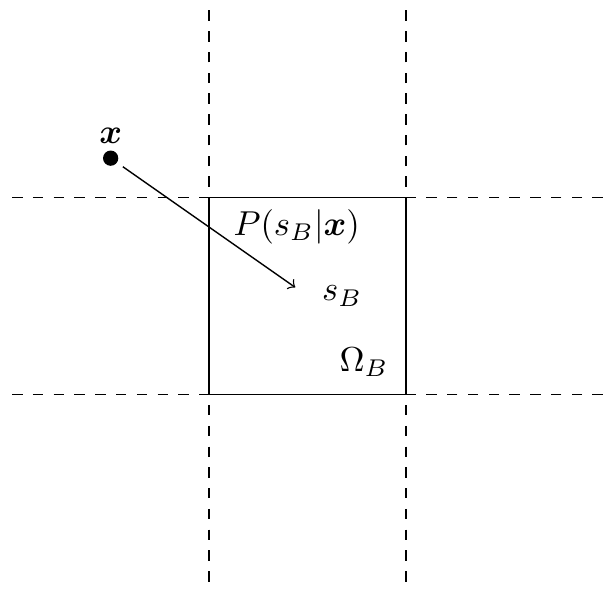}
         \caption{A single source located at $\v{x}$ contributes $s_\bin$ counts to bin $\bin$ as defined by the bin extents $\Omega_\bin$ (solid square). The number of counts is determined by the Poisson distribution $P(s_\bin | \v{x})$ as defined in the text.}
         \label{fig:terms:single_source}
     \end{subfigure}
     \hfill
     \begin{subfigure}[t]{0.495\textwidth}
         \centering
            \includegraphics[page=2]{figures/diagrams.pdf}
         \caption{Multiple sources located at $\{\v{x}_l\}$ each contribute $\{s_{i\rightarrow\bin}^{(j)}\}$ counts to bin $\bin$ creating a total of $k_\bin$ counts in the bin.}
         \label{fig:terms:multi_source}
     \end{subfigure}
     \begin{subfigure}[t]{0.495\textwidth}
         \centering
            \includegraphics[page=3]{figures/diagrams.pdf}
         \caption{A Poisson process describes sources spatially distributed according to $T(\v{x})$ (sketched by shaded region). Each source may contribute multiple photons to the bin, which are recorded as counts at locations $\{\v{y}_j\}$.}
         \label{fig:terms:poisson_process}
     \end{subfigure}
     \hfill
     \begin{subfigure}[t]{0.495\textwidth}
         \centering
            \includegraphics[page=4]{figures/diagrams.pdf}
         \caption{A single photon created by a source at $\v{x}_1$ is converted to an expected number of counts using the following detector effects: the detector response (effective area and exposure time), $\kappa(\v{x}_1)$, that is evaluated at the source location; the PSF distribution $\phi(\v{y}_2 | \v{x}_1)$ evaluated at the location where the photon lands, $\v{y}_2$, and conditioned on the source location; the detector efficiency, $\eta(\v{y}_2)$, that is evaluated where the photon lands.}
         \label{fig:terms:detector_effects}
     \end{subfigure}
    \caption{A relational diagram of the terms defined and used in Sec.~\ref{sec:cpg} and App.~\ref{app:gen}, demonstrating the relationship between the terms graphically.}
    \label{fig:terms}
\end{figure*}

\subsection{Single Source Generating Function\label{sub:cpg:single}}

To begin with, consider the emission from a single source located at a position $\v{x}$.
The number of detected photons from a time-integrated observation of that source follows a Poisson distribution.
This is a result of the exponential nature of the inter-detection time distribution, and the statistical independence of multiple detections conditioned on the mean number of detected photons, $S_\bin$.
Here the index $\bin$ denotes the spatial bin in which the counts are detected.
Continuing, let $P_P(s_\bin | S_\bin)$ be the probability mass function for the Poisson distribution describing $s_\bin$ detected photons with mean $S_\bin$.
Then given the true spatial location of the source, $\v{x}$, the probability mass function for $s_\bin$ at this location is
\begin{equation}
    P(s_\bin | \v{x}) = \int \d S_\bin P_P(s_\bin|S_\bin) p(S_\bin | \v{x}).
\label{eq:pSb1}
\end{equation}
Here we introduced $p(S_\bin | \v{x})$, which is the probability distribution for the mean, $S_\bin$, for a given source location of $\v{x}$, i.e. given a point source at $\v{x}$, it is the probability that it produces a mean number of counts $S_\bin$ in the bin $\bin$.
As the exact value of $S_\bin$ is unknown, we marginalise over it in the above expression.
The exact configuration is shown in Fig.~\ref{fig:terms:single_source}, where a source at location $\v{x}$ contributes $s_\bin$ detected photons via the distribution $P(s_\bin | \v{x})$ to bin $\bin$, which has a spatial size $\Omega_\bin$.

We can move from the expected number of counts from a source, $S_{\bin}$, to the physical flux $F$, by combining a conditional distribution $p(S_\bin|F,\v{x})$ and a distribution over source flux, $p(F)$, as follows,
\begin{equation}
    p(S_\bin | \v{x}) = \int \d F p(S_\bin|F, \v{x})\, p(F).
\label{eq:pSb2}
\end{equation}
This construction explicitly assumes that the flux distribution is isotropic, such that $p(F | \v{x}) = p(F)$, an assumption which holds when the sources are themselves identically distributed -- a key property that will be needed in the next section, and we will assume throughout.
The flux distribution itself has already been defined: it is given by the differential source-count function, $\d N/\d F$, in Eq.~\ref{eq:dndf_def}.
The conditional distribution $p(S_\bin|F, \v{x})$, however, will depend centrally on the detector effect correction -- as discussed already, the conversion from flux to counts intimately depends on the detector response -- and it will determine the amount of flux $F$ that contributes to the bin $\bin$.
For the moment we will leave it unspecified, postponing a definition until Sec.~\ref{sub:cpg:detector}.

Combining Eq.~\ref{eq:pSb1} and \ref{eq:pSb2}, we have
\begin{equation}\begin{aligned}
P(s_\bin | \v{x}) = &\int \d S_\bin P_P(s_\bin|S_\bin) \\
\times &\int \d F p(S_\bin|F, \v{x})\, p(F)\,,
\end{aligned}\end{equation}
which fully specifies the probability of observing $s_\bin$ counts from a single point source in terms of the differential source-count function, in $p(F)$, the instrumental response, in $p(S_\bin|F, \v{x})$, and of course the Poisson distribution, $P_P(s_\bin|S_\bin)$.
From this we can immediately determine the generating function for a single source located at a given position $\v{x}$, $G_{s_\bin|\v{x}}(z)$,
\begin{equation}\begin{aligned}
    G_{s_\bin | \v{x}}(z) &= \E_F\left[ \E_{S_\bin|F,\v{x}}\left[ \E_{s_\bin|S_\bin}[z^n] \right] \right] \\
        &= \int \d S_\bin e^{S_\bin(z-1)} \int \d F p(S_\bin|F, \v{x}) p(F), \label{eq:source_gen}
\end{aligned}\end{equation}
which follows as the generating function for the Poisson distribution $P_P(s_\bin|S_\bin)$, that is $\E_{s_\bin|S_\bin}[z^n] = e^{S_\bin(z-1)}$.

\subsection{Multiple Source Generating Function\label{sub:cpg:multi}}

We next extend the discussion to account for the emission detected from a population of sources.
Statistically, the locations of these point sources follows a spatial Poisson point process.
This follows from the observation that:
\begin{itemize}
    \item The number of sources can be any non-negative integer.\footnote{In reality, there will be a physical upper bound to the number of sources one can find in most populations. Nevertheless, the number is typically sufficiently large that an unbounded process is an excellent approximation.}
    \item The occurrence of each source is independent from other sources. The presence of a point source does not have an effect on the probability of a new source forming.
    \item Two sources never occupy the same spatial location, an infinitesimal area on the sky has only zero or one sources in it.
\end{itemize}
Spatial clustering of sources may violate the last two assumptions. If the clustering is known in advance (e.g., it is known that the sources cluster around the galactic center), then the construction presented here will render the source locations independent by conditioning on the spatial distribution of the sources. Some clustering processes cannot be made conditionally independent in this way; e.g., if the presence of one source increases the probability of further sources forming nearby. In this case, the construction presented here does not apply for counting sources. Instead, the method can count entire clusters as single sources, and the flux distribution, $p(F)$, would describe the flux of clusters. The size of this effect will depend on the degree to which the spatial distribution assumption is violated by the clustering. As an explicit example, the presence of binary emitters would violate these assumptions, and the CPG will count binary systems, rather than sources.\footnote{For this caveat to be relevant, both objects must be emitters that can be detected by the instrument. If only one object in a binary is an emitter, the presence of the other is irrelevant to these assumptions.}

To proceed, let $T(\v{x})$ denote the spatial distribution for the location of a source (or system) in the population.
In terms of $T(\v{x})$, often referred to as the spatial template of the point sources, the source intensity function is $N T(\v{x})$, so that the integral of the source intensity function is equal to the mean number of sources, $N$.\footnote{We leave the region over which $T(\v{x})$ is normalised unspecified. In general it need not correspond to the region within which the analysis is being performed. For example, when studying an extragalactic source-class, it may be convenient to normalise $T(\v{x})$ over the full sky, even if this is larger than the region over which the data is collected.}
In more detail, $T(\v{x})$ carries units of $[\text{sr}^{-1}]$, so that $N T(\v{x})$ has dimension $[\text{sources }\text{sr}^{-1}]$.

Now, let the template
\begin{equation}
    T_\bin = \int_{\Omega_\bin} \d\v{x}\, T(\v{x})
\end{equation}
be the fraction of the spatial distribution in bin $\bin$ with bin extents $\Omega_\bin$.
We will use this to construct a generating function that accounts for sources located in surrounding bins which contribute flux to the current bin.

We define $k_{i\rightarrow\bin}$ as the number of counts that all sources located in bin $i$ contributes to bin $\bin$.
As there can be more than one source in bin $i$, $k_{i\rightarrow\bin}$ is a sum of $M_{i}$ random variates,\footnote{Here we are calculating the likelihood exclusively for the bin $\bin$, and as such $M_i$ is formally the number of sources in bin $i$ that contribute counts to $\bin$. Accordingly, $M_i$ is implicitly defined as a random variable in the context of calculating the likelihood for $\bin$. As this particular approach to the construction does not take into account correlations, there is a separate and independent random variate of $M_i$ for each $\bin$.} $s_{i\rightarrow\bin}^{(j)}$:
\begin{equation}
    k_{i\rightarrow\bin} = \sum_{j=1}^{M_{i}} s_{i\rightarrow\bin}^{(j)},
\end{equation}
as shown in Fig.~\ref{fig:terms:multi_source}.

Now, note that the spatial distribution for a source, when conditioned on it being located in bin $i$, is
\begin{equation}
    T(\v{x}|i) = \begin{cases} \frac{T(\v{x})}{T_i} & \v{x} \in \Omega_i \\ 0 & \text{otherwise.} \end{cases}
\end{equation}

Then, each $s_{i\rightarrow\bin}^{(j)}$ is identically distributed according to the generating function $G_{s_{i\rightarrow\bin}}$. This generating function is formed by taking the expectation of the generating function\footnote{The index $j$ does not parameterise this generator, as each $s_{i\rightarrow\bin}^{(j)}$ is a specific realisation of the random variate $s_{i\rightarrow\bin}$.} $G_{s_{i\rightarrow\bin}|\v{x}}$ over $T(\v{x}|i)$ --- the spatial distribution conditioned on the source being located in bin $i$:
\begin{align}
    G_{s_{i\rightarrow\bin}}(z)
        &= \int \d\v{x}\,T(\v{x}|i) G_{s_{i\rightarrow\bin}|\v{x}}(z) \\
        &= \int_{\Omega_i} \d\v{x}\,\frac{T(\v{x})}{T_i} G_{s_{i\rightarrow\bin}|\v{x}}(z) \\
        &= \int \d S_{i\rightarrow\bin}\, e^{S_{i\rightarrow\bin}(z-1)} \nonumber \\ &\times \int \d F\, p_{i\rightarrow\bin}(S_{i\rightarrow\bin}|F) p(F),
        \label{eq:GsiB}
\end{align}
where
\begin{equation}
    p_{i\rightarrow\bin}(S_{i\rightarrow\bin}|F) = \int_{\Omega_i} \d\v{x}\, \frac{T(\v{x})}{T_i} p(S_{i\rightarrow\bin}|F,\v{x})
\end{equation}
is now the instrument response for bin $i$ contributing to bin $\bin$. This object can be intuitively thought of as the average response in bin $\bin$, from sources located in bin $i$, marginalised over the probability of a source appearing at any specific location within bin $i$.

As emphasised already, the number of sources, $M_{i}$, is itself a random variate distributed according to a Poisson distribution with mean $N T_i$, recalling that $N$ is the mean number of sources for the total population.
Thus, the generating function for $M_{i}$ is
\begin{equation}
    G_{M_{i}}(z) = e^{N T_i (z-1)}.
\end{equation}

Now, the generating function for the sum $k_{i\rightarrow\bin}$ is the composition of the generating functions for $M_{i}$ and $s_{i\rightarrow\bin}^{(j)}$:
\begin{equation}
    G_{k_{i\rightarrow\bin}}(z) = G_{M_{i}}\left(G_{s_{i\rightarrow\bin}}(z)\right).
\end{equation}
The total number of counts in bin $\bin$, denoted by $k_\bin$, is then the sum of all the $k_{i\rightarrow\bin}$ across all bins:
\begin{equation}
    k_\bin = \sum_i k_{i\rightarrow\bin}.
\end{equation}
Note that this sum must range over the full domain of the template, $T_i$.
Primarily, this ensures that the recovered value of $N$ is correctly normalised to the template, but it can also be understood to ensure that the contribution from any source -- no matter where it may be in the spatial distribution of the population -- is counted correctly, as the PSF generally allows contributions to $k_\bin$ from anywhere in the spatial distribution.

Thus, the generating function for $k_\bin$ is the product of the generating functions for $k_{i\rightarrow\bin}$:
\begin{equation}\begin{aligned}
    G_{k_\bin}(z)
        &= \prod_i G_{k_{i\rightarrow\bin}} \\
        &= \exp{\left[ N \left( \left[ \sum_i T_i G_{s_{i\rightarrow\bin}}(z) \right] - 1\right) \right]},
\end{aligned}\label{eq:Gkb_sub}\end{equation}
as $\sum_i T_i = 1$.

At this stage, we could substitute into this expression $G_{s_i\to\bin}(z)$ as given in Eq.~\ref{eq:GsiB}.
Before doing so, however, let us capture into a single function the detector response as it would appear in Eq.~\ref{eq:Gkb_sub},
\begin{align}
    p(S_\bin | F) &= \sum_i T_i p_{i\rightarrow\bin}(S_{i\rightarrow\bin}|F) \\ 
        &= \int \d\v{x}\, T(\v{x}) p(S_\bin|F,\v{x}).
\end{align}
Using this, we can write the full generating function for $k_\bin$ as 
\begin{widetext}
\begin{equation}
    G_{k_\bin}(z) = \exp{\left[ N \left(\int \d S_\bin\, e^{S_\bin(z-1)} \int \d F\, p(S_\bin | F) p(F) - 1  \right) \right]}. \label{eq:cpg_def}
\end{equation}
\end{widetext}
The distribution for $k_\bin$ is known as a compound Poisson distribution \cite{IntroductionTheoryPoint2003}, and so $G_{k_\bin}$ is a compound Poisson generator.
As shown in App.~\ref{app:gen}, this compound Poisson distribution actually describes a compound Poisson point process.
The process gives the probability density of finding a detected count at location $\v{y}$ as shown in Fig.~\ref{fig:terms:poisson_process}.
Eq.~\ref{eq:cpg_def} generates the probability of finding $k_\bin$ such detected counts, $\{\v{y}_1,\ldots,\v{y}_{k_\bin}\}$, anywhere in $\Omega_\bin$.

\subsection{Detector Effect Correction\label{sub:cpg:detector}}

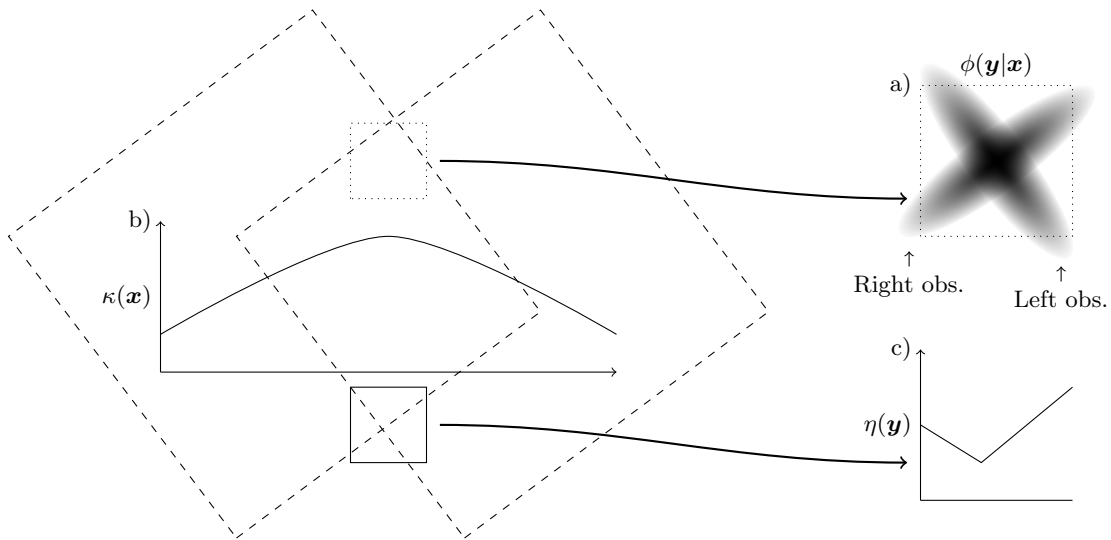
\begin{figure*}
    \centering
    \begin{tikzpicture}[]
        \begin{scope}[cm={1.0,0,0,1.0,(1.5,4.5)}]
            \draw[dotted] (0,0) rectangle ++(1,1);
            \coordinate (psf) at (1, 0.5);
            \node[rotate=36.9] at (0.5, 0.5) {\tikz{\shade[path fading=fade out] (0,0) ellipse (0.8 and 0.2)}};
            \node[rotate=-53.1] at (0.5, 0.5) {\tikz{\shade[path fading=fade out] (0,0) ellipse (0.8 and 0.2)}};
        \end{scope}
        \begin{scope}[cm={1.0,0,0,1.0,(0,0)}]
            \draw[dashed] (0,0) -- ++(4, 3) -- ++(-3, 4) -- ++(-4, -3) -- cycle;
        \end{scope}
        \begin{scope}[cm={1.0,0,0,1.0,(3,0)}]
            \draw[dashed] (0,0) -- ++(4, 3) -- ++(-3, 4) -- ++(-4, -3) -- cycle;
        \end{scope}
        \begin{scope}[cm={1.0,0,0,1.0,(1.5,1)}]
            \draw (0,0) rectangle ++(1,1);
            \coordinate (bin) at (1, 0.5);
        \end{scope}
        \begin{scope}[cm={1.0,0,0,1.0,(9,0.5)}]
            \draw[->] (0,0) -- ++(0, 2) node[midway, left] {$\eta(\v{y})$};
            \draw (0,0) -- ++(2, 0);
            \draw plot coordinates { (0,1) (0.8,0.5) (2, 1.5)};
            \coordinate (etaplot) at (0, 0.5);
            \node[left] at (0, 2) {c)}; 
        \end{scope}
        \begin{scope}[cm={1.0,0,0,1.0,(-1,2.2)}]
            \draw[->] (0,0) -- ++(0, 2) node[midway, left] {$\kappa(\v{x})$};
            \draw[->] (0,0) -- ++(6, 0);
            \draw plot [smooth] coordinates { (0,0.5) (3,1.8) (6, 0.5)}; 
            \node[left] at (0, 2) {b)};
        \end{scope}
        \begin{scope}[cm={1.0,0,0,1.0,(9,4)}]
            \draw[dotted] (0,0) rectangle (2, 2);
            \coordinate (zoomplot) at (0, 0.5);
            \node[rotate=36.9] at (1.0, 1.0) {\tikz{\fill[black, path fading=fade out] (0,0) ellipse (1.6 and 0.4)}};
            \node[rotate=-53.1] at (1.0, 1.0) {\tikz{\fill[black, path fading=fade out] (0,0) ellipse (1.6 and 0.4)}};
            \node[above] at (1, 2) {$ \phi(\v{y} | \v{x}) $};
            \draw[->] (-0.15, -0.4)-- ++(0, 0.2) node[at start, below] {Right obs.};
            \draw[->] (1.85, -0.6) -- ++(0, 0.2) node[at start, below] {Left obs.};
            \node[left] at (0, 2) {a)};
        \end{scope}
        \draw[thick, ->, shorten >=5pt, shorten <=5pt] (bin) to[out=0, in=180] (etaplot);
        \draw[thick, ->, shorten >=5pt, shorten <=5pt] (psf) to[out=0, in=180] (zoomplot);
    \end{tikzpicture}
    \caption{Two separate observations with dashed boundaries form a mosaic. \textbf{a)} In areas where the observations overlap, the PSF will be an appropriate mixture of the PSF for each observation. Near the edge of the observation, we show an example of how the PSF can be highly distorted in the radial direction, as expected for NuSTAR observations. The mixture forms a cross with the PSF from the left observation lying diagonally NW-SE, and the PSF from the right observation lying diagonally SW-NE. \textbf{b)} The detector response, $\kappa(\v{x})$, is a sum of the response for each observation, and varies smoothly across the mosaic as the response extends outside the observation boundaries. \textbf{c)} In contrast, the detector efficiency, $\eta(\v{y})$, has a sharp transition from unity to zero at the observation boundary, and will generally be non-smooth across any bin that straddles these boundaries.}
    \label{fig:sim}
\end{figure*}

In the discussion thus far, we have determined the generating function associated with a population of sources, specified by a differential source-count function.
The final aspect of Eq.~\ref{eq:cpg_def} we have avoided confronting is the instrument response, which we turn to now.
Recall, we codified the instrumental effects as follows,
\begin{equation}
    p(S_\bin | F) = \int \d\v{x}\, T(\v{x}) p(S_\bin|F, \v{x}),
    \label{eq:Sb-marginal}
\end{equation}
where, as defined above, $T(\v{x})$ is the spatial point-source template, and $p(S_\bin|F, \v{x})$ accounts for how the instrument converts a flux $F$ from a source at location $\v{x}$ into an expected number of counts in bin $\bin$.
In general, the instrument response and the spatial template cannot be fully factorised, as they are in the NPTF approach to the problem.
We will see this explicitly in the discussion that follows.

Our treatment will consider four separate detector effects: the exposure time, which converts flux to time-integrated flux; the effective area, which converts time-integrated flux to expected photons incident on the detector; the PSF, which gives the probability density for the deviation of a photon's recorded direction of arrival from its true incident direction; and the detector efficiency, which gives the probability of a single incident photon being detected.
Both the exposure time and effective area are merged into a single detector response value, $\kappa(\v{x})$ which converts flux to mean counts for a point-source at location $\v{x}$, and was discussed in Sec.~\ref{sec:ppi}.
The PSF is a probability density, $\phi(\v{y} | \v{x})$, for a count detected at location $\v{y}$ conditioned on its parent point-source at location $\v{x}$.
The detector efficiency, $\eta(\v{y})$ is a probability of a count being detected conditioned on the location of detection $\v{y}$.
The geometry of each function is shown in Fig.~\ref{fig:terms:detector_effects}, and examples of how these quantities combine for mosaiced images is shown in Fig.~\ref{fig:sim}.

To give a concrete example of these terms in the context of NuSTAR (\emph{Fermi}, IceCube): a source is located in the direction of $\v{x}$ and emits a primary X-ray ($\gamma$-ray, neutrino) from the direction of $\v{x}$.
This primary interacts with the optics (detector, ice) and produces a secondary X-ray (electron, lepton) that is scattered in the direction of $\v{y}$ according to the PSF $\phi(\v{y}|\v{x})$.
This secondary interacts with the active detector causing a count to be recorded with probability $\eta(\v{y})$.

In terms of these individual detector responses, the expected number of incident photons produced by a point-source at location $\v{x}$ with flux $F$ is $S_{\v{x}, F} = \kappa(\v{x}) F$,
and then the mean detected photon density at location $\v{y}$ will be $S_{\v{x}, F}(\v{y}) = \eta(\v{y}) \phi(\v{y} | \v{x}) S_{\v{x}, F}$.
As a result, the mean number of detected photons in bin $\bin$ is
\begin{equation}
    S_\bin = \int_{\Omega_\bin} \d\v{y}\, S_{\v{x}, F}(\v{y}),
\end{equation}
and the associated distribution, $p(S_\bin | F, \v{x})$, must be
\begin{equation}
    \hspace{-0.08cm}p(S_\bin | F, \v{x}) = \delta\left(S_\bin - F \kappa(\v{x}) \int_{\Omega_\bin} \d\v{y} \eta(\v{y}) \phi(\v{y} | \v{x}) \right),
\end{equation}
which then determines the marginal form as given in Eq.~\ref{eq:Sb-marginal}.

Direct substitution of this result into Eq.~\ref{eq:cpg_def} is not immediately helpful, as it results in a double integral over spatial coordinates that must be evaluated during the computation of the likelihood.
Instead, all of the detector effects can be encoded into a single measure, which we denote $\mu_\bin(\varepsilon)$, over an effective detector response variable, $\varepsilon$, such that $S_\bin = \varepsilon F$.
In particular, we define
\begin{equation}
    \mu_\bin(\varepsilon) = \int d\v{x} T(\v{x}) \delta\left(\varepsilon - \kappa(\v{x}) \int_{\Omega_\bin} \hspace{-0.1cm}\d\v{y} \eta(\v{y}) \phi(\v{y} | \v{x}) \right). \label{eq:mu_eps_def}
\end{equation}
In terms of the new measure, we can restate Eq.~\ref{eq:Sb-marginal} as
\begin{equation}
    p(S_\bin | F) = \int \d\varepsilon \mu_\bin(\varepsilon) \delta(S_\bin - \varepsilon F).
\end{equation}
Then, substitution of this form of $p(S_\bin | F)$ into Eq.~\ref{eq:cpg_def} reframes the CPG as
\begin{align}
    &G_{k_\bin}(z) = \label{eq:cpg_mueps} \\ &\exp{\left[ N \left(\int \d F \int \d\varepsilon e^{\varepsilon F (z-1)} \mu_\bin(\varepsilon) p(F) - 1  \right) \right]}, \nonumber
\end{align}
which only involves integrals over the effective detector response $\varepsilon$ and source flux $F$.

Importantly, the detector correction function, $\mu_\bin(\varepsilon)$, can be pre-calcuated thus saving considerable computation time during likelihood evaluation.
As it is rare to have closed form expressions for $\kappa(\v{x})$ and $\phi(\v{y} | \v{x})$ in most experiments, $\mu_\bin(\varepsilon)$ will almost always need to be numerically estimated.
This estimation can proceed by effectively evaluating the integrals over $\v{x}$ and $\v{y}$ in Eq.~\ref{eq:mu_eps_def} through Monte-Carlo integration.
Samples are drawn from $T(\v{x})$, and then -- conditioned on these values of $\v{x}$ -- samples a drawn from $\phi(\v{y} | \v{x})$.
The values of $\v{y}$ are accumulated to create a value of $\varepsilon$, and the resulting $\varepsilon$ values are histogrammed to create a density estimate over $\varepsilon$ --- which is $\mu_\bin(\varepsilon)$.
This process, and an explicit algorithm for construction $\mu_\bin(\varepsilon)$, is detailed in App.~\ref{app:mu_eps}.
As emphasised at the outset, the detector response involves the source template intimately.
This is distinct to the handling of the detector response in NPTF, which occurs through the function $\rho(f)$, and we will explore the differences between these two approaches in Sec.~\ref{sec:nptf}.

\subsection{Calculation of the Single-Pixel Likelihood\label{sub:cpg:likelihood}}

Once $\mu(\varepsilon)$ has been constructed, evaluation of the CPG in Eq.~\ref{eq:cpg_mueps} requires the integrals over $\varepsilon$ and $F$ to be performed.
As $\mu(\varepsilon)$ is numerically constructed, the integral over $\varepsilon$ will also be performed numerically.
The remaining integral over $F$ may then be performed numerically, or analytically, if the assumed form of $p(F)$ is amenable to such treatment.
For the examples in this investigation, $p(F)$ is assumed to follow a broken power-law distribution, in which case the evaluation can be performed analytically as detailed in App.~\ref{app:series}.
In this section, the only assumption required is that this evaluation produces a power series in $z$:
\begin{equation}
    G_{k_\bin}(z) = \exp{\left[\sum_{m=0}^\infty \frac{a_\bin^{(m)}}{m!} z^m \right]}. \label{eq:exp_series}
\end{equation}
This is a fairly mild assumption, as it only requires that the expression within the square brackets of Eq.~\ref{eq:cpg_mueps} be an analytic function of $z$. This is easily satisfied when the moments of $p(F)$ and $\mu_\bin(\varepsilon)$ are finite.
For now, this power series will be assumed to be infinite in order; later, it will be shown that only a finite order is required in practice.

The goal is to find $P(k_\bin)$, the probability of measuring $k_\bin$ detected photons, from this generating function.
Recall that a generating function is defined as
\begin{equation}
    G(z) = \sum_{k=0}^\infty P(k) z^k. \label{eq:pwr_series}
\end{equation}
The relationship between the power series in Eq.~\ref{eq:exp_series} and the power series in Eq.~\ref{eq:pwr_series} is given by the Bell polynomials \cite{comtetAdvancedCombinatorics1974}
\begin{equation}
    \exp{\left[\sum_{m=0}^\infty \frac{a_\bin^{(m)}}{m!} z^m \right]} = \sum_{k=0}^\infty \frac{B_k(a_\bin^{(0)},\ldots,a_\bin^{(k)})}{k!} z^k. \label{eq:pk_relation}
\end{equation}
Thus, by inspection we have in this case $P(k_\bin) = B_k(a_\bin^{(0)},\ldots,a_\bin^{(k)})/k!$, and note that from Eq.~\ref{eq:pwr_series}, for a bin with $k$ counts, only the first $k$ terms of the power series in Eq.~\ref{eq:exp_series} need to be calculated.
The evaluation of the Bell polynomials can be performed using recurrence relations, as we demonstrate in App.~\ref{app:bell}.

\subsection{Whole Image Likelihood\label{sub:cpg:whole_lh}}

Through Eq.~\ref{eq:pk_relation} and the results preceding it, we have achieved our aim of writing the single-bin likelihood, $P(k_\bin | \v{\theta})$, where $\v{\theta}$ are the model parameters for the population, as encoded in $\d N/\d F$.
A likelihood for the whole image, $\mathcal{I} = \{ k_\bin : \forall \bin\}$, can be constructed as a simple product over the bins,
\begin{equation}
    P(\mathcal{I} | \v{\theta}) = \prod_{k_\bin \in \mathcal{I}} P(k_\bin | \v{\theta}). \label{eq:whole_image_lh_def}
\end{equation}
Importantly though, this construction does not take into account the correlations between the bins, which may be induced by the PSF.
Indeed, it assumes that the $k_\bin$ are statistically independent, which will be not be true unless the PSF is a delta-function distribution. Such a delta-distribution ensures that sources at the edge of a pixel only deposit flux in a single bin.

The effect of this broken assumption is that the resulting posterior distribution on $\v{\theta}$ will be narrower than the true posterior if these correlations were accounted for --- by treating every pixel as independent, we have assumed there is more available information than is actually present in the image.
This will underestimate the uncertainty on the model parameters, as sources which overlap bins are effectively counted as multiple independent observations of the same source, instead of the true single observation --- a similar effect to double counting data.
The degree to which the posterior is narrowed will depend on the chosen bin size, as smaller bins -- relative to the PSF -- will be more strongly correlated.
The test cases in Sec.~\ref{sec:nptf} show that this is not a significant effect for the bin sizes chosen in this investigation, which are several times larger than the PSF.

In principle there are other ways correlations between pixels can be induced that would invalidate the factorization in Eq.~\ref{eq:whole_image_lh_def} --- for instance an incorrect template for one of the Poisson models, or perhaps other instrumental effects.
The reason we single out the PSF is that it is never truly a delta-function distribution in any real instrument, and this represents the most significant correlation that can only be mitigated by appropriate choice of bin sizes.

While this is an unambiguous deficiency of the above derivation, so far no computationally feasible method has been proposed to account for the correlations in a binned analysis.
Indeed, the most obvious extensions of the above construction require the computation of every possible combination of how a source can distribute its counts to all bins in the image.
As stated, in the present work we will work with a binning such that this shortcoming is suppressed, but we caution that for a general binning the biases associated with this effect must be considered.

\subsection{Multiple Emission Models\label{sub:cpg:multimodel}}

It is a common occurrence for an image to have contributions from both populations of point-sources as well as purely Poisson emission or background components. As such, it is important to be able to accommodate this reality in the likelihood, as we do so in this subsection.
Let $k_{\bin,j}$ be the number of counts that population $j$ contributes to bin $\bin$, and $\varpi_{\bin,l}$ be the number of counts that Poisson component $l$ contributes to the same bin.
Then the total number of counts in this bin is simply a sum over the contribution from each component,
\begin{equation}
    K_\bin = \sum_j k_{\bin,j} + \sum_l \varpi_{\bin,l}.
\end{equation}

In turn, the generating function for the combined emission, $K_\bin$, is
\begin{equation}
    G_{K_\bin}(z) = \left(\prod_j G_{k_{\bin,j}}(z) \right) \left( \prod_l G_{\varpi_{\bin,l}}(z) \right).
\end{equation}
For point-source populations, the generating function is as derived earlier in this section, whereas $G_{\varpi_{\bin,l}}(z) = \exp{[\lambda_{\bin,l}(z-1)]}$ is the generating function for $\varpi_{\bin,l}$ and
\begin{equation}
    \lambda_{\bin,l} = \int_{\Omega_\bin} \d\v{y}\, I_l(\v{y}),
\end{equation}
is the integral of the intensity function $I_l(\v{y})$ for Poisson component $l$ in bin $\bin$, which parameterises the mean of $\varpi_{\bin,l}$.
Recall that a single point source in the population is specified by a flux $F$, which carried dimensions of $[\text{photons }\text{cm}^{-2}\text{ s}^{-1}]$.
In comparison, the Poisson diffuse-emission component is an extended source and has a differential flux of $F_P T_P(\v{x})$ with dimensions $[\text{photons }\text{cm}^{-2}\text{ s}^{-1}\text{ sr}^{-1}]$.
Here $T_P(\v{x})$ is the template for the Poisson emission, which may or may not be the same as the spatial distribution of the source population, $T(\v{x})$.
It does, however, have the same units of $[\text{sr}^{-1}]$, so that $F$ and $F_P$ will also carry the same dimensions.
In terms of these quantities, the Poisson intensity is given by
\begin{equation}
    I(\v{y}) = \eta(\v{y}) \int \d\v{x} \phi(\v{y} | \v{x}) \kappa(\v{x}) F_P T_P(\v{x}),
\end{equation}
as photons from the diffuse component are also scattered by the PSF.
The mean can then be written more compactly as
\begin{equation}
    \lambda_{\bin,l} = F_{P,l} \int \d\varepsilon~ \varepsilon~ \mu_{\bin,l}(\varepsilon),
\end{equation}
where $F_{P,l}$ is the flux for Poisson component $l$, and $\mu_{\bin,l}$ is the detector correction function for template $T_{P,l}$.
As for the point-source population, we envision the spatial template as fixed, which leaves a single model-parameter for the emission, $F_P$.

Combining these results, the generating function for $K_\bin$ can be written in the same form as Eq.~\ref{eq:exp_series}:
\begin{widetext}
\begin{equation}
    G_{K_\bin}(z) = \exp{\left[ \left(\sum_j a_{\bin,j}^{(0)} - \sum_l \lambda_{\bin,l}\right) + \left(\sum_j a_{\bin,j}^{(1)} + \sum_l \lambda_{\bin,l} \right) z + \sum_{m=2}^\infty \left( \sum_j \frac{a_{\bin,j}^{(m)}}{m!} \right) z^m \right]},
    \label{eq:full_cpg}
\end{equation}
\end{widetext}
and the same likelihood evaluation method of Sec.~\ref{sub:cpg:likelihood} can be employed, thereby completely specifying the CPG likelihood.

\section{Biases Induced by Common Prior Parameterisations\label{sec:priors}}

In this section, the priors on the population model parameters -- necessary for a Bayesian analysis -- are discussed.
Our focus will be to demonstrate that poorly chosen priors on a common combination of the population flux and background flux can lead to misleading posterior distributions.
We further introduce a set of priors where these issues can be reduced, and advocate for their use more generally in population studies.

Let us make a general point at the outset. In Bayesian analyses, one is free to choose any set of priors. Priors can be adopted which reflect a preference towards either hypothesis. The Poisson hypothesis may be preferred for its simplicity, or alternatively one may wish the results to reflect an underlying bias towards the point-source model given that in many situations it is known that unresolved point-sources must be present. Whatever set of priors is adopted, however, the preference they reflect should be considered. As we will show in this section, taking simple priors on the parameters that describe the point-source model can induce a complex bias in the question of which model is generating the flux. The priors we will introduce instead place the Poisson and point-source models on equal footing at the outset, and if not adopted directly, at the very least represent a starting point for adopting a principled set of priors for Bayesian point-source inference.

For the remainder of this investigation, we will restrict the model under question to at most one population of point sources and one Poisson component, that both share a common template, i.e. $T(\v{x}) = T_P(\v{x})$.
Realistic analyses are more complicated than this restriction; for instance, existing NPTF \emph{Fermi} analyses involve two (or three) point-source population models and multiple Poisson components, while the NPTF IceCube analysis used one population with multiple Poisson components.
However, common to both analyses was the use of a point-source population model and Poisson component with an identical spatial distributions.
This is the situation where the particular bias we will discuss can emerge, justifying our restriction.

Fundamentally, such a scenario arises when the underlying nature of flux distributed according to $T(\v{x})$ is unknown, and the question of interest is whether it is due to a measurable population of sources, purely diffuse emission, or instead a mixture of the two.
By using a common template for the two possible emission sources, the flux could be assigned to either, or some fraction to both.
For example, in the case of \emph{Fermi}, the fundamental question was determining whether the anomalous flux at the Galactic Center was attributable to a population of sources, such as millisecond pulsars, or instead to diffuse dark-matter emission that would be Poisson distributed.
For the example of IceCube, the goal has been to determine whether a measurable fraction of the astrophysical neutrino flux can be assigned to a population of sources.

To begin the investigation, the differential source-count function must be parameterised in order to evaluate the power series terms, $a_\bin^{(m)}$ as given in Eq.~\ref{eq:full_cpg}, and accordingly the CPG likelihood.
Note that while we will use the CPG likelihood in order to demonstrate the potential prior bias, the effects we reveal are equally applicable to other point-source likelihoods such as the NPTF.
The use of the CPG also furnishes us with an example to demonstrate the application of the likelihood.
In the \emph{Fermi} and IceCube analyses, the following standard form was used for the source-count distribution:
\begin{widetext}
\begin{equation}
    \frac{\d N}{\d F} = A \begin{cases}
            \left[\prod\limits_{j=2}^m \left( \frac{F_{b(j+1)}}{F_{b(j)}} \right)^{-n_j}\right]\left(\frac{F}{F_{b(m)}}\right)^{-n_m} & F \in [0, F_{b(m)}] \\
            \vdots & \\
            \left[\prod\limits_{j=2}^i \left( \frac{F_{b(j+1)}}{F_{b(j)}} \right)^{-n_j}\right]\left(\frac{F}{F_{b(i)}}\right)^{-n_i} & F \in (F_{b(i+1)}, F_{b(i)}] \\
            \vdots & \\
            \left(\frac{F}{F_{b(2)}}\right)^{-n_2} & F \in (F_{b(3)}, F_{b(2)}] \\
            \left(\frac{F}{F_{b(2)}}\right)^{-n_1} & F > F_{b(2)} 
        \end{cases}. \label{eq:full_broken_plaw_def}
\end{equation}
\end{widetext}
This defines a broken power-law with $m-1$ breaks in flux, specified by $F_{b(m)}, \ldots, F_{b(2)}$; $m$ power-law indices, parameterised by $n_m,\ldots,n_1$; and a scale factor $A$, that is related to the expected number of sources in the population.
The break parameters have the same units as $F$, while $A$ has units of sources per inverse units of $F$.
Note this is a generalisation of the singly broken power-law given in Eq.~\ref{eq:singlebreak-PL}.

With this parameterisation, \emph{Fermi} and IceCube analyses -- see, for example, Refs.~\cite{Lee:2015fea} and \cite{Aartsen:2019mbc}, respectively -- have used a Bayesian inference framework.
In both cases, uniform priors on the indices, $n_i$, and log-uniform priors on $A$ were chosen.
For the \emph{Fermi} analysis, the flux breaks, $F_{b(2)}$, and Poisson component flux $F_P$ were given uniform and log-uniform priors, respectively, whereas in the IceCube analysis, $F_{b(2)}$ was given a log-uniform prior, while $F_P$ was given a uniform prior.

\subsection{A Sketch of Poor Prior Parameterisation}

Let us firstly outline where the issues associated with the poor prior parameterisation originate.
For simplicity, we will concentrate on a single break differential source-count function; however, the arguments here generalise to multiple breaks.
It is important to note that the total flux of the point-source population, $F_{PS}$, is not proportional to $F_{b(2)}$, and also depends on $A$,
\begin{equation}
    F_{PS} = A F_{b(2)}^2 \left(\frac{1}{n_1 - 2} + \frac{1}{2 - n_2}\right).
    \label{eq:FPS}
\end{equation}
Already from this expression we can make the following observation: \emph{a uniform prior chosen for $F_{b(2)}$ will not, in general, uniformly weight values of $F_{PS}$ after the change in coordinates.}

Consider a situation in which this model is used to analyse data that has no distinct population of sources, such that the data is entirely consistent with a Poisson distribution.
Clearly, the model can explain this data by assigning the entirety of the flux to the Poisson component; however, this is not the only solution for this inference problem.
In particular, if we have a population of dim sources, there is a limit in which the sources are so dim that this distribution becomes indistinguishable from Poisson emission.
To provide an explicit example, if we had a population with $N$ expected sources, each of which produces $\mu$ counts on average, then the mean number of counts produced by the population is $\lambda = N \mu$.
In the limit where $\mu \ll 1$, such that all sources produce either 0 or 1 counts only, then the point-source likelihood exactly reduces to the Poisson distribution with mean $\lambda$, and the two hypotheses are formally indistinguishable in the data.
Note, that for $\lambda$ to stay finite, we require a large $N$ when $\mu \ll 1$. 
Returning to our single-break scenario, note that
\begin{equation}
    N = A F_{b(2)} \left(\frac{1}{n_1 - 1} + \frac{1}{1 - n_2}\right),
\end{equation}
and so for this point-source-Poisson degeneracy regime to be achieved here, the prior on $A$ must be sufficiently large.
But so long as it is, then in principle data associated with Poisson emission can be equally well described by the diffuse or point-source hypothesis.

Ideally, in this situation, the posterior should show complete uncertainty on $F_P$ and $F_{PS}$, with a perfect anti-correlation that corresponds to the total flux in the image.\footnote{Of course, in principle one may wish to choose a prior that reflects a preference for one of the two hypotheses. Nevertheless, this preference should be placed in the priors in a principled manner---the point we are seeking to emphasize in the present discussion is that existing priors adopted in the literature represent a complicated transformation away from the natural coordinate system we introduce, thereby introducing biases that could be unintended.}
However, if the same kind of prior is chosen for $F_{b(2)}$ and $F_P$ -- for example, both log-uniform -- then the corresponding prior on $F_{PS}$ may have a different form to the prior on $F_P$.
In this case, the posterior will show a preference to assigning flux to either the population or the Poisson component.
If the difference between the priors is substantial, essentially all of the flux will be assigned to one of the two components, despite the data having no power to distinguish them.

Unless this subtlety in choice of prior is accounted for, the result may be unexpected, potentially leading an experimenter to erroneously conclude that the data supports a population of sources where there are none, or vice-versa.
This is the bias we will explore in more detail below, and then outline priors that can be chosen such that the two hypotheses remain indistinguishable given uninformative data.

\subsection{Prior Effect Demonstration\label{sub:prior_demo}}

\begin{figure*}
    \centering
    \begin{subfigure}[t]{0.495\textwidth}
         \centering
            \includegraphics[page=1]{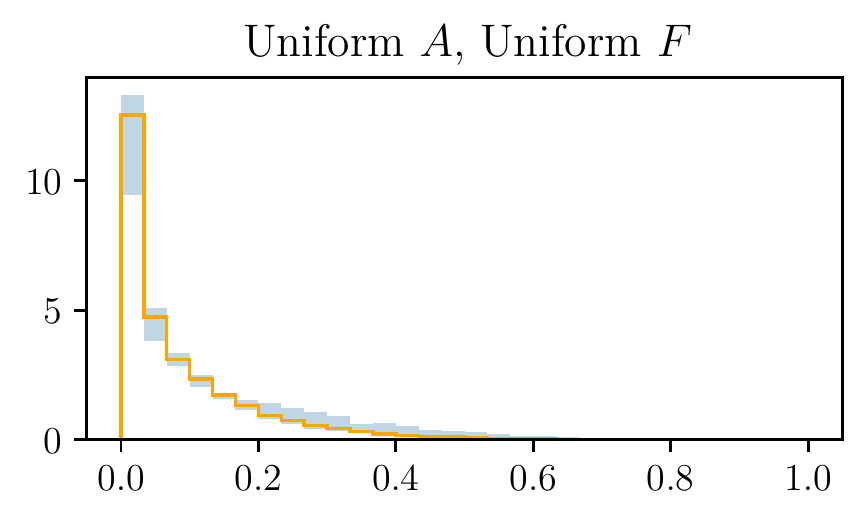}
     \end{subfigure}
    \hfill
    \begin{subfigure}[t]{0.495\textwidth}
         \centering
            \includegraphics[page=1]{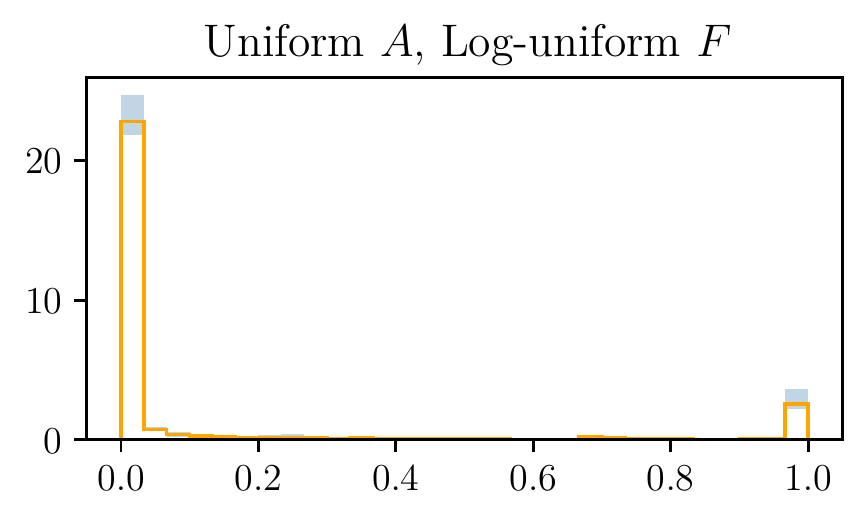}
     \end{subfigure}
    \begin{subfigure}[t]{0.495\textwidth}
         \centering
            \includegraphics[page=1]{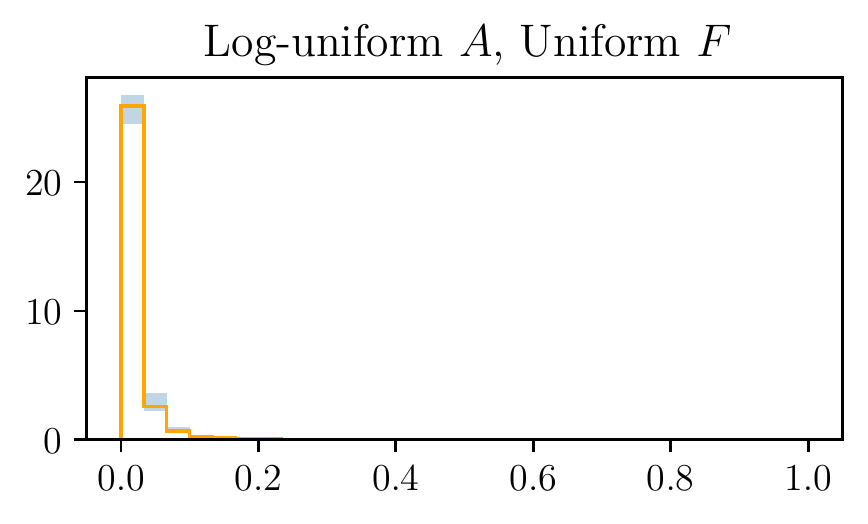}\vspace{-1em}
            \hphantom{~~~~~~}\large{$\omega$}
     \end{subfigure}
    \hfill
    \begin{subfigure}[t]{0.495\textwidth}
         \centering
            \includegraphics[page=1]{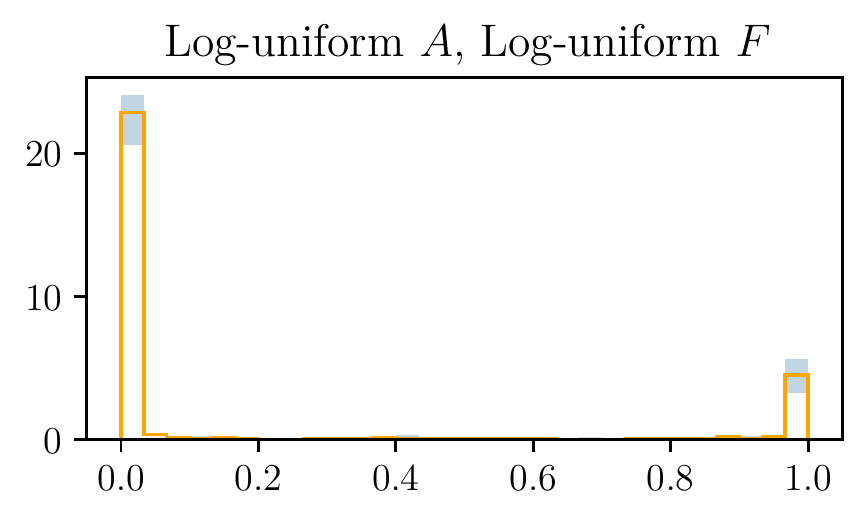}\vspace{-1em}
            \hphantom{~~~~~~}\large{$\omega$}
     \end{subfigure}
    \caption{The posterior for the fraction of flux assigned to the point-source population, $\omega = F_{PS}/(F_P + F_{PS})$, when analysing data sets generated from purely a Poisson distribution, but analysed with both a Poisson and point-source template using the CPG likelihood. The median posterior (over multiple random trials, for each bin) is shown by the solid line, while the 16\% and 84\% quantile range is shown by the fill. The four panels correspond to different prior choices,  representative of common choices in the literature. None of the choices of prior result in a uniform posterior for $\omega$, instead the posterior is biased toward assigning all flux to one model component or another, which could easily be misleading unless accounted for.}
    \label{fig:priors_omega}
\end{figure*}

In order to investigate in detail the potential bias induced by the choice of prior parameterisation as discussed in the previous subsection, we consider simulated NuSTAR data sets (for details of the NuSTAR simulation see App.~\ref{app:nustar_sim}).
For this scenario, the vignetting is disabled so that the detector response is uniform across the image.
In addition, the PSF is locked to the on-axis PSF of NuSTAR, so that the PSF does not change as a function of source location.
These simplifications are chosen to ensure that the effect of prior parameterisation is not obscured.

The spatial distribution for both the population model and the Poisson component is specified as a uniform distribution concentric with the image and with a width and height twice that of the field of view.
As this demonstration requires data that is indistinguishable from a Poisson distribution, no point-sources are injected and only a uniform Poisson background is used to generate the image.
Further details are given in Tab.~\ref{tbl:all_config} (all tables are located in the appendices).

Given this scenario, we perform a Bayesian analysis of the resulting images using the CPG, but taking four variations on the choice of prior, considering variations for the prior of the amplitude $A$ and the flux parameters $F_{b(2)}$ and $F_P$.
The same form of prior is used for $F_{b(2)}$ and $F_P$.
We consider the four combinations resulting from a uniform linear or log-uniform prior on the amplitude and flux parameter.
The detailed prior ranges are given in Tab.~\ref{tbl:priors_priors}.
We take the same lower limit for both flux priors.
The upper limit of the $F_P$ prior is extended as this parameter captures the total flux of the Poisson component, while $F_{b(2)}$ is more closely related to the average flux of a single source in the population, so we allow for a lower value.
The upper and lower limits for the priors on each parameter are identical across variations.
Uniform priors are chosen for the power-law indices, $n_i$.
Of course, the results from running point-source inference on one simulated image may not be representative, as the simulation is a Monte-Carlo procedure that may produce an outlier image.
To capture the variation in the simulation, for each choice of priors, six images are generated and a posterior was sampled for each of these trials.
The posteriors were sampled using the \texttt{emcee} Affine Invariant MCMC \cite{emcee}, discarding the first 90\% of samples as burn-in.

From the resulting posteriors, we focus our attention on a single parameter of interest, the proportion of the flux that is assigned to the population model.
In particular, we define the fraction of flux assigned to the population as $\omega = F_{PS}/(F_P + F_{PS})$.
For each of the trial images, a coordinate transformation is applied to the posterior samples to yield samples in $\omega$.
These samples are then histogrammed, and from the set of trials the median, 16\% and 84\% quantiles over the histogram bins are computed and shown in Fig.~\ref{fig:priors_omega}.
The clear trend observed here is a highly bimodal posterior in $\omega$ --- the model assigns essentially all of the flux to only the population or the Poisson component in a situation where from the perspective of the data, the two are indistinguishable.
Unless this behavior was anticipated, it could generate misleading conclusions.
Again, the data supports both the population model and the Poisson component, so one expects that the posterior on $\omega$ will be close to uniform.

Recently, in the context of application of NPTF to the \emph{Fermi} GCE, concerns have been raised on the possibility that flux from a diffuse dark-matter component could be misattributed to the point-source population model.
\textcite{Leane:2019xiy} raised these concerns in relation to mismodelling of the spatial distribution of sources, and the authors show that a strong flux misattribution effect can result from spatial mismodelling.
The authors also examine the attribution of flux when the spatial distribution is correctly modelled.
In particular, Fig.~S7 shows that even with correct modelling, some diffuse dark-matter flux is attributed to the point-source model --- although it should be clarified that the effect is considerably weaker than the spatial mismodelling effect observed in the rest of the study.
Figure~S7 also shows that when the true flux of the diffuse dark-matter component exceeds the flux of the point-source population, the flux is then attributed to the diffuse component of the model.
We can understand that this behaviour is very likely impacted by the choice of priors in that work, given their importance as demonstrated above.
In particular, \textcite{Leane:2019xiy} placed a log-uniform prior on the diffuse emission and $A$, but a uniform prior on $F_{b(2)}$.
The use of a uniform prior for the point-source model flux break will place greater weight on this model to describe the diffuse flux --- when the $\d N/\d F$ for this model can accommodate both the diffuse flux and the brighter point-source emission favoured by the data.
Accordingly, when small amounts of diffuse flux are injected we would expect it to be absorbed by the point-source model for this choice of priors.
However, once the diffuse flux is comparable or larger than the point-source flux, it becomes difficult to explain both the population \emph{and} the diffuse flux using the same power-law $\d N/\d F$.
Thus, the model reverts to attributing the diffuse flux to the Poissonian model.

The prior effect can be observed more clearly in \textcite{Chang:2019ars}.
Figure~3 of this study shows the posterior distribution for the flux of a point-source and Poissonian component for a dark-matter GCE scenario.
Although this study concludes that all flux is correctly attributed to the dark-matter component in the posterior, in fact, based on the previous arguments and Fig.~\ref{fig:priors_omega}, the unbiased posterior would uniformly assign the flux between the point-source and dark-matter components.
The observed posterior is instead likely driven entirely by the chosen priors and the constraint that the flux of the population and Poisson component must sum to the total flux in the image.

More generally, if we denote the total flux by $F_T$, so that $F_T = F_{PS} + F_P$ -- a diagonal line on the plane of $F_{PS}$ and $F_P$ -- it is clear that the only choice of flux priors that will result in the expected behaviour are ones that assign equal probability to all values of $F_{PS}$ and $F_P$ that lie on this diagonal.
Two choices of priors satisfy this requirement: either both uniform priors or both exponential priors on $F_{PS}$ and $F_P$.\footnote{As for an exponential prior, $p(F)\propto \exp{(-F)}$, so the product of such priors for $F_{PS}$ and $F_P$ is $\propto\exp{(-F_{PS}-F_P)} = \exp{(-F_T)}$.}
This may appear to suggest that a flat posterior on $\omega$ should be observed in Fig.~\ref{fig:priors_omega} for the ``Uniform F'' cases.
The non-uniform posterior in these cases comes from a second effect: although the prior on the flux break, $F_{b(2)}$, is specified to be uniform, this does not mean the prior on the total point-source flux is uniform, as emphasised below Eq.~\ref{eq:FPS} above.
Recall, the total flux is proportional to $A F_{b(2)}^2$, so a uniform prior on $A$ and $F_{b(2)}$ is effectively a $F_T^{-1}$ prior on the total flux, resulting in the observed concentration toward $\omega=0$.

A log-uniform prior on $A$ with a uniform prior on $F$ effectively generates an $N^{-2}$ prior on the mean number of sources.
Although this appears to be a uniform prior on the total flux, one must consider how the boundary of the prior transforms.
The effect of the prior boundaries can be incorporated by marginalising out $N$ from the joint prior, which results again in an $F_T^{-1}$ prior on the total flux.
Accordingly, none of the commonly existing prior choices result in the desired flat $\omega$ distributions. Of course, the particular priors existing choices achieve may be desirable in certain circumstances. Nevertheless, when considering the question of whether emission is fundamentally Poisson or point source, it is undoubtedly useful to have a system of priors that generates a posterior where both are treated equally when the data is uninformative. With this in mind in the next section we will introduce a new approach which involves reparameterising the priors entirely with a new coordinate system.

\subsection{Reparameterising Priors in a Natural Coordinate System}

\begin{figure*}
    \centering
    \begin{subfigure}[t]{0.495\textwidth}
         \centering
            \includegraphics[page=1]{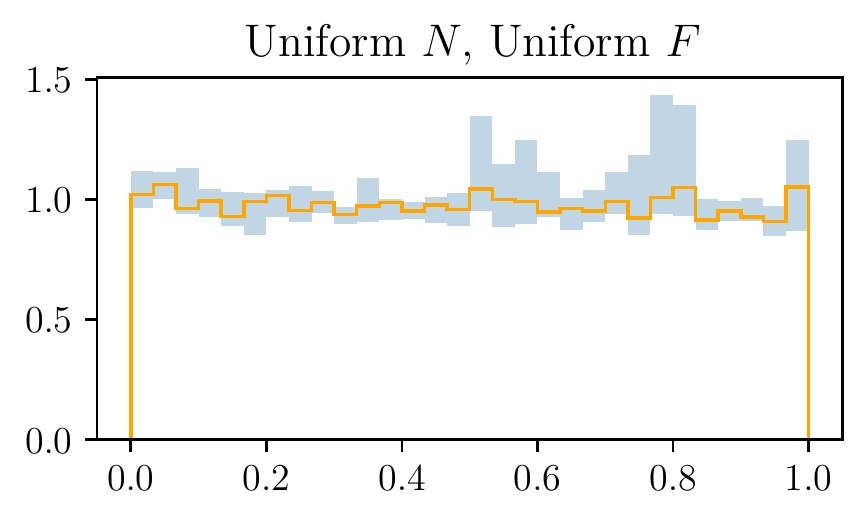}
     \end{subfigure}
    \hfill
    \begin{subfigure}[t]{0.495\textwidth}
         \centering
            \includegraphics[page=1]{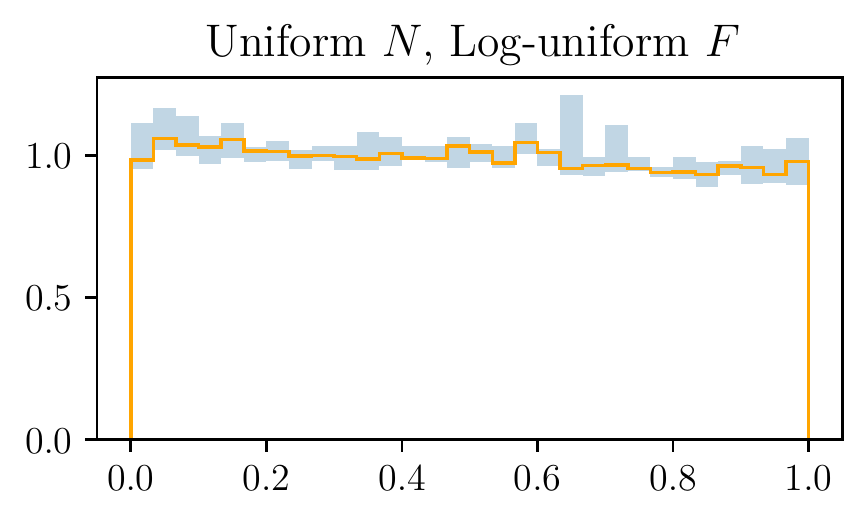}
     \end{subfigure}
    \begin{subfigure}[t]{0.495\textwidth}
         \centering
            \includegraphics[page=1]{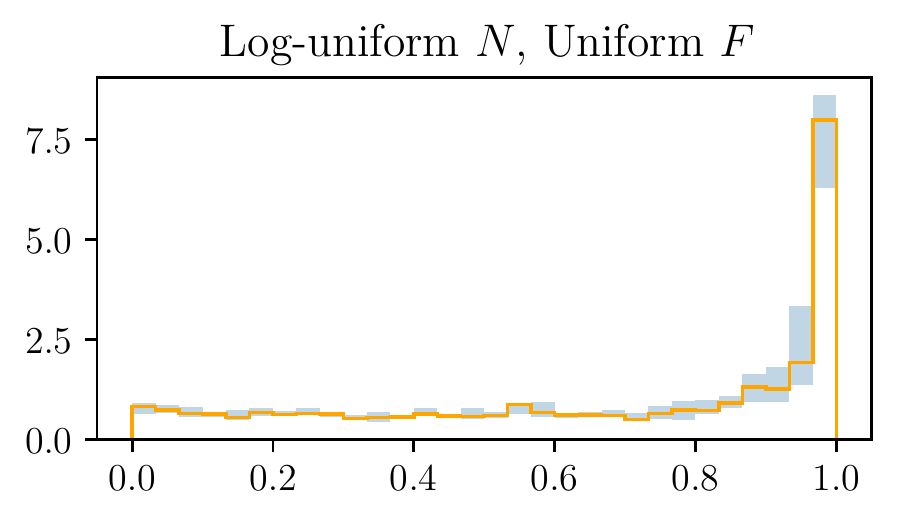}\vspace{-1em}
            \hphantom{~~~~~~}\large{$\omega$}
     \end{subfigure}
    \hfill
    \begin{subfigure}[t]{0.495\textwidth}
         \centering
            \includegraphics[page=1]{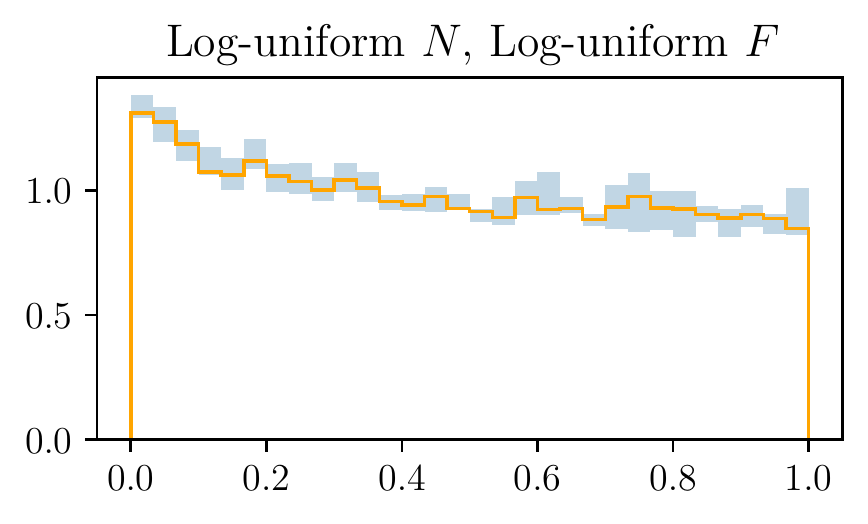}\vspace{-1em}
            \hphantom{~~~~~~}\large{$\omega$}
     \end{subfigure}
    \caption{The posterior for the fraction of flux assigned to the point-source population, $\omega = F_{PS}/(F_P + F_{PS})$, under the natural coordinate system. The median posterior (over multiple random trials, for each bin) is shown by the solid line, while the 16\% and 84\% percentile range is shown by the fill. Ideally, the posterior for $\omega$ should be uniform -- as the diffuse emission cannot be distinguished from a below-threshold point-source population -- and, unlike the standard system shown in Fig.~\ref{fig:priors_omega}, the new natural coordinate system recovers a uniform posterior; with the exception of the log-uniform $N$, uniform $F$ priors which is discussed further in the text.}
    \label{fig:natural_omega}
\end{figure*}

\begin{figure}
    \centering
    \includegraphics[page=2]{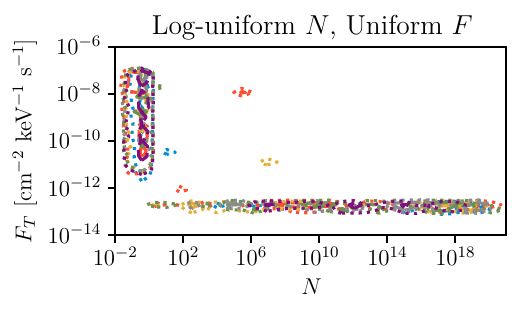}\caption{The posterior for the mean number of sources $N$, and total flux $F_T$, under the natural coordinate system for the Log-uniform $N$, Uniform $F$ prior scenario. The total flux is tightly constrained for $N>1$; however, it becomes completely unconstrained once the mean number of sources is less than one in the whole population ($N<1$).}
    \label{fig:logN_linF_NF_post}
\end{figure}

Much of the previous discussion already suggests an alternative approach.
A natural way to describe the combination of both population and Poisson component is through the coordinates of $\omega$ and $F_T$, the relative and total flux, respectively, so that the population model is specified in terms of $F_{PS}$.
From $\omega$ and $F_T$, the flux of either component is straightforward to calculate, although we caution that given the inherent degeneracy, the individual fractions must be interpreted carefully.
Nevertheless, as it is considerably easier to define and compute the likelihood using the $A$, $n_i$, and $F_{b(i)}$ coordinates, a coordinate transform is needed from $F_{PS}$.

First, the complete coordinate system for the population model must be defined.
A natural complement to $F_{PS}$ is $N$, the expected number of sources.
From this, the average flux per source can be readily determined.
Next, the position of the breaks must be defined.
For $m-1$ breaks, $m-1$ coordinates are required.
The $F_{PS}$ coordinate is necessarily involved, leaving $m-2$ remaining coordinates to specify.
These are defined as a series of fractions, $\beta_i$, that give the location of each break relative to the previous break:
\begin{equation}
    \beta_i = \frac{F_{b(i+1)}}{F_{b(i)}}.
\end{equation}
These coordinates define a system of equations from which the break locations can be solved.
The full transformation is detailed in App.~\ref{app:coords}.

Finally, the power-law indices must be specified.
These could be left as the $\{n_i\}$ in Eq.~\ref{eq:full_broken_plaw_def}; however, we take the opportunity to correct another subtle issue with the priors.
The most common choice of prior on $n_i$ is a uniform prior.
Suppose that such a uniform prior is chosen for $n_1$ in the range of $2$ to $100$.
The uniform prior assigns nearly ten times as much prior probability to $10 < n_1 < 100$ as it does to $2 < n_1 < 10$.
This is despite the fact that most observed power-law indices are less than $10$; thus, a uniform prior is contrary to our prior knowledge of these physical systems.
Instead, the index can be specified as an angle, $\psi_i$, and the index defined as $n_i = \tan{\psi_i}$.
A uniform prior on $\psi_i$ now places as much probability to $2 < n_i < 4$ as it does to $4 < n_i < 38$.
The intuition is also clear, on a log-log plot of the source-count function the prior is uniform on the angle of the line formed by the power-law.
This should not be taken as an objectively better choice, and there may be scenarios where a uniform prior on the power-law index is a preferable; yet, the uniform prior is all too often used without much consideration, and the intention here is to provide a principled alternative.

In summary, we propose defining the priors on the broken power-law with $m-1$ flux breaks, as defined in Eq.~\ref{eq:full_broken_plaw_def}, using the coordinate system $\{N, F_{PS}, \beta_2,\ldots,\beta_{m-1},\psi_1,\ldots, \psi_m\}$, rather than $\{A, F_{b(2)}, ,\ldots,F_{b(m)},n_1,\ldots,n_m\}$. A Poisson component may be added to these coordinates through the $\omega$ flux fraction parameter.
The $F_{PS}$ parameter is then replaced by a $F_T$ parameter which represents the combined flux of the point-source population and the Poisson component.
Then, during the likelihood evaluation, a coordinate transform is applied using $F_{PS} = \omega F_T$ and $F_{\text{Poiss}} = (1 - \omega) F_T$.
When considering a population model plus Poisson component, the new coordinate system we advocate for has parameters $\{\omega, N, F_T, \beta_2,\ldots,\beta_{m-1},\psi_1,\ldots, \psi_m\}$, which may be compared to the equivalent in the standard coordinate system: $\{F_{\text{Poiss}}, A, F_{b(2)}, ,\ldots,F_{b(m)},n_1,\ldots,n_m\}$.

In the next subsection we will demonstrate that within this arguably more natural coordinate system for point-source distributions, priors can be chosen where the degeneracy inherent in the physics is faithfully represented in the posteriors.

\subsection{Demonstration of Bias Removal\label{sub:priors:nat_demo}}

We consider an identical simulation scenario to that defined in Sec.~\ref{sub:prior_demo}, except we now approach it using priors defined in the natural coordinate system.
A unit uniform prior was chosen for $\omega$, uniform priors are chosen for the $\psi_i$ coordinates.
As before, we consider four prior variations, which result from considering combinations of linear or log-flat priors on $N$ and $F_T$.
Specifics are provided in Tab.~\ref{tbl:natural_priors}.

The results are shown in Fig.~\ref{fig:natural_omega}, in the same format to those in Fig.~\ref{fig:priors_omega}.
As the prior on $\omega$ is uniform, we observe that the posterior on $\omega$ is now generally uniform when the data has no preference for the population or Poisson component.

There is, however, one exception to this behavior that occurs for the combination of a log-uniform prior on $N$ and a uniform prior on $F$.
In that case, the results demonstrate a clear bias in the posterior toward assigning all of the flux to the point-source template.
The cause is a combination of effects from both priors.
The log-uniform prior on $N$ allows for small values of $N$.
In particular, when $N<1$ the probability that there will be no sources in the image becomes significant.
If there are no sources, the flux on the source population cannot be constrained, and any value on $F$ is allowed.
In such a case, if the prior on $F$ is also log-uniform, then large values of $F$ are relatively less weighted than they would be with a uniform prior, resulting in this lack of constraint having little effect on the posterior.
However, when the prior on $F$ is uniform, large values of $F$ are encouraged, and the posterior assigns significant probability to $N<1$ and $F_{PS} \gg F_P$, as shown by Fig.~\ref{fig:logN_linF_NF_post}.
This issue may be avoided in two ways: choose either a uniform prior on $N$ or a log-uniform prior on $F$, or alternatively set the lower bound of the prior on $N$ to be larger than one.

Regardless, beyond this specific case, the desired diffuse and point source degeneracy can be readily be achieved in this coordinate system, and as such we advocate for its use generally over the existing choices.

\subsection{Degeneracies between Multiple Components}

This section has concentrated on the effect of the prior parameterisation on a Poisson and point-source population component with identical spatial distributions.
Certainly one can expect similar problems if two Poisson components have identical spatial distributions, or if the spatial distributions for two point-source population components are identical, and we briefly comment on these scenarios here.

Caution should even be taken even for components that do not share a spatial distribution.
If the spatial distributions for two components -- Poisson or point source -- are similar to the degree that each distribution cannot be distinguished from each other given the available data, then we can also expect a prior effect to manifest.
Even if the spatial distributions for each component in the model are highly distinct, a degeneracy between the distributions can arise if the set of distributions is not linearly independent.
This degeneracy allows multiple solutions for the given data, and so the prior effect may also manifest.
If the spatial distributions are nearly linearly dependent -- as measured by the statistical power to distinguish them -- then they should also be considered potentially problematic.

Therefore, when constructing a model, care should be taken to avoid near linear-dependence between the spatial distributions.
If the hypothesis in question requires such linear dependence, then the prior parameterisation we have introduced provides a solution that ensures that any physical degeneracy is faithfully represented in the posterior for the flux assigned to each source.

\section{CPG Performance and Comparison with existing methods\label{sec:nptf}}

In this section, a mathematical connection is drawn between the CPG construction and previous approaches to the problem of parametric point-source inference.
In particular we will consider a number of scenarios that highlight expected problems with the existing methods, and a performance comparison with the CPG is made using simulations.
To highlight that these limitations arise from the likelihood construction, the natural coordinate system described in Sec.~\ref{sec:priors} is employed for all simulations shown here.

As the NPTF method is the current leading parametric point-source inference method in high-energy astrophysics, this comparison will focus on the essentials of how the NPTF likelihood relates to the CPG construction.
For a complete explanation of the NPTF method, we refer the reader to \textcite{Mishra-Sharma:2016gis}.
The NPTF likelihood is specified in \textcite{Mishra-Sharma:2016gis} as a generating function.
The NPTF generator, $\hat{G}$, is written as an exponential of a power series, but can be equivalently written as
\begin{align}
    &\hat{G}_{k_\bin}(z) = \label{eq:NPTF_gen} \\ &\exp{\left[ N T_\bin \left( \int \d F \int_0^1 \d f e^{\bar{\kappa}_\bin f F (z-1)} \rho(f) p(F) - 1 \right) \right]}. \nonumber
\end{align}
written in terms of
\begin{align}
    \bar{\kappa}_\bin &= \frac{1}{|\Omega_\bin|} \int_{\Omega_\bin} \d\v{x} \kappa(\v{x}), \\
    T_\bin &= \int_{\Omega_\bin} \d\v{x} T(\v{x}),
\end{align}
which are the average detector response and template in the bin of interest, respectively.

The NPTF generator in Eq.~\ref{eq:NPTF_gen} also depends on $\rho(f)$, a function introduced to correct for the PSF.
This PSF correction encodes the fraction, $f$, of point-source flux that migrates out-of or in-to a bin due to the finite angular resolution of the instrument.
To build intuition for the effect this correction function aims to address, consider a single point source that is located in a bin, $\bin$.
Not all of the flux generated with this point source will fall within $\bin$: one can find the captured flux by integrating the PSF over the extent of the bin.
The expected observed counts from the point source within $\bin$ is then the product of this flux fraction and the total expected counts from the source.
If another point source is in an adjacent bin, then the PSF can cause a fraction of the flux to leak into $\bin$.
The leakage would cause the flux of the original point source to be overestimated unless accounted for, so the PSF needs to be integrated again -- now centered somewhere in the adjacent bin -- to find the fraction of flux that leaks into $\bin$.
This fraction is again multiplied by the appropriate counts of this second source -- as determined by $\d N/\d F$ that is specified for the entire population -- so that this extra flux is properly accounted for.
The value of $\rho(f)$ is proportional to the frequency of occurrence for a point-source contributing the fraction, $f$, of flux to a bin.
The correction function is not a probability, however.
The normalisation of $\rho(f)$ is fixed so that $\int_0^1 f \rho(f) \d f = 1$, which ensures that both the flux of the population is conserved, and the number of sources in the population is not overestimated.

The statistical justification for $\rho(f)$ is not given in \textcite{Malyshev:2011zi}, where it was introduced, nor any subsequent works.
Instead, $\rho(f)$ is defined as an infinitesimal limit of a numerical estimation for the fraction of flux that a point source will contribute to a pixel after accounting for the PSF.
This procedure is constructed as a simulation, and the complete details of the algorithm are given in App.~\ref{app:rho_f}.
To provide a brief description, the algorithm places a point source somewhere on the sky, and counts associated with this source are drawn from the PSF and placed in the image bins.
All bins are then divided by the total number of simulated counts, so that each bin now contains the fraction of flux that the source contributes to that bin.
These flux fractions are then histogrammed, and the final estimate is an average over multiple repetitions of this procedure, so that the histogrammed flux fractions captures multiple possible source locations.
The final result is a numerical estimate of $\rho(f)$ that is normalised so that $\int_0^1 f \rho(f) \d f = 1$.

A direct comparison of the CPG and NPTF generating functions, as given in Eqs.~\ref{eq:cpg_mueps} and \ref{eq:NPTF_gen}, reveals that a clear difference lies in the quantities $\bar{\kappa}_\bin$, $T_\bin$, and $\rho(f)$, which only appear for the NPTF.
In the NPTF construction, the detector effects are captured in $\bar{\kappa}$ and $\rho(f)$.
One might imagine that a connection to $\mu_\bin(\varepsilon)$ of Eq.~\ref{eq:cpg_mueps} could be made through $\d\varepsilon = \bar{\kappa}_b \d f$; however, $\rho(f)$ is not specified per bin, instead it is an average over all bins.
Indeed, the CPG construction reveals that the detector effects cannot, in general, be averaged over bins in this way.

As an example, consider data that is binned irregularly (for instance bins of significantly differing size), the construction of $\rho(f)$ does not produce a correction function that represents any bin under consideration, and as a result, the likelihood does not describe the statistics of the data.
In the case of NuSTAR, the field of view for the instrument is narrow enough that a Euclidian coordinate system can be used as a good approximation to the angular sky coordinates -- allowing a regular binning.
However, instruments such as \emph{Fermi} and IceCube measure the entire sky.
As a regular tiling of a sphere is impossible, data from these experiments must be irregularly binned, usually with the use of a \texttt{HEALPix} map \cite{2005ApJ...622..759G}.
As NuSTAR does not suffer from this issue, an in-depth examination of it is outside the scope of this investigation.
However, one should not expect this to have a particularly large effect on the \emph{Fermi} GCE analysis or most of the results from the recent IceCube analysis. 
Both analyses generally prioritise the region of the sky around the Galactic Center, using templates that place the most weight in this region.
The \texttt{HEALPix} maps employed were centered on the galactic coordinate system, placing the Galactic Center at the center of the \texttt{HEALPix} maps, where the maps have the most regular tiling.\footnote{The \texttt{HEALPix} tiling is most irregular near the poles of the maps, and has greater regularity near the equator, which contains the Galactic Center when galactic coordinates are mapped onto the \texttt{HEALPix} coordinate system.}
Thus, for both analyses, the tiling is close to regular where the templates under consideration have the greatest weight. 
(One IceCube analysis considered a uniform all-sky template, for which the irregular \texttt{HEALPix} binning could produce issues.)
In the case of the IceCube analysis, the construction of the $\rho(f)$ was further weighted by the spatial template under consideration, ensuring it more closely reflected the binning in this region.
It should be noted that the irregulatity of the HEALPix binning is reduced -- both in terms of average bin shape and average number of neighbours -- for large numbers of bins; however, the potential for the regularity to be an issue must still be considered --- especially for analyses where important information is present in the poles, or where bin sizes are large.

Extending NPTF to use a per-bin PSF correction, $\rho_\bin(f)$, would address this problem, but limitations in such an NPTF-like method remain.
The primary limitation of the NPTF construction is the use of an integrated value for the spatial distribution --- the template.
The CPG construction shows that the spatial distribution and detector response cannot, in general, be factorised into two separate terms.
We will demonstrate this explicitly by considering a number of examples in the following subsections.

\subsection{Non-uniform Spatial Distribution \label{sub:nptf:nonuniform}}

\begin{figure*}
    \centering
    \includegraphics[page=6]{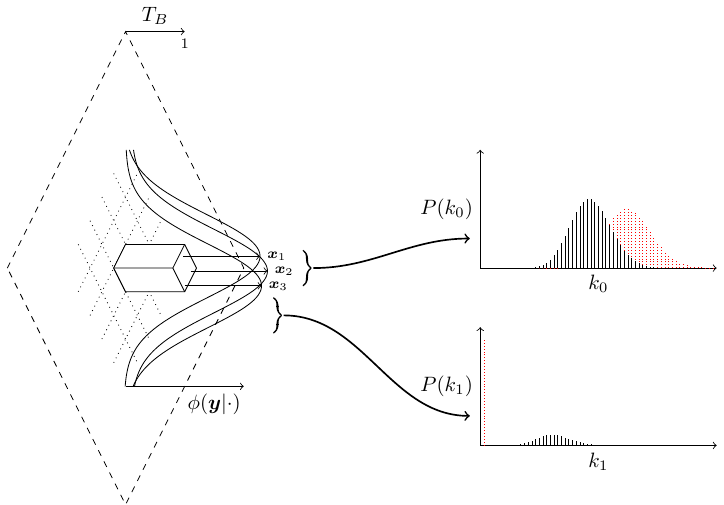}
    \caption{\textbf{Left:} A template has all of its probability concentrated in a single bin, shown as the 2D histogram on the left. As such, all three sources, $\{\v{x}_1, \v{x}_2, \v{x}_3\}$, are located within this bin. Each has an associated PSF shown as the solid distribution centered on the source location. \textbf{Above Right:} The probability distribution for the number of counts in the central bin. The distribution for conserved flux, as in the NPTF construction, is shown in dotted red. The actual distribution is shown in black, with a smaller mean. \textbf{Below Right:} The distribution for an adjacent bin. The red dotted line shows the distribution for the NPTF construction with a zero-mean, the black shows the actual distribution which has a non-zero mean.}
    \label{fig:delta_template_diag}
\end{figure*}

\begin{figure*}
    \centering
    \begin{subfigure}{0.495\textwidth}
    \begin{subfigure}{\textwidth}
         \centering
            \includegraphics[page=2]{figures/delta_template/mu_eps.pdf}
     \end{subfigure}
    \begin{subfigure}{\textwidth}
         \centering
            \includegraphics[page=2]{figures/delta_template/rho_eps.pdf}
     \end{subfigure}
    \caption{The recovered posterior for both the CPG $\mu_\bin(\varepsilon)$ construction and the NPTF $\rho(f)$ construction. The solid red line shows the (true) injected source-count function. Each coloured dashed and dotted line are the $1$-$\sigma$ and $2$-$\sigma$ HPD regions respectively for separate trials. }
    \label{fig:template}
    \end{subfigure}
    \begin{subfigure}{0.495\textwidth}
    \begin{subfigure}{\textwidth}
         \centering
            \includegraphics[page=1]{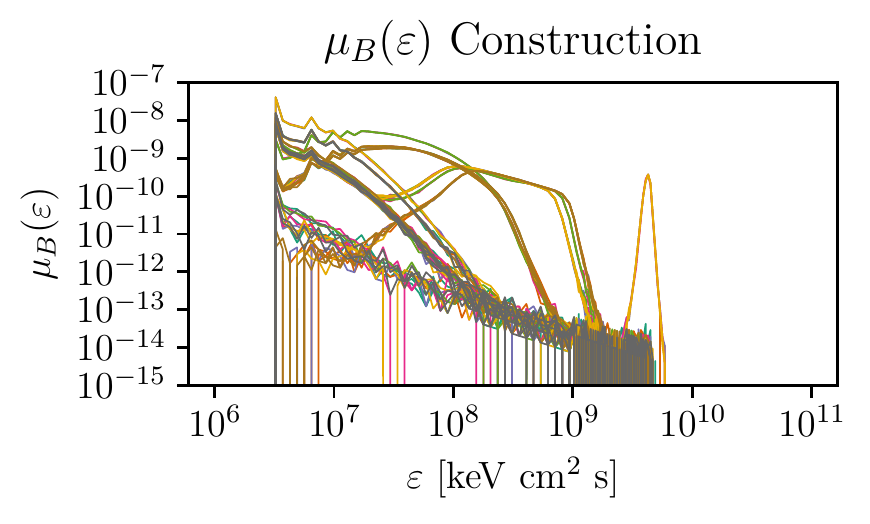}
     \end{subfigure}
    \begin{subfigure}{\textwidth}
         \centering
            \includegraphics[page=1]{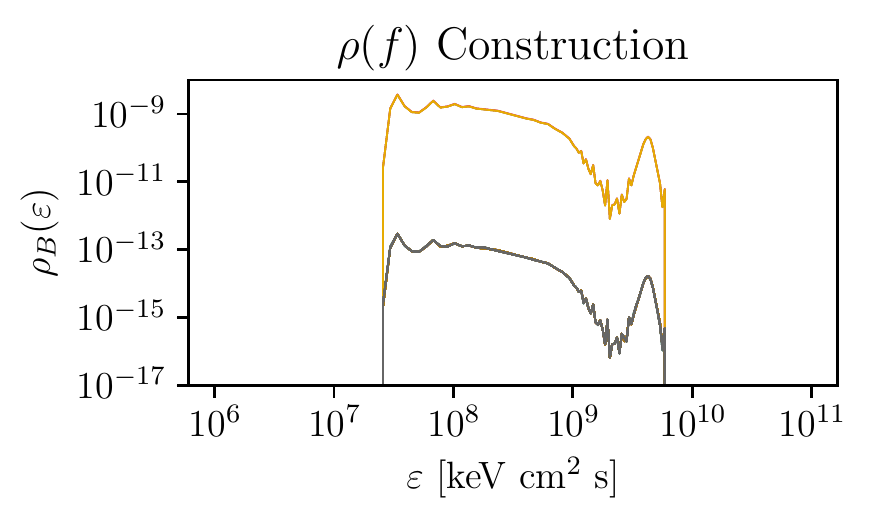}
     \end{subfigure}
    \caption{The detector effect correction function for both the CPG $\mu_\bin(\varepsilon)$ construction and the NPTF $\rho(f)$ construction (converted to $\rho_\bin(\varepsilon)$ for comparison as described in the text). Here, each coloured line corresponds to the function for a bin in the image.}
    \label{fig:template_mueps}
    \end{subfigure}
    \caption{The results of applying the CPG $\mu_\bin(\varepsilon)$ construction and the NPTF $\rho(f)$ construction to the non-uniform spatial distribution scenario. The CPG posteriors are clustered around the injected population, while the NPTF posteriors show a bias toward higher $N$. The averaging procedure in constructing $\rho(f)$ results in a detector correction that is not representative for any of the bins in the image, as shown by the $\mu_\bin(\varepsilon)$ functions.}
    \label{fig:template_all}
\end{figure*}

\begin{figure*}
    \centering
    \begin{subfigure}{0.495\textwidth}
    \begin{subfigure}{\textwidth}
         \centering
            \includegraphics[page=2]{figures/noniso_psf/mu_eps.pdf}
     \end{subfigure}
    \begin{subfigure}{\textwidth}
         \centering
            \includegraphics[page=2]{figures/noniso_psf/rho_eps.pdf}
     \end{subfigure}
    \caption{The recovered posterior for both the CPG $\mu_\bin(\varepsilon)$ construction and the NPTF $\rho(f)$ construction. }
    \label{fig:anisoPSF}
    \end{subfigure}
    \begin{subfigure}{0.495\textwidth}
    \begin{subfigure}{\textwidth}
         \centering
            \includegraphics[page=1]{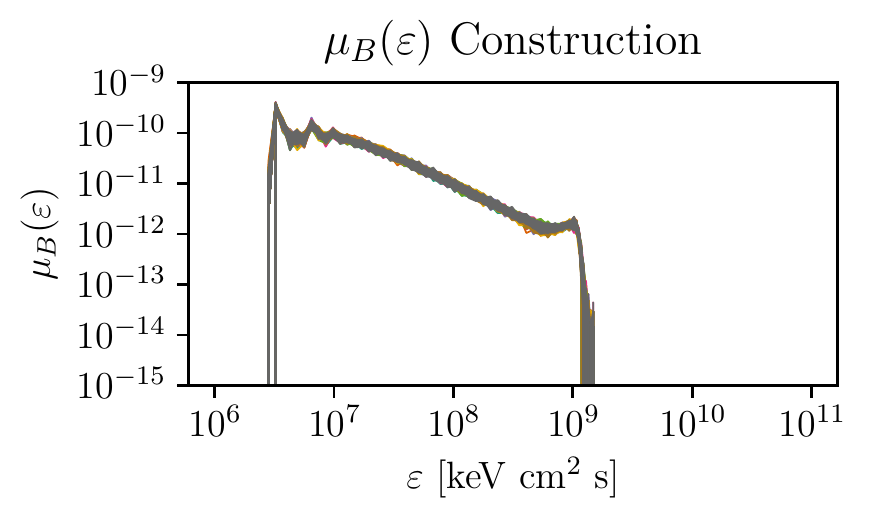}
     \end{subfigure}
    \begin{subfigure}{\textwidth}
         \centering
            \includegraphics[page=1]{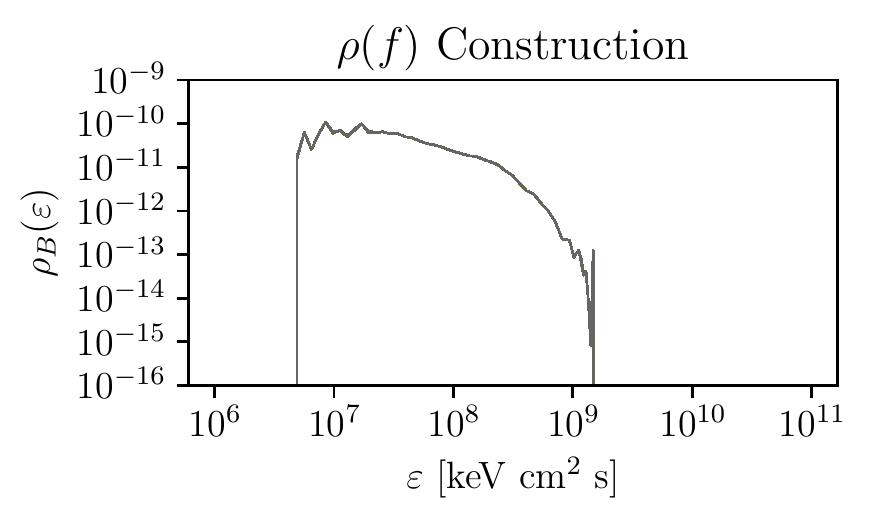}
     \end{subfigure}
    \caption{The detector effect correction function for both the CPG $\mu_\bin(\varepsilon)$ construction and the NPTF $\rho(f)$ construction. }
    \label{fig:anisoPSF_mueps}
    \end{subfigure}
    \caption{As in Fig.~\ref{fig:template_all}, but for the anisotropic PSF scenario. The injected population is well-recovered by CPG, while the NPTF posteriors are clearly biased towards low $N$. The cusp structure visible in each $\mu_\bin(\varepsilon)$ is smeared out in $\rho_\bin(\varepsilon)$.}
\end{figure*}

To begin with, we consider an edge case which illustrates the problems that arise from the factorisation of the template from the detector effects.
Consider a spatial population that occupies only a single bin, $l=0$, such that $T_l = \delta_{l0}$.
In addition, suppose the PSF is broad enough that a non-negligible amount of flux migrates out of bin $0$.
As the population does not exist outside of bin $0$, out-migration is the only effect of the PSF for this bin.
This implies that flux is not conserved for this bin: the flux captured by this bin is always less than the flux of the population.
The other bins, in turn, are only affected by in-migration.
According to the spatial template, $T_l = 0$ for these bins and so the population flux for these bins are also zero.
Thus, flux is not conserved for these bins either, as they receive flux from adjacent bins.
Both of these effects are illustrated in Fig.~\ref{fig:delta_template_diag}.

Conservation of flux is only a property of the entire image, and cannot be enforced on a bin-by-bin basis through $\rho(f)$.
The distribution of flux fractions that these bins receive is clearly a function of the bin in question, and so the average over all bins, $\rho(f)$, will not describe any bin in the image.
In addition, the distribution of flux fractions for each bin clearly depends on the distance of that bin from bin $0$, demonstrating manifestly why the spatial distribution of the population cannot be factorised out of the detector response.

Although this edge case is artificially extreme -- in order to bring out the effect -- it is not uncommon for scenarios to have much of the spatial distribution concentrated in a single bin.
Any sharply peaked spatial distribution will suffer this problem to an extent; for example, the GCE spatial profile has a sharp peak at the Galactic Center.

A scenario was constructed in the NuSTAR simulation (described in detail in App.~\ref{sub:app:sim}) in order to investigate the potential bias this may induce.
In this scenario, five bins have equally non-zero share of the population.
Multiple bins are necessary for point-source population reconstruction, and the bins are spaced far enough apart that the demonstrated effect will be largely similar to the single-bin scenario.
The edge case as described above can only be meaningfully analysed with the CPG likelihood.
The NPTF method will fail entirely and result in zero probability for all parameter values.
The reason for this is that from Eq.~\ref{eq:NPTF_gen} it can be seen that the NPTF predicts zero flux in any bin with $T_\bin = 0$, for any value of the model parameters.
If any of these bins record counts due to the finite PSF, then these are events the NPTF simply cannot reconstruct.
In order to allow for a comparison with the NPTF, we rescale the template so that no bins are exactly zero, but rather a small non-zero value of $T_\bin$ which cumulatively add up to no more than 1\% of the template.
Further details on the scenario setup are given in Tab.~\ref{tbl:all_config}.

The NPTF method was implemented by estimating $\rho(f)$ according to the algorithm in App.~\ref{app:rho_f}, and then transforming it to an equivalent CPG representation through $\rho_\bin(\varepsilon) = T_\bin \rho(\varepsilon/\bar{\kappa}_\bin) / \bar{\kappa}_\bin$.
In this way, the exact same computational implementation of the likelihood and Bayesian analysis procedure can be used to evaluate both the NPTF and CPG methods.
Thus, any differences observed can be entirely attributed to the NPTF $\rho(f)$ construction, and are not due to any subtle differences in the software implementation of the likelihood and MCMC.
The generation of $\mu(\varepsilon)$ required approximately 100 CPU hours;\footnote{Under 2 hours wall-time using a AMD Ryzen Threadripper 3990X 64-Core processor.} however, in practice, excellent results can be achieved with significantly less time --- as little as 5 CPU hours. As $\rho(f)$ is shared between bins, it requires substantially less time to generate; in this case, less than one CPU hour total. These times are highly application dependent, as they are almost an entirely a function of the computational complexity of the detector simulation, and in the case of $\mu(\varepsilon)$, the chosen binning scheme.

The natural coordinate system employed in the prior effect demonstrations of Sec.~\ref{sub:priors:nat_demo} was used, with the exception of the Poisson component.
Thus, the priors used here are also described by Tab.~\ref{tbl:natural_priors}, other than for $\omega$ which is not needed, and that the flux parameter is referred to as $F_{PS}$ instead of $F_T$.
The posteriors were sampled using the \texttt{emcee} Affine Invariant MCMC \cite{emcee}. A total of 200,000 steps were taken with 64 walkers, and the first 90\% of samples were discarded as burn-in.

The results are shown in Fig.~\ref{fig:template}.
The true source-count function parameters are shown by a solid red line.
The dotted and dashed lines describe the recovered posterior, where each colour represents one trial image.
The dashed lines are the $1$-$\sigma$ Highest Posterior Density (HPD) regions, while the dotted lines define the $2$-$\sigma$ HPD regions.
The NPTF posterior is biased towards a high number of sources with an approximately correct total flux.
In comparison, CPG does not exhibit the same bias.
Here bias is defined as an overall shift in the recovered parameters across multiple trials.
We can see that the yellow CPG trial is well outside the true parameters, while the purple NPTF trial is inside the true parameters.
This is to be expected due to the random variations between trials.
However, as a group, the NPTF trials are clearly recovering a larger number of sources than the true $N$.
In Fig.~\ref{fig:template_mueps} we further show the difference between the CPG $\mu_\bin(\varepsilon)$ and the equivalent NPTF $\rho_\bin(\varepsilon)$.
Each colour represents the $\mu_\bin(\varepsilon)$ measure for a bin in the scenario, as NPTF averages the PSF correction over bins and the effective area is isotropic the equivalent $\rho_\bin(\varepsilon)$ is the same for all bins up to the two unique template values of $T_\bin$.

The averaging approximation used in the NPTF construction is clearly deficient in this edge case.
The inference is driven by the five bins that contain the population.
The $\mu_\bin(\varepsilon)$ for these bins is heavily weighted toward high $\varepsilon$.
The NPTF $\rho(\varepsilon)$ is heavily weighted toward low $\varepsilon$, due to averaging over the larger number of bins that do not contain the population.
The normalisation condition, $\int_0^1 \d f\,f \rho(f) = 1$, imposed on $\rho(f)$ modifies the overall normalisation of the function so that flux is conserved in all bins.
However, as described above, the flux is manifestly not conserved on a bin by bin basis in this scenario.
In Eq.~\ref{eq:NPTF_gen}, a change in the normalisation of $\rho(f)$ is equivalent to a change in $N$ as this generator can also be written as
\begin{align}
    &\hat{G}_{k_\bin}(z) = \\ &\exp{\left[ N T_\bin \int_0^1 \d f \rho(f) \left( \int \d F  e^{\bar{\kappa}_\bin f F (z-1)} p(F) - f \right) \right]}. \nonumber
\end{align}
Thus, the normalisation effectively drives $\rho_\bin(\varepsilon)$ at high $\varepsilon$ down, causing the posterior on $N$ to be driven up to compensate.
This leads to the observed overestimation bias in the number of sources.

The posteriors in Fig.~\ref{fig:template} show a between-trial variation that is on the same scale as the posteriors themselves. We can rule out statistical fluctuations as the dominant cause, as the variations are an order of magnitude in $N$ for the NPTF posteriors. Instead, these variations are due to small amount of information that is present in only five bins, which leads to poor recovery of the total number of sources. We should expect the posteriors to overlap (and they mostly do for CPG), but as the NPTF likelihood does not describe the data, we cannot expect the posteriors to be well-behaved.

\subsection{Anisotropic PSF\label{sub:nptf:anisoPSF}}

In the standard NPTF construction, $\rho(f)$ is common to all bins and thus it can, at best, capture the average PSF throughout the image.
The following scenario examines the bias that this approximation may introduce to the inference of the source-count function model parameters when an anisotropic PSF is present, as arises, for instance, in NuSTAR.

We consider a scenario largely identical to that in Sec.~\ref{sub:prior_demo}, with details given in Tab.~\ref{tbl:all_config}.
In this scenario, multiple exposures are overlaid into a mosaic, forming a single binned image.
This increases the amount of anisotropy in the PSF, as sources can bleed across the boundaries of the exposures.
When a source contributes to two or more exposures, an appropriate mix of PSFs is required, leading to more complicated PSF distributions.
For this scenario, we wish to only test the effect of an anisotropic PSF.
Thus, the exposures are perfectly aligned edge to edge in a grid with no gaps or overlaps.
This, along with the lack of vignetting, ensures that the effective area is uniform across the entire image.
We note that this is not a natural arrangement --- most mosaics involve a degree of overlap between exposures.

The results of analysing the resulting data sets are shown in Fig.~\ref{fig:anisoPSF}.
A high degree of bias toward a low number of sources is observed in the NPTF posterior, while CPG is largely consistent with the true population parameters.
Figure~\ref{fig:anisoPSF_mueps} further shows that the equivalent NPTF $\rho_\bin(\varepsilon)$ is much closer to the CPG $\mu_\bin(\varepsilon)$ in this scenario as compared to the previous non-uniform spatial template scenario.
However, the shape of $\rho_\bin(\varepsilon)$ is significantly different from $\mu_\bin(\varepsilon)$ near the highest $\varepsilon$.
Unlike before, the normalisation of $\rho_\bin(\varepsilon)$ will be correct here: the uniform spatial distribution ensures that flux is conserved on a bin-by-bin basis.
While this normalisation ensures the average $\varepsilon$ is correct, the distribution of $\varepsilon$ is not.
Observe that $\mu_\bin(\varepsilon)$ generally has a cusp at high $\varepsilon$, while $\rho_\bin(\varepsilon)$ shows no cusp.
This lack of cusp causes the NPTF likelihood to be driven by lower $\varepsilon$ in comparison, and so the posterior drives up the average flux per source, and drives down the number of sources in order to maintain the conservation of flux.
Note that the quantity shown in Fig.~\ref{fig:anisoPSF} is the total flux of the entire population, so when the number of sources, $N$, reduces, it is the average flux per source, roughly proportional to $F_{PS}/N$, that increases.

\subsection{Sub-bin Effective Area\label{sub:nptf:effarea}}

\begin{figure}
    \centering
    \includegraphics[page=7]{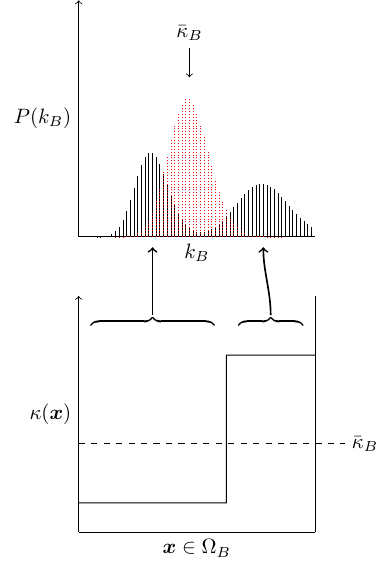}
    \caption{\textbf{Below:} The detector response, $\kappa(\v{x})$, varies sharply over bin $b$. The average detector response for this bin, $\bar{\kappa}_b$, is shown by the dashed line. \textbf{Above:} If the average detector response is used to convert flux to counts, the probability distribution follows the red dotted comb; however, the within-bin variation of the detector response must create two modes in the distribution, shown by the black comb.}
    \label{fig:subpixel_eff_diag}
\end{figure}

\begin{figure*}
    \centering
    \begin{subfigure}{0.495\textwidth}
    \begin{subfigure}{\textwidth}
         \centering
            \includegraphics[page=2]{figures/subpixel_effarea/mu_eps.pdf}
     \end{subfigure}
    \begin{subfigure}{\textwidth}
         \centering
            \includegraphics[page=2]{figures/subpixel_effarea/rho_eps.pdf}
     \end{subfigure}
    \caption{The recovered posterior for both the CPG $\mu(\varepsilon)$ construction and the NPTF $\rho(f)$ construction.}
    \label{fig:subpix}
    \end{subfigure}
    \begin{subfigure}{0.495\textwidth}
    \begin{subfigure}{\textwidth}
         \centering
            \includegraphics[page=1]{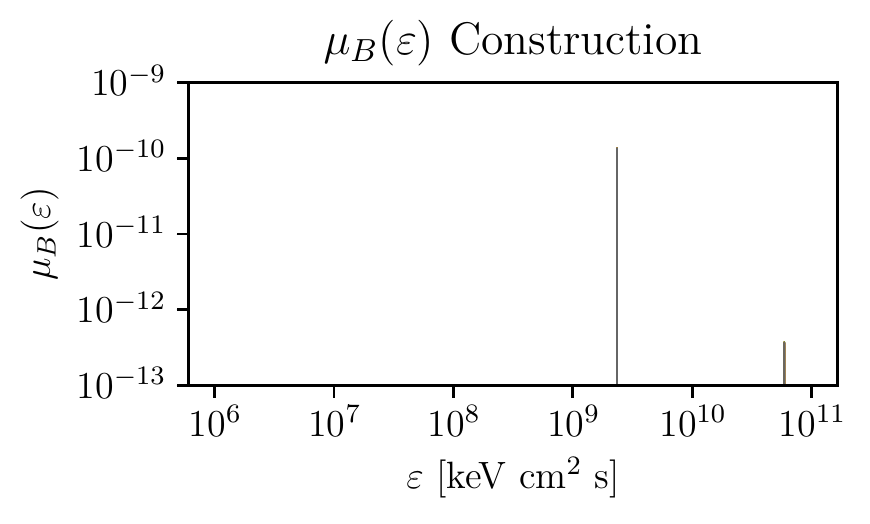}
     \end{subfigure}
    \begin{subfigure}{\textwidth}
         \centering
            \includegraphics[page=1]{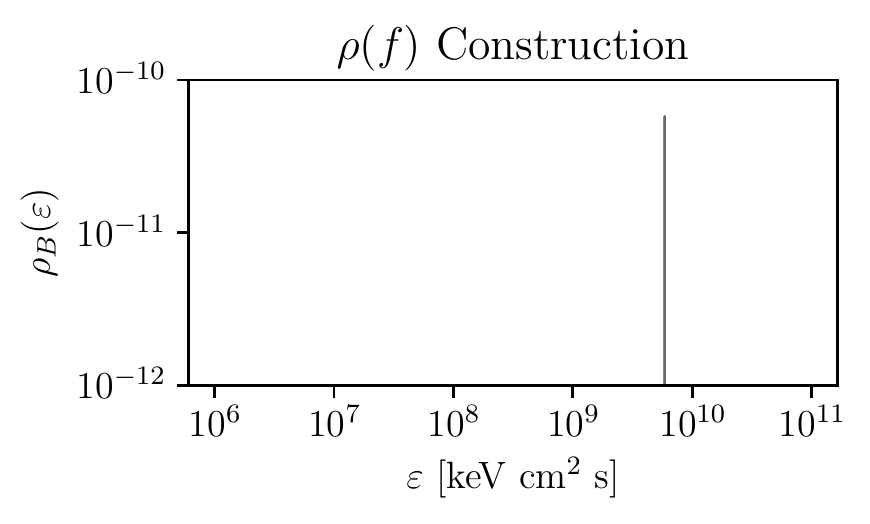}
     \end{subfigure}
    \caption{The detector effect correction function for both the CPG $\mu(\varepsilon)$ construction and the NPTF $\rho(f)$ construction. }
    \label{fig:subpix_mueps}
    \end{subfigure}
    \caption{As in Fig.~\ref{fig:template_all}, but for the variation in sub-bin effective area scenario. The CPG posteriors are clustered around the injected population, while the NPTF posteriors exhibit a small bias towards low $N$. The bimodal effective area is shown clearly by $\mu_\bin(\epsilon)$ while for $\rho_\bin(\varepsilon)$ this structure is removed by the averaging the effective area, resulting in a single mode.}
\end{figure*}

The prescribed construction of $\rho(f)$ in the NPTF method does not include any accounting for the effective area, exposure time, or detector efficiency of the instrument and observation.
This may lead to an incorrect construction of the likelihood distribution, for which the following scenario illustrates.

Consider an instrument where the effective area varies sharply within a bin, and a population which is spatially uniform.
Thus, the distribution of fluxes for a source is not a function of position within this bin.
However, the distribution of expected counts will be a function of position within the bin, as the flux must be converted to a number of photons using the effective area.
In particular, take a scenario where there is no PSF, $\phi(\v{y} | \v{x}) = \delta(\v{y} - \v{x})$; and the injected source population has no variation in flux, $p(F) = \delta(F - \bar{F}_{PS})$.
In order to generate the effect of interest, we then let the detector response change discontinuously at some point within the bin, between the values of $\kappa_0$ and $\kappa_1$.
If a source is located where $\kappa(\v{x}) = \kappa_i$, then the distribution of the detected number of counts at this location is $p(S_\bin | \v{x}) = \delta(S_\bin - \kappa_i \bar{F}_{PS})$, with $i=1$ or $2$.
The distribution for the whole bin is found by integrating the position dependent distribution over $T(\v{x})$.
Let the spatial template be uniform, then the whole bin distribution is a mixture of the previous two distributions:
\begin{equation}
    p(S_\bin) = C \delta(S_\bin - \kappa_0 \bar{F}_{PS}) + (1 - C) \delta(S_\bin - \kappa_1 \bar{F}_{PS}),
\end{equation}
where $C$ is some mixture fraction that depends on what area of the bin has a detector response of $\kappa_0$.

For this scenario, the NPTF calculates the average detector response across the bin, $\bar{\kappa}$, and uses this to find $p(S_\bin) = \delta(S_\bin - \bar{\kappa} \bar{F}_{PS})$.
This will not result in the correct distribution of photon number; for example, if $\kappa_0 = 0$, then the NPTF construction preserves the total flux of the bin, but will overestimate the flux of sources within that bin, as $\bar{\kappa} < \kappa_1$.
A comparison between the exact and mean distributions are shown in Fig.~\ref{fig:subpixel_eff_diag}.
It should be noted here that, due to computational constraints, the effective area is rarely considered on a bin-by-bin basis in the NPTF construction, and is instead averaged over larger ``exposure regions'' as the computational complexity of the NPTF power series calculation depends on the number of bins.
Although these non-contiguous regions are chosen by the similarity of the effective area of each bin within the regions, this necessarily reduces the accuracy of the likelihood evaluation further.
In contrast, CPG requires no exposure regions, as the computational complexity of the CPG power series calculation does not depend on the number of bins and is computed only once for each likelihood evaluation.
As exposure regions only enter into NPTF due to computational constraints, they are not, by themselves, an intrinsic limitation of NPTF; therefore, we do not consider the effect of taking a number of exposure regions smaller than the number of pixels in this investigation.

To test this scenario, the simulated vignetting was modified into a checkerboard pattern, taking values of either $\kappa_0$ or $\kappa_1$.
Each square of the checkerboard was defined to be equal to the bin size, but with an offset so that the checkerboard boundaries lie within the bins of the image.
This causes the effective area to change discontinuously across most bins of the image.
The remaining scenario parameters are largely identical to those in Sec.~\ref{sub:prior_demo}, with details given in Tab.~\ref{tbl:all_config}.

The results are shown in Fig.~\ref{fig:subpix}.
Here, NPTF fares significantly better when compared to the previous scenarios.
However, a recognisable bias is still present in the recovered posteriors.
CPG, on the other hand, is considerably closer to the true population parameters.
The detector effect correction functions are shown in Fig.~\ref{fig:subpix_mueps}.
Clearly, the CPG $\mu_\bin(\varepsilon)$ has the two expected modes corresponding to the sub-bin effective area variation.
In contrast, the equivalent NPTF $\rho_\bin(\varepsilon)$ has a single mode corresponding to the average effective area.
Thus the likelihood is driven by a lower $\varepsilon$, and so the average flux per source is driven up while the number of sources is driven down to compensate.

The CPG posteriors in Fig.~\ref{fig:subpix} display a long tail toward high $N$.
This same behaviour can be observed to a lesser extent in Fig.~\ref{fig:anisoPSF} (anisotropic PSF scenario).
The summary Fig.~\ref{fig:N_summary} shows this effect most clearly.
As discussed in Sec.~\ref{sec:priors}, the discrimination ability of the CPG likelihood is reduced at high $N$ as the population model approaches the diffuse limit.
In contrast, population models with $N$ less than the true value will be highly disfavoured, as they produce sources with high flux.
In these scenarios, the prior on $N$ is log-uniform, and does not discourage models with high $N$.
Therefore, an asymmetry in the posterior toward high $N$ is expected.
This effect is also present in Fig.~\ref{fig:template} (non-uniform template scenario), however the small number of bins with useful data causes a wide variation in the posteriors that largely hides the effect.

\subsection{A Scenario where $\rho(f)$ is Valid}

\begin{figure}
    \centering
    \begin{subfigure}[t]{0.495\textwidth}
         \centering
            \includegraphics[page=2]{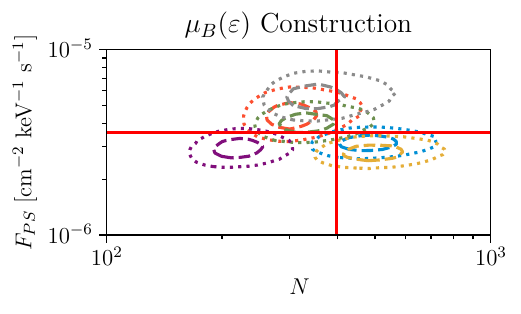}
     \end{subfigure}
    \hfill
    \begin{subfigure}[t]{0.495\textwidth}
         \centering
            \includegraphics[page=2]{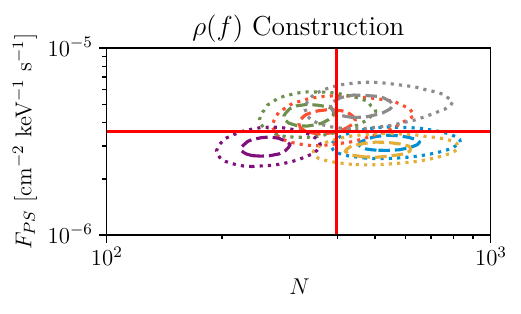}
     \end{subfigure}
    \caption{The recovered source-count function for both the CPG $\mu_\bin(\varepsilon)$ construction and the NPTF $\rho(f)$ construction in the $\rho(f)$ scenario, described in the text. When all conditions for the construction of $\rho(f)$ are met, both CPG and NPTF produce posteriors clustered around the injected population.}
    \label{fig:nptf_conds}
\end{figure}

The three previous scenarios demonstrate the biases introduced into NPTF by the failure to correctly model the effective area, PSF, and the spatial distribution of sources, as well as how the comprehensive approach of CPG resolves these issues.
When these complications are not present, however, it is expected that the NPTF construction will be equivalent to CPG.
To confirm this, we consider a scenario where the population is spatially uniform, the PSF is isotropic and the effective area is constant throughout the image.
The details are given in Tab.~\ref{tbl:all_config}, and the results are shown in Fig.~\ref{fig:nptf_conds}.
Both NPTF and CPG display no bias, and each trial is equivalent -- up to slight variations -- between the two methods.
This shows that under these conditions, the more general CPG reduces to NPTF, and NPTF does indeed work correctly.

\subsection{Realistic Scenario \label{sub:nptf:nustar}}

\begin{figure*}
    \centering
            \includegraphics[page=1,width=0.45\textwidth]{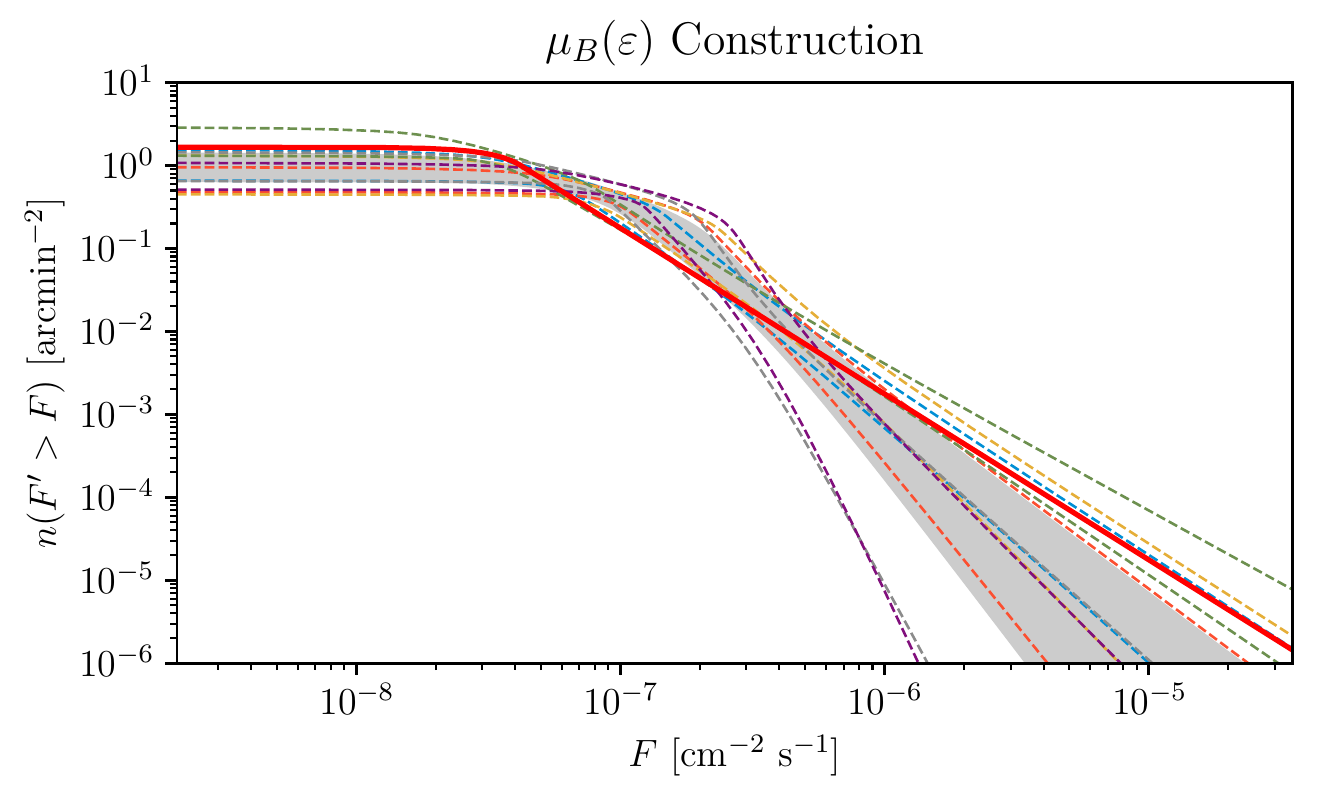}
            \hfill
            \includegraphics[page=1,width=0.45\textwidth]{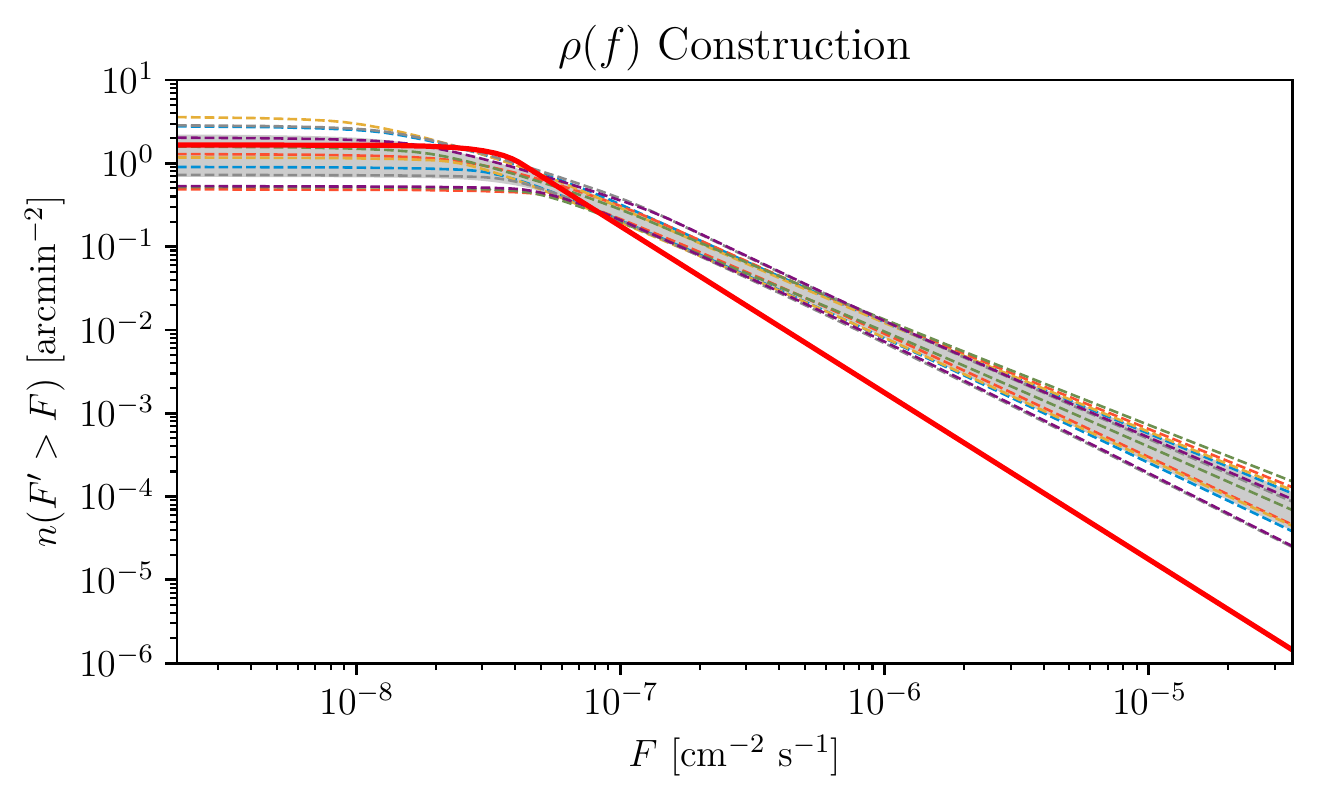}
            \vfill
            \includegraphics[page=2,width=0.45\textwidth]{figures/nustar_obs/mu_eps.pdf}
            \hfill
            \includegraphics[page=2,width=0.45\textwidth]{figures/nustar_obs/rho_eps.pdf}
    \caption{The recovered source-count function (as defined in Eq.~\ref{eq:source_count_def}) for the CPG $\mu_\bin(\varepsilon)$ construction (top left) and the NPTF $\rho(f)$ construction (top right) in the realistic NuSTAR scenario. Each colour represents a different random trial, with the 16\% and 84\% percentiles shown as dashed lines derived from the posterior. The true source-count function is shown in solid red. The grey fill shows the average percentile band, as a guide. The $\rho(f)$ construction clearly recovers the incorrect source-count function, as the slope of the power-law is well outside all percentile bands. In comparison, the CPG construction is generally more accurate. The underestimation of sources in both methods is further discussed in the text. The same results, with the posterior conditioned to $\omega > 0.9$, are shown in the bottom left and bottom right for CPG and NPTF respectively. With this conditioning, the recovery of the total number of sources is drastically improved for the CPG.}
    \label{fig:realistic}
\end{figure*}

Finally, we turn to a realistic scenario, where all NuSTAR detector effects are enabled, and a Poisson background is injected.
This results in a realistic simulation of a NuSTAR observation.
Eight observations are combined into a mosaic that exhibits the overlapping diamond pattern that is common to composite NuSTAR images --- see \textcite{Hong:2016qjq} for one such instance.

For this scenario, there are two Poisson background components: the intrinsic detector background, which is uniform within each observation, and the the Cosmic X-ray Background (CXB), which is a flux that is multiplied by the effective area of the instrument.
Both components scale with the exposure time, and their respective rate and flux is detailed in Tab.~\ref{tbl:all_config}.
The background rate was chosen to be representative of an observation with NuSTAR. The CXB flux was derived from \textcite{Krivonos:2020qvl} by integrating over the $3$ -- $10$ keV energy window.

The CXB is incorporated into the model using a shared flux parameter in the natural coordinate system discussed in Sec.~\ref{sec:priors}.
The intrinsic background, in contrast, is not specified as part of the natural coordinate system.
The natural coordinate system divides a total amount of flux between multiple models, but this background component is not described in terms of a flux, as it is not astrophysical in origin.
Thus, it was given a separate uniform prior around the known background rate, and is parameterised by $\lambda_{\text{bkg}}$, a count rate for the entire detector, which is multiplied by the exposure time and divided by the relative bin size to get a mean number of counts per bin.
The injected source-count function was selected to be similar to the observed population of X-ray sources near the Galactic Center \cite{Hong:2016qjq}.
Further details are given in Tab.~\ref{tbl:all_config}, and the prior ranges are shown in Tab.~\ref{tbl:realistic_priors}.

The results -- derived with both the CPG $\mu_\bin(\varepsilon)$ and NPTF $\rho(f)$ constructions -- are shown in the upper half of Fig.~\ref{fig:realistic} as a source-count density function.
Each colour represents one of six trial images, and the dashed lines show the 16\% and 84\% percentiles for the source-count function recovered from the posterior.
The solid red line shows the true population source-count function.
The top of the grey fill is the mean of the 84\% percentiles for each trial, while the bottom of the grey fill is the mean of 16\% percentiles.

For the CPG construction, there is a large uncertainty on the location of the flux break itself; and, below the break, there is an underestimation in the number of sources.
This is unsurprising, as the uniform prior on the flux fraction parameter between the point-source and CXB component, $\omega$, gives weight to models where a significant amount of flux is carried by the CXB, which corresponds to the below threshold point-source flux being assigned to the CXB component.

This effect can be reduced by placing a more informative prior on $\omega$, but this is not as easy at it may first appear.
Even if the CXB flux is well-known, placing a prior on $\omega$ requires prior knowledge on the relative total flux between the population and the CXB.
If the prior is informed only by an estimate of this relative flux, then an incorrect value could exacerbate the effect.
Alternatively, the prior on $\omega$ could be conditioned on the total flux $F_T$ so that the inferred total flux informs the prior on the estimated flux of the point-source population.

We can also imagine a similar problem occurring for the intrinsic detector background, in that below threshold sources could be confused with the intrinsic background model.
The solution here, which has been employed in this scenario, is to use a tight prior on the intrinsic detector background based on previous measurements.
This is simple to implement, as the intrinsic detector background is not specified in the natural coordinate system and thus has an independent prior.

In the end, this is not an issue with the CPG construction, it is simply the nature of Bayesian analyses. We can see this more clearly by conditioning the posterior on $\omega > 0.9$. This, in effect, largely removes the CXB from the model. The result is shown in the lower half of Fig.~\ref{fig:realistic}. The uncertainty on the flux break is greatly reduced, and the total number of sources is now accurately recovered to within the percentiles.

However, the power index, $n_1$, sits just outside the percentiles. This turns out to be a generally unavoidable effect. The large bin sizes cause the brightest sources to be washed out by the large number of dimmer sources. For this reason, the model cannot distinguish between a few bright sources and many dim ones, and so the posterior is extended to larger $n_1$, thereby pushing the true $n_1$ outside of the percentile interval. This effect can be reduced by reducing the bin size, so that bright sources make up a greater fraction of the total flux in each bin. Unfortunately, as discussed in Sec.~\ref{sub:cpg:whole_lh}, smaller bins will cause the statistics to be over-counted, making the posteriors artificially small.

Thus, when applying CPG to a realistic scenario with a Poisson background that has a poorly informed prior, care must be taken in interpreting the number of very low flux sources.
In addition, the prior on the detector background should be chosen to incorporate as much information as possible that is known about the background rate.
Finally, one must take care in the selection of the bin size. When the goal is to recover the population parameters, smaller bins should be chosen to retain more information on the high flux sources. Any reported results should then caution of the between-bin correlation effect elaborated on in Sec.~\ref{sub:cpg:whole_lh}. If model selection is the aim, and marginal likelihoods are in use, it may be better to use larger bins, as a high degree of over-counting of statistics may result in an erroneous marginal likelihood.

In comparison, the NPTF $\rho(f)$ construction has poorer performance. The same underestimation of the number of sources below the break is observed, but to a larger degree.
In addition, the location of the break and index of the power-law above the break is incorrectly estimated, and the true source-count function lies well outside all of the recovered source-count functions in the region above the break.
Conditioning the posterior on $\omega > 0.9$ does not improve the situation appreciably: the total number of sources now lies within the percentiles (although the uncertainty is much larger than when using CPG), but the region above the break is still incorrectly recovered. In this instance, while the total number of sources is approximately correct, the total flux of the population is overestimated.

\section{Conclusion}

In this paper we have introduced a new approach to the age-old problem of point-source population inference.
In particular, we constructed a new likelihood using Compound Poisson Generator (CPG) functionals.
To complement this likelihood, a new natural coordinate system has been developed for parametrising Bayesian priors on point-source population models where a Poisson background component is also present.
In particular, we revealed that existing prior choices often result in posteriors that assign all flux to either the point-source population, or to the Poisson background model.
The new natural coordinate system produces posterior distributions that correctly capture the uncertainty on the population model in this circumstance.
Combined with this new prior formulation, the CPG likelihood has been shown to correctly handle a series of test scenarios that each exhibit a particular edge case that is common in astronomical instrumentation.
For these scenarios, a comparison was made to Non-Poissonian Template Fitting (NPTF) --- a current leading parametric point-source inference method.
In contrast to CPG, the NPTF method has been shown to produce biases in the recovered posterior distributions, specifically the mean number of sources.
These biases are attributed to limitations in the construction of the NPTF likelihood, which factorises the spatial distribution, PSF, and effective area.
The CPG derivation shows how these contributions to the likelihood cannot be factorised in general.
An implementation of the CPG is made publicly available \href{https://github.com/ghcollin/cpg_likelihood}{here}.

Our focus in the present work has been on the construction of the CPG and on demonstrating the improvements it brings.
However, given the general importance of point-source studies applying this new method to actual data is an important open direction, and may even be able to finally shed light on open questions such as the nature of the GCE.

\section*{Acknowledgments}

We thank M. Lisanti, F. List, S. Mishra-Sharma, B. M. Roach, B. Safdi, and T. Slatyer for useful discussions, as well as the wider community in the NSF AI Institute for Artificial Intelligence and Fundamental Interactions, the Laboratory for Information and Decision Systems, and the MIT Statistics and Data Science Center. 
N.L.R. is supported by the Miller Institute for Basic Research in Science at the University of California, Berkeley.
K.P. and G.H.C. were supported by the Cottrell Scholar Award, Research Corporation for Science Advancement (RCSA), ID 25928. 
This work made use of resources provided by the National Energy Research Scientific Computing Center, a U.S. Department of Energy
Office of Science User Facility supported by Contract No. DE-AC02-05CH11231.

\bibliography{bib}

\onecolumngrid
\appendix

\section{NuSTAR Observatory and Simulation Procedure\label{app:nustar_sim}}

The NuSTAR observatory comprises of two adjacent aligned telescopes.
Each telescope consists of a grazing incidence X-ray optics module and a Focal Plane Module (FPM) that houses four pixelated CdZnTe semiconductor detectors.
For both telescopes, these elements are separated by a single shared extended mast.
For this investigation, only one of these telescopes will be considered in order to simplify the simulation. 

The optics focus X-rays via grazing reflections from two sets of concentric shells arranged in an approximate Wolter-I geometry.
When an X-ray is reflected by both shells, it will be correctly focused on to the detector plane.
An X-ray may only undergo one reflection in the optics.
These X-rays are poorly focused and are called ghost-rays.
The observatory does not have an enclosed barrel, and the majority of the length of the telescope is open to space.
A series of aperture stops collimate incoming X-rays, preventing most X-rays from entering the FPM through the sides of the telescope.
However, a small window remains allowing X-rays that pass close to the optics modules to strike the detector without passing through the optics.
Known as stray-light, X-rays that enter the FPM this way are unfocused.

The four detectors in each FPM are arranged in a $2\times 2$ grid. Each detector is composed of $32 \times 32$ pixels, and sub-pixel positional information is available.
The gap between the detectors is small and is ignored in this study.
Further details on the observatory can be found in Refs.~\cite{Harrison:2013md,Wik:2014boa,Madsen:2015jea,madsenObservationalArtifactsNuSTAR2017}.

\subsection{Simulation\label{sub:app:sim}}

Use of NuSTAR as a test case requires a simulation of the detector in order to create observations where the parameters of the injected point-source population are known exactly.
The simulation draws a number of point sources from a given differential source-count function using a Poisson distribution with mean $N$.
The flux of each source is also drawn from the $p(F)$, and the location of each source is drawn from a given spatial distribution specified by the template.
The simulation constructs an image according to the NuSTAR field of view, and with an adjustable bin size.
The detector response is extracted from the NuSTAR CALDB~\cite{Harrison:2013md} FPM data using the vignetting corrected effective area.
The energy spectrum is assumed to take the power-law form given in Eq.~\ref{eq:energy_spectrum} with $E_0 = 1$ keV and $\gamma = 1.5$, whereas the energy range and exposure time are scenario dependent.

Once the mean counts for a given source, $S$, has been determined, a number of photons is then drawn from a Poisson distribution with mean $S$, and these photons are placed into the image according to locations drawn from the CALDB PSF.
This PSF is generated by a physics based simulation of the X-ray optics.
If a photon lies outside the field of view, the photon is discarded.
The NuSTAR PSF is non-isotropic, and varies according to the distance of the source from the optical axis of the telescope.
At the optical axis, the PSF is radially symmetric.
As the source moves towards the edge of the image, it is distorted radially creating a fish-eye effect.

Apart from the point-source population, multiple other effects can cause the real or apparent detection of photons.
Astrophysical backgrounds and foregrounds may include extended sources such as gas clouds, or very bright point-sources that are not part of the population such as magnetars.
These can be modelled as Poisson distributions with appropriate flux maps.
As they are highly scenario dependent, these effects are not modelled in this investigation.
Ghost-rays and stray-light are usually only visible when bright sources are in or around the FOV, assuming the sources can be located, their contribution can be modelled by a Poisson distribution with an appropriate flux map.
The effect of ghost rays from population sources is incorporated into the PSF by construction.
The CXB is a significant source of stray-light induced background.
This stray-light contribution can be modelled much like other astrophysical backgrounds --- with a Poisson distribution using an appropriate CXB flux.
As with the other sources of astrophysical background, the effect of stray-light CXB is not considered in this investigation.
NuSTAR also suffers from a detector background caused by space-borne radiation.
This background enters through the FPM shielding, and is not associated with the X-ray optics.
Thus, the background is uniformly distributed across each detector, and varies across each of the detectors in the FPM.
It can be simulated by drawing counts from a Poisson distribution according to an appropriate background count rate.

Position information of photons beyond the detector pixelisation is not available; however, in this simulation, the effect of pixelation is not considered and the photons are directly processed into bins.
The size of these bins is scenario dependent, but they are generally much larger than the detector pixels, thus the effect of pixelation is considered small and the additional complications involved in the pixel subdivisions are avoided.

If the scenario calls for multiple observations, then the above procedure is repeated for each observation minus the final binning.
This creates a list of photons from all observations, which are then transformed into a shared coordinate system for all of the observations and then binned into a shared binning scheme.
An example of this is shown in Fig.~\ref{fig:sim_obs}, with a binning scheme chosen to be approximately the same as the NuSTAR sub-pixelation, in order to show the PSF.

\section{Generating Functions and Functionals\label{app:gen}}

In this appendix, a full derivation of the CPG generating function without approximations will be presented. 
From this generating function, an unbinned likelihood and a binned likelihood can be derived, both taking into account correlations between pixels.
Unfortunately these likelihoods are intractable, and so the derivation in Sec.~\ref{sec:cpg} is shown to be a tractable mean-field approximation for the likelihoods here.

Generating functions are a useful representation for a discrete probability distribution when complex constructions are required.
In this appendix we provide a more expansive discussion of these objects than appeared in the main body.
The generating function, $G(z)$, is defined as the $z$-transform of a discrete distribution, $P(n)$, over its support $\mathcal{N}$:
\begin{equation}
    G(z) = \sum_{n\in\mathcal{N}} P(n) z^n = \E_{P}[z^n]. 
\end{equation}
Typically, $\mathcal{N} \subset \mathbb{N}_0$, with $\inf \mathcal{N} = 0$.
The probability distribution can be recovered from the generating function via higher order derivatives evaluated at $z=0$:
\begin{equation}
    P(n) = \frac{1}{n!} \frac{{\d}^n G}{\d z^n}(0).
\end{equation}

Chief among the generating function's useful properties is the sum-of-random-variates rule.
Let each $X_i$ be one of $M$ random variates, drawn from the distribution $P_i(x)$ with generating function $G_i(z)$.
The generating function for the sum of these variates,
\begin{equation}
    K = \sum_{i=1}^M X_i,
\end{equation}
is given by
\begin{equation}
    G_K(z) = \prod_{i=0}^M G_i(z).
\end{equation}
This property has an analogy in the characteristic function of continuous distributions; where the distribution for a sum of two random variates is the convolution of the two distributions, the characteristic function is defined as the Fourier transform of the distribution function, so that the characteristic function for the sum of the variates is the product of the characteristic functions via the convolution theorem.

If the $X_i$ are identically distributed -- that is, drawn from the same distribution $P_X(x)$ -- and the number of variates to sum, $M$, is itself a random variate drawn from a distribution $P_M(m)$ with generating function $G_M(z)$, then the generating function for the sum
\begin{equation}
    K = \sum_{i=1}^M X_i,
\end{equation}
is provided by the nested expression,
\begin{equation}
    G_K(z) = G_M(G_X(z)).
\end{equation}
When $P_M(m)$ is a Poisson distribution, $G_K(x)$ is known as a compounded generator.

As a simple example, for the Poisson distribution,
\begin{equation}
    P(n | \lambda) = \frac{\lambda^n e^{-\lambda}}{n!}
\end{equation}
with $\lambda$ the mean of the distribution, the generating function is
\begin{equation}
    G(z) = e^{\lambda (z-1)}.
\end{equation}
Poisson distributions count the number of events that occur in a defined interval --- most often in time, but we are also interested in space intervals.
The above properties work well when the underlying Poisson process -- the mechanism that generates these events -- is homogeneous in this interval.

When the Poisson process is inhomogeneous, a generating functional can prove more useful.
An inhomogeneous Poisson process is defined by an intensity function $\Lambda(\v{x})$ with support $\Xi$.
The integral of this intensity function over the support gives the expected number of events, $\lambda = \int_\Xi \d\v{x}\, \Lambda(\v{x})$.
Thus, the distribution for the number of events is a Poisson distribution with mean $\lambda$.
The generating functional for the process in this case is defined as
\begin{equation}
    G[f] = e^{\int_\Xi \d\v{x}\, \Lambda(\v{x}) (f(\v{x}) - 1)},
\end{equation}
where $f(\v{x})$ is a functional argument defined over the same support $\Xi$.
The probability density function for finding an event at $\v{y}$ is given by the variational derivative of this functional:
\begin{equation}
    p(\v{y}) = \frac{\d G}{\d f(\v{y})}[\mathbb{0}] = \Lambda(\v{y}) e^{- \int_\Xi \d\v{x}\, \Lambda(\v{x})},
\end{equation} 
where $\mathbb{0}(\v{x}) = 0$.
The probability density for a set of events $\{\v{y}_i\}$ is the higher order variational derivative
\begin{equation}
    p(\v{y}_i) = \left( \left[ \prod_{i} \frac{\d}{\d f(\v{y}_i)}\right] G \right)[\mathbb{0}] = \left[\prod_i \Lambda(\v{y}_i) \right] e^{-\int_\Xi \d\v{x}\, \Lambda(\v{x}) }. 
\end{equation}
The Poisson process is one kind of process that generates events -- known generally as a point process.
The concept of a generating functional can be extended to all point processes.

The generating functional has a similar compounding property to the generating functions.
Let $G_X$ be the generating functional for a point process, while $G_M$ is the generating functional for another point process.
If $G_M$, rather than directly generating events, instead generates point processes according to $G_X$, then draws from $G_M$ will generate multiple events according to $G_X$.
The generating functional for this compounded process is $G_K[f] = G_M[G_X[f]]$.

For a point source population, the intensity function is
\begin{equation}
    \Lambda(\v{x}) = N T(\v{x}),
\end{equation}
such that the intensity is distributed according to the spatial distribution, $T(\v{x})$, and the total of the intensity function over the support of the spatial distribution is exactly the mean number of sources:
\begin{equation}
    N = \int \d\v{x} \Lambda(\v{x}).
\end{equation}
Now, let $G_{\v{x}|N,T}$ be the generating functional for sources from the population:
\begin{equation}
    G_{\v{x}|N,T}[f] = \exp{\left(N \int \d\v{x} T(\v{x}) (f(\v{x}) - 1) \right)},
\end{equation}
so that events drawn from this process define individual sources located at location $\v{x}$.
Each source, accordingly, defines its own generating functional:
\begin{equation}
    G_{\v{y}|F,\v{x}}[h] = \exp{\left(\kappa(\v{x}) F \int \d\v{y} \eta(\v{y}) \phi(\v{y} | \v{x}) (h(\v{y}) - 1) \right)},
\end{equation}
from which events define counts detected at location $\v{y}$, given a source of flux $F$ located at $\v{x}$.
The population averaged source functional is thus
\begin{equation}
    G_{\v{y}|p(F),\v{x}}[h] = \E_{p(F)}[G_{\v{y}|F,\v{x}}[h]] = \int \d F p(F) \exp{\left(\kappa(\v{x}) F \int \d\v{y} \eta(\v{y}) \phi(\v{y} | \v{x}) (h(\v{y}) - 1) \right)}.
\end{equation}
The total count generating functional is now a compound of these two generators \cite{IntroductionTheoryPoint2003}:
\begin{equation}\begin{aligned}
    G_{\v{y}|N,T,p(F)}[h(\v{y})] &= G_{\v{x}|N,T}[ G_{\v{y}|p(F),\v{x}}(h(\v{y})) ] \\ &= \exp{\left[ N \int \d\v{x} T(\v{x}) \left(\int \d F p(F) \exp{\left[\kappa(\v{x}) F \int \d\v{y} \eta(\v{y}) \phi(\v{y} | \v{x}) (h(\v{y}) - 1) \right]} - 1\right)  \right]}.
\end{aligned}\end{equation}
The unbinned likelihood can be calculated using $G_{\v{y}|N,T,p(F)}$; here, however, we are interested in the binned likelihood.

Consider the trivial generating function $G(z) = z$ --- a distribution with $p(0) = 0$ and $p(1) = 1$.
Notice that
\begin{equation}
    G_{\v{x}|N,T}[G(z)] = e^{N (z - 1) };
\end{equation}
that is, the compound of the generating functional $G_{\v{x}|N,T}$  with the generating function $G(z)$ gives the generating function for the number of events generated by the Poisson process.
We can interpret this as a compounding of a point process with a counting process which records 100\% of events, as $p(1) = 1$.

Direct substitution of this trivial generating function into $G_{\v{y}|N,T,p(F)}$ would give the total number of counts in the entire image, as the integral $\int \d\v{y}$ runs over the entire support of $\eta(\v{y})$.
To get the number of counts, $k_B$, in bin $B$, we could perform this substitution, and then alter the support of $\eta(\v{y})$ to $\Omega_B$.
An alternative, is to consider the slightly less trivial generating function
\begin{equation}
    G_{k_B|\v{y}}(z) = 1 + \mathbf{1}_{\Omega_B}(\v{y})(z-1),
\end{equation}
where $\mathbf{1}$ is the indicator function, that is equal to one when $\v{y}\in\Omega_B$ and zero otherwise. 
Thus the probability of recording an event conditioned on the photon location $\v{y}$ is $p(1|\v{y}) = \mathbf{1}_{\Omega_B}(\v{y})$ which is equal to one when the photon is inside the bin and zero otherwise. 
The conditional probability of discarding an event is $p(0|\v{y}) = 1 - \mathbf{1}_{\Omega_B}(\v{y})$ as expected.
Using this, the generating function for the number of counts in bin $B$ is
\begin{equation}\begin{aligned}
    G_{k_B|N,T,p(F)}(z) &= G_{\v{y}|N,T,p(F)}[G_{k_B|\v{y}}(z)] \\ &= \exp{\left[ N \int \d\v{x} T(\v{x}) \left(\int \d F p(F) \exp{\left[\kappa(\v{x}) F \int_{\Omega_B} \d\v{y} \eta(\v{y}) \phi(\v{y} | \v{x}) (z - 1) \right]} - 1\right)  \right]},
\end{aligned}\end{equation}
and after rearranging,
\begin{equation}
    G_{k_B|N,T,p(F)}(z) = \exp{\left[ N \left(\int \d F p(F) \int \d\v{x} T(\v{x}) \exp{\left[\kappa(\v{x}) F \int_{\Omega_B} \d\v{y} \eta(\v{y}) \phi(\v{y} | \v{x}) (z - 1) \right]} - 1\right)  \right]}.
\end{equation}

From here, the $\mu_B(\varepsilon)$ measure defined in Eq.~\ref{eq:mu_eps_def} can be used to bring this into the expected form:
\begin{equation}
    G_{k_B}(z) = \exp{\left[ N \left(\int \d F p(F) \int \d\varepsilon \mu_B(\varepsilon) e^{\varepsilon F (z - 1) } - 1\right)  \right]}.
\end{equation}
For multiple bins, the indicator generating function is
\begin{equation}
    G_{\{k_{\bin}\}|\v{y}}(\{z_\bin\}) = 1 + \sum_\bin \mathbf{1}_{\Omega_{\bin}}(\v{y})(z_\bin-1).
\end{equation}
Substitution of this into $G_{\v{y}|N,T,p(F)}$ gives the multiple bin generating function.
However, evaluation of the probabilities for such a generating function would not only require a multidimensional $\mu(\{\varepsilon_i\})$ with an effective detector response for each bin, but would also require computing many cross terms between the $z_i$ variables.
For more than a few bins, this quickly becomes computationally untenable.
For this reason, the whole image likelihood is constructed by taking the product of the one bin likelihoods as in Eq.~\ref{eq:whole_image_lh_def} --- essentially a mean field approximation, where each bin is treated as statistically independent, removing the correlations.

\section{Evaluation of the Power Series\label{app:series}}

Here we demonstrate how to analytically evaluate the CPG generating function when the source-count function takes the form of a multiply broken power-law.
Firstly, we write the generating function for a point-source population as
\begin{equation}
    G_{k_B}(z) = \exp{\left[ \left( \int \d\varepsilon g(\varepsilon, z) \mu_B(\varepsilon) - N \right) \right]}, \label{eq:gen_mueps_g}
\end{equation}
where
\begin{equation}
    g(\varepsilon, z) = \int \d F e^{\varepsilon F (z-1)} N p(F).
\end{equation}
As stated, we assume a broken power-law differential source-count function:
\begin{equation}\begin{aligned}
	\frac{dN}{dF} &= A \begin{cases} 
			c_m \left(\frac{F}{F_{b(m)}}\right)^{-n_m} & F < F_{b(m)} \\
			c_i \left(\frac{F}{F_{b(i)}}\right)^{-n_i} & F_{b(i+1)} < F < F_{b(i)} \\
			c_1 \left(\frac{F}{F_{b(2)}}\right)^{-n_1} & F_{b(2)} < F
		\end{cases}.
\end{aligned}\end{equation}
This is equivalent to the broken power-law definition in Eq.~\ref{eq:full_broken_plaw_def}, with the $c_i$ introduced to account for the normalisation factors.

From the relation $N p(F) = \d N/\d F$, we have
\begin{equation}
    g(\varepsilon, z) = \int \d F e^{\varepsilon F (z-1)} \frac{\d N}{\d F}.
\end{equation}
The solution to this integral has a closed form for this choice of source-count function:
\begin{equation}
    g(\varepsilon, z) = A \left( c_1 F_{b(2)} \sigma^0_1(\varepsilon,z) + \sum_{i=2}^{m-1} c_i F_{b(i)} \sigma^0_i(\varepsilon,z) + c_m F_{b(m)} \sigma^0_m(\varepsilon,z) \right),
\end{equation}
where, using $J_i(\varepsilon,z) = - \varepsilon F_{b(i)}(z-1)$ to shorten the notation,
\begin{align}
    \sigma_1^j(\varepsilon,z) &= J_2(\varepsilon,z)^{n_1-1} \Gamma(1 + j - n_1, J_2(\varepsilon,z)), \\
    \sigma_i^j(\varepsilon,z) &= J_{i+1}(\varepsilon,z)^{n_i-1} \Gamma(1 + j - n_i, J_{i+1}(\varepsilon,z)) \left(\frac{F_{b(i)}}{F_{b(i+1)}}\right)^{n_i-1} - J_i(\varepsilon,z)^{n_i-1} \Gamma(1 + j - n_i, J_i(\varepsilon,z)), \\
    \sigma_m^j(\varepsilon,z) &= J_m(\varepsilon,z)^{n_m-1} \gamma(1+j-n_m, J_m(\varepsilon,z)), \label{eq:sigma_m}
\end{align}
where $\Gamma(n, x)$ and $\gamma(n, x)$ are the upper and lower incomplete gamma functions.
The form of these $\sigma$ functions has been chosen such that the following identities hold:
\begin{align}
    \frac{\d \sigma_1^j}{\d z}(\varepsilon,0) &= \sigma_1^{j+1}(\varepsilon,0), \\
    \frac{\d \sigma_i^j}{\d z}(\varepsilon,0) &= \sigma_i^{j+1}(\varepsilon,0), \\
    \frac{\d \sigma_m^j}{\d z}(\varepsilon,0) &= \sigma_m^{j+1}(\varepsilon,0).
\end{align}

The above results imply that an identification can be made with the power series coefficients in Eq.~\ref{eq:exp_series}.
For $j>0$, we have
\begin{equation}
    a_\bin^{(j)} = \int \d\varepsilon g^j(\varepsilon, 0) \mu_b(\varepsilon),
\end{equation}
where
\begin{equation}
    g^j(\varepsilon, 0) = A \left( c_1 F_{b(2)} \sigma^j_1(\varepsilon,0) + \sum_{i=2}^{m-1} c_i F_{b(i)} \sigma^j_i(\varepsilon,0) + c_m F_{b(m)} \sigma^j_m(\varepsilon,0) \right).
\end{equation}
We postpone a discussion of the case of $j=0$ for now.
Note that $g^j(\varepsilon, 0)$ is not a function of $\bin$, the bin number.
This allows the $z$-series of $g^j$ to be computed once for the entire image, all of the bin specific detector effects are contained in $\mu_\bin$ which requires a simple numerical integration against this series for each bin.
This can be many orders of magnitude more computationally efficient than NPTF, which requires the $z$-series to be computed for every bin.

The evaluation of upper and lower incomplete gamma functions is computationally expensive, and must be performed using numerical approximations.
Thus, it is desirable to avoid evaluating these special functions for every term in the $z$-series.
There exists a recurrence relation for the incomplete gamma functions, which is best exploited in terms of the scaled $\sigma$ functions, $\hat{\sigma}$:
\begin{align}
    \hat{\sigma}_1^{j+1}(\varepsilon) = \frac{\sigma_1^{j+1}(\varepsilon, 0)}{(j+1)!} &= \left(1 - \frac{n_1}{j+1}\right)\hat{\sigma}_1^j(\varepsilon) + \exp{\left[ j \ln{J_2(\varepsilon,0)} - J_2(\varepsilon,0) - \ln{\Gamma(j+2)} \right]}, \\
    \hat{\sigma}_i^{j+1}(\varepsilon) = \frac{\sigma_i^{j+1}(\varepsilon, 0)}{(j+1)!} &= \left(1 - \frac{n_i}{j+1}\right)\hat{\sigma}_i^j(\varepsilon) + \exp\Bigg[ j\ln{J_i(\varepsilon,0)} - J_i(\varepsilon,0) - \ln{\Gamma(j+2)} \Bigg]  \nonumber \\ 
        & \quad\quad \times \left( \left[\frac{F_{b(i)}}{F_{b(i+1)}}\right]^{n_i - 1 - j} \exp{\left[ J_i(\varepsilon,0) - J_{i+1}(\varepsilon,0) \right]} - 1 \right), \\
    \hat{\sigma}_m^j(\varepsilon) = \frac{\sigma_m^j(\varepsilon, 0)}{j!} &= \left(1 - \frac{n_m}{j}\right)^{-1} \left( \hat{\sigma}_m^{j+1}(\varepsilon) + \exp{\left[ (j-1)\ln{J_m(\varepsilon,0)} - J_m(\varepsilon,0) - \ln{\Gamma(j+1)} \right]} \right).
\end{align}
For $\hat{\sigma}_1$ and $\hat{\sigma}_i$, the base case of $j=0$ is first computed from
\begin{align}
    \sigma_1^0(\varepsilon,z) &= J_2(\varepsilon,z)^{n_1-1} \Gamma(1 - n_1, J_2(\varepsilon,z)) = E_{n_1}(J_2(\varepsilon,z)), \\
    \sigma_i^0(\varepsilon,z) 
        &= J_{i+1}(\varepsilon,z)^{n_i-1} \Gamma(1 - n_i, J_{i+1}(\varepsilon,z)) \left(\frac{F_{b(i)}}{F_{b(i+1)}}\right)^{n_i-1} - J_i(\varepsilon,z)^{n_i-1} \Gamma(1 - n_i, J_i(\varepsilon,z)) \nonumber\\
        &= E_{n_i}(J_{i+1}(\varepsilon, z)) \left(\frac{F_{b(i)}}{F_{b(i+1)}}\right)^{n_i-1} - E_{n_i}(J_i(\varepsilon, z)),
\end{align}
where the exponential integral, $E_n(x)$, is used to stabilise the calculation.
Then, the higher order $\hat{\sigma}_1$ and $\hat{\sigma}_i$ terms are found using the above recurrence relations.

For $\hat{\sigma}_m$, the initial case starts from the highest order term required, which for the whole image is $j = \max_\bin k_\bin$.
This is computed using Eq.~\ref{eq:sigma_m}, then the lower order terms are found using the recurrence relations.
The power series coefficient for $j=0$ also includes the constant factor of $N$ which appeared in Eq.~\ref{eq:gen_mueps_g}:
\begin{equation}
    a_\bin^{(0)} = \int \d\varepsilon g^0(\varepsilon, 0) \mu_b(\varepsilon) - N.
\end{equation}
If $g^0$ is calculated using the above procedure, this factor of $N$ will cause a catastrophic cancellation for very dim (i.e. below threshold) populations.
This cancellation arises from the left hand term of the right hand side of this equation taking a value very close to $N$ such that it is equal to $N$ up to the machine precision of the floating point data-type used in the computation.
A resolution to this cancellation is achieved by incorporating the constant factor into the calculation of the exponential integral.
The expected number of sources, $N$, is first evaluated in closed form in terms of the source-count function parameters:
\begin{equation}
    N = A \left( \frac{c_1 F_{b(2)}}{n_1 - 1} + \sum_{i=2}^{m-1} \frac{c_i F_{b(i)}}{n_i - 1} \left[\left(\frac{F_{b(i)}}{F_{b(i+1)}}\right)^{n_i-1} - 1\right] + \frac{c_m F_{b(m)}}{1 - n_m} \right).
\end{equation}
Each term in this equation is incorporated into the corresponding $\hat{\sigma}^0$ function, to give the modified $\sigma$ terms
\begin{align}
    \tilde{\sigma}_1^0(\varepsilon) &= E_{n_1}(J_2(\varepsilon,0)) - \frac{1}{n_1 - 1}, \\
    \tilde{\sigma}_i^0(\varepsilon) &= \left(E_{n_i}(J_{i+1}(\varepsilon, 0)) - \frac{1}{n_i - 1} \right) \left(\frac{F_{b(i)}}{F_{b(i+1)}}\right)^{n_i-1} - \left(E_{n_i}(J_i(\varepsilon, 0)) - \frac{1}{n_i - 1}\right).
\end{align}
In the very dim below threshold regime, $J_i(\varepsilon, 0) \ll 1$, and the exponential integral becomes dominated by a leading constant factor:
\begin{equation}
    E_{n}(x) = \frac{1}{n - 1} + \mathcal{O}(x).
\end{equation}
Thus, the cause of the numerical instability is revealed.
Define a modified exponential integral function, $\tilde{E}_{n}(x, y) = y + E_{n}(x)$, so that
\begin{equation}
    \tilde{E}_{n}\left(x, -\frac{1}{n - 1}\right) = \mathcal{O}(x).
\end{equation}
Now the modified $\sigma$ terms can be written
\begin{align}
    \tilde{\sigma}_1^0(\varepsilon) &= \tilde{E}_{n_1}\left(J_2(\varepsilon,0), -\frac{1}{n_1 - 1}\right), \\
    \tilde{\sigma}_i^0(\varepsilon) &= \tilde{E}_{n_i}\left(J_{i+1}(\varepsilon,0), -\frac{1}{n_i - 1}\right) \left(\frac{F_{b(i)}}{F_{b(i+1)}}\right)^{n_i-1} -\tilde{E}_{n_i}\left(J_i(\varepsilon,0), -\frac{1}{n_i - 1}\right).
\end{align}

The numerical algorithm for approximating the exponential integral is based on \textcite{navas-palenciaFastAccurateAlgorithm2018}.
This algorithm was modified to calculate $\tilde{E}_{n}(x, y)$.
For $x \ll 1$, series calculations are used by \textcite{navas-palenciaFastAccurateAlgorithm2018}; during evaluation of the first series term, $y$ is added so that the modification occurs before subsequent terms are added.
The result is that the first term of the series is cancelled, allowing subsequent terms to accumulate without being lost in the machine precision.
For other scales of $x$, non-series calculations are used and so $y$ is simply added to the final result -- an acceptable solution as catastrophic cancellation does not occur in these regimes.

When $J_i(\varepsilon, 0) \gg 1$, the exponential integral algorithm may overflow.
This corresponds to populations with unrealistically high flux per source.
If this happens, a NaN is propagated through the likelihood evaluation algorithm and is returned as the final probability.
This allows this failure mode to be caught and reported.
In these investigations, the correct course of action was to simply reduce the range of the flux prior, as the problem was always caused by the MCMC initialisation drawing these unrealistically large flux values from the uniform prior.

With these modifications in place, the scaled power series coefficients can be defined as
\begin{align}
    \nu_B^{(0)} &= \int \d\varepsilon A \left( c_1 F_{b(2)} \tilde{\sigma}^0_1(\varepsilon) + \sum_{i=2}^{m-1} c_i F_{b(i)} \tilde{\sigma}^0_i(\varepsilon) + c_m F_{b(m)} \tilde{\sigma}^0_m(\varepsilon) \right) \mu_B(\varepsilon), \\
    \nu_B^{(j)} &= \int \d\varepsilon A \left( c_1 F_{b(2)} \hat{\sigma}^j_1(\varepsilon) + \sum_{i=2}^{m-1} c_i F_{b(i)} \hat{\sigma}^j_i(\varepsilon) + c_m F_{b(m)} \hat{\sigma}^j_m(\varepsilon) \right) \mu_B(\varepsilon).
\end{align}
We will use these expressions in App.~\ref{app:bell}.

\section{Evaluation of Bell Polynomials\label{app:bell}}

The probability for $k_b$ counts from the CPG likelihood was shown to be given by
\begin{equation}
    P(k_B) = \frac{ B_{k_B}(a_B^{(1)}, \ldots, a_B^{(k_\bin)}) }{ k_B! }.
\end{equation}
This expression is written in terms of the Bell polynomials, which obey the recurrence relation
\begin{equation}
    B_i(x_0,\ldots,x_i) = \sum_{j=0}^{i-1} \frac{(i-1)!}{j!(i-j-1)!} B_{i-j-1}(x_1,\ldots,x_{i-j-1}) x_{j+1},
\end{equation}
with $B_0=1$.
Given this, we can write
\begin{equation}\begin{aligned}
    P(k_B) 
        &= \sum_{j=0}^{k_B-1} \frac{(k_B-1)!}{j!k_B!} \frac{ B_{k_B-j-1}(a_B^{(1)},\ldots, a_B^{(k_B-j-1)} }{(k_B-j-1)!} a_B^{(j+1)} \\
        &= \sum_{j=0}^{k_B-1} a_B^{(j+1)} \frac{(k_B-1)!}{j!k_B!}  P(k_B-j-1).
\end{aligned}\end{equation}
Now let $l=k_B-j-1$ so that
\begin{equation}\begin{aligned}
    P(k_B) 
        &= \sum_{l=0}^{k_B-1} \frac{a_B^{(k_B-l)}}{(k_B-l)!} \frac{(k_B-l)}{k_B} P(l) \\
        &= \sum_{l=0}^{k_B-1} \nu_B^{(k_B-l)} \frac{(k_B-l)}{k_B} P(l),
\end{aligned}\end{equation}
where $\nu_B^{(k_b-l)}$ are the scaled power series coefficients from App.~\ref{app:series}.
This recurrence relation is equivalent to that found in \textcite{Mishra-Sharma:2016gis}.

The numerical calculation of $P(k_B)$ is performed logarithmically,
\begin{equation}
    \ln{P(k_B)} = \ln{ \sum_{l=0}^{k_B-1} \exp{\left[ \ln{(\nu_B^{(k_B-l)})} + \ln{(k_B-l)} - \ln{k_B} + \ln{P(l)} \right]} },
\end{equation}
so as to ensure that numerical underflow does not occur if the scale of the likelihood is below that of the machine precision of the floating point data-type used.
A standard exponential sum algorithm is employed, which determines the scale of the sum terms in advance; then, the terms are rescaled to avoid underflow before the summation is computed; lastly, the final result is scale corrected.

\section{Numerical Evaluation of $\mu_\bin(\varepsilon)$\label{app:mu_eps}}

In this appendix we outline how to numerically compute $\mu_\bin(\varepsilon)$. The process is detailed in algorithm~\ref{alg:mu_eps}, which constructs a histogram density estimate $(\mu_\bin)_i$.
In detail, a simulated source is drawn from $T(\v{x})$, the location $\v{x}$ of this source is used to find the detector response $\kappa(\v{x})$ and to select the appropriate PSF, $\phi(\v{y} | \v{x})$.
This draw will form a sample from $\mu_\bin(\varepsilon)$, but as $\mu_\bin$ is a function of $\varepsilon$ the corresponding value of $\varepsilon$ must be determined.

According to Eq.~\ref{eq:mu_eps_def}, this requires an integral of the PSF and detector efficiency over the bin extents.
This integral will also be performed through Monte-Carlo sampling: a number, $N_{\text{sim counts}}$, of samples are drawn from the PSF.
Let $\{\v{y}_l\}$ be the set of these samples, the detector response for bin $b$ is calcuated as
\begin{equation}
    \varepsilon = \kappa(\v{x}) \frac{1}{N_{\text{sim counts}}} \sum_{\v{y} \in \{\v{y}_l\}} \eta(\v{y}) \mathbf{1}_{\Omega_\bin}(\v{y}),
\end{equation}
where $\mathbf{1}_{\Omega_\bin}(\v{y}) = 1$ if $\v{y}$ is within the bin extent $\Omega_\bin$ and is zero otherwise.

Thus, a set $\{\v{x}_j\}$ of $N_{\text{sim sources}}$ simulated sources yields a set of detector responses, $\{\varepsilon_j\}$.
The empirical estimate of $\mu_b(\varepsilon)$ can now be written as
\begin{equation}
    \mu_\bin(\varepsilon) = \frac{1}{N_{\text{sim sources}}} \sum_{\varepsilon' \in \{\varepsilon_j\}} \delta(\varepsilon - \varepsilon').
\end{equation}
Although not strictly required, conversion of this empirical measure to a histogram, $(\mu_\bin)_i$, will substantially speed up calculation of the likelihood when $N_{\text{sim sources}}$ and $N_{\text{sim counts}}$ are large.
For histogram bin $i$, the value of the bin is
\begin{equation}
    (\mu_\bin)_i = \int_{\hat{\varepsilon}_i, \hat{\varepsilon}_{i+1}} \d\varepsilon \mu_\bin(\varepsilon),
\end{equation}
where $\hat{\varepsilon}_i$ and $\hat{\varepsilon}_{i+1}$ are the histogram bin edges.

\begin{algorithm}[H]
\begin{algorithmic}
\State{$(\mu_\bin)_i \leftarrow 0 \quad \forall \bin, i$}
\Loop{ $N_{\text{sim sources}}$ times}
    \State{$\varepsilon_{\bin} \leftarrow 0 \quad \forall \bin$}
    \State{$\v{x} \leftarrow \text{sample from } T(\cdot)$}
    \Loop{ $N_{\text{sim counts}}$ times}
        \State{$\v{y} \leftarrow \text{sample from } \phi(\cdot | \v{x})$}
        \State{$\bin \leftarrow \bin' \text{ such that } \v{y} \in \Omega_{\bin'}$}
        \State{$\varepsilon_\bin \leftarrow \varepsilon_\bin + \kappa(\v{x}) \eta(\v{y}) / N_{\text{sim counts}}$}
    \EndLoop
    \State{$i_\bin \leftarrow {i_\bin}' \text{ such that } \varepsilon_\bin \in [\hat{\varepsilon}_{i_\bin'}, \hat{\varepsilon}_{i_\bin'+1}) \quad \forall \bin $}
    \State{$(\mu_\bin)_{i_\bin} \leftarrow (\mu_\bin)_{i_\bin} + 1 / N_{\text{sim sources}} \quad \forall \bin $}
\EndLoop
\end{algorithmic}
\caption{Numerical $\mu_\bin(\varepsilon)$}\label{alg:mu_eps}
\end{algorithm}

\section{Numerical estimation of $\rho(f)$\label{app:rho_f}}

For completeness, we also provide the algorithm for computing $\rho(f)$ as it appears in the NPTF. Again this is computed numerically and not analytically, using algorithm~\ref{alg:rho_f}. 
This algorithm creates a density estimate for $\rho(f)$ in the form of a histogram $\rho_i$, with bin edges $\hat{f}_i$.

\begin{algorithm}[H]
\begin{algorithmic}
\State{$\rho_i \leftarrow 0 \quad \forall i$}
\Loop{ $N_{\text{sim sources}}$ times}
    \State{$f_\bin \leftarrow 0 \quad \forall \bin$}
    \State{$\v{x} \leftarrow \text{sample from } \Omega$}
    \Loop{ $N_{\text{sim counts}}$ times}
        \State{$\v{y} \leftarrow \text{sample from } \phi(\cdot | \v{x})$}
        \State{$\bin \leftarrow  \bin' \text{ such that } \v{y} \in \Omega_{\bin'}$}
        \State{$f_\bin \leftarrow f_\bin + 1/N_{\text{sim counts}}$}
    \EndLoop
    \State{$i_\bin \leftarrow {i_\bin}' \text{ such that } f_\bin \in [\hat{f}_{i_\bin'}, \hat{f}_{i_\bin'+1}) \quad \forall \bin$}
    \State{$\rho_{i_\bin} \leftarrow \rho_{i_\bin} + 1 / N_{\text{sim sources}} \quad \forall \bin $}
\EndLoop
\State{$\tilde{f}_i \leftarrow (\hat{f}_{i} + \hat{f}_{i+1})/2 \quad \forall i$}
\State{$\Delta_i \leftarrow (\hat{f}_{i+1} - \hat{f}_{i}) \quad \forall i$}
\State{$\rho_i \leftarrow \rho_i / \sum_{j} \tilde{f}_j \rho_j \delta_j \quad \forall i$}
\end{algorithmic}
\caption{Numerical $\rho(f)$}\label{alg:rho_f}
\end{algorithm}

\section{Natural Coordinate System\label{app:coords}}

In this appendix we provide additional details regarding the natural coordinate system introduced in Sec.~\ref{sec:priors} of the main text.
The mean number of sources is taken to be $N$, while, here, $F_{PS}$ will be the total amount of flux emitted by the population of sources.
There will be $m-1$ breaks, the locations of which will be defined by $m-2$ numbers $\beta_i \in (0,1)$, where
\begin{equation}
    \beta_i = \frac{F_{b(i+1)}}{F_{b(i)}},
\end{equation}
is the ratio of flux location of the $(i+1)$th break to same for the $i$th break.

The conversion between this coordinate system and the standard coordinate system of Eq.~\ref{eq:full_broken_plaw_def} requires finding $A$ and all $F_{b(i)}$ in terms of $N$, $F_{PS}$, and all $\beta_i$. To do so, let
\begin{equation}
    \alpha_i = \frac{F_{b(i)}}{F_{b(2)}} = \prod_{j=2}^{i-1} \beta_j, \label{eq:nat_coord_alpha_def}
\end{equation}
be a similar ratio to $\beta_i$ but to the flux location of the first break, $F_{b(2)}$, instead. Now we can write the mean number of sources as
\begin{align}
    N &= A \left( \frac{c_1 F_{b(2)}}{n_1 - 1} + \sum_{i=2}^{m-1} \frac{c_i F_{b(i)}}{n_i - 1} \left[\left(\frac{F_{b(i)}}{F_{b(i+1)}}\right)^{n_i-1} - 1\right] + \frac{c_m F_{b(m)}}{1 - n_m} \right), \\
    A F_{b(2)} &= N \left( \frac{c_1 \alpha_2}{n_1 - 1} + \sum_{i=2}^{m-1} \frac{c_i \alpha_i}{n_i - 1} \left[\left(\beta_i^{-1}\right)^{n_i-1} - 1\right] + \frac{c_m \alpha_m}{1 - n_m} \right)^{-1}. \label{eq:nat_coord_soln1}
\end{align}
There is a similar expression for the total flux of the population:
\begin{align}
    F_{PS}
        &= A \left( \frac{c_1 F_{b(2)}^2}{n_1 - 2} + \sum_{i=1}^{m-1} \frac{c_i F_{b(i)}^2}{n_i - 2} \left[\left(\frac{F_{b(i)}}{F_{b(i+1)}}\right)^{n_i-2} - 1\right] + \frac{c_m F_{b(m)}^2}{2 - n_m} \right), \\
    A F_{b(2)}^2
        &= F_{PS} \left( \frac{c_1 \alpha_2^2}{n_1 - 2} + \sum_{i=1}^{m-1} \frac{c_i \alpha_i^2}{n_i - 2} \left[\left(\beta_i^{-1}\right)^{n_i-2} - 1\right] + \frac{c_m \alpha_m^2}{2 - n_m} \right)^{-1}. \label{eq:nat_coord_soln2}
\end{align}

From the ratio of Eqs.~\ref{eq:nat_coord_soln2} and \ref{eq:nat_coord_soln1}, we can find $F_{b(2)}$, then from either of these equations we can find $A$.
Finally, all remaining flux break locations, $F_{b(i)}$, can be found through Eq.~\ref{eq:nat_coord_alpha_def}.

\begin{table}[b]
    \centering
    \begin{tabular}{c|c|c|c|c|c|c}
        Setting         & Prior demo 				& Non-uniform dist 	& Anisotropic PSF		& Sub-bin eff. area		& $\rho(f)$ scenario	& Realistic \\ \hline \hline
        Spatial dist.   & Uniform 				& 5 Delta		& Uniform 			& Uniform 			& Uniform		& Uniform \\
        Vignetting      & Off 					& Off			& Off				& Checkerboard	& Off			& On \\
        PSF             & Isotropic 				& Isotropic		& NuSTAR 			& Delta distribution		& Isotropic		& NuSTAR \\
        Binning         & $9\times 9$ 				& $9\times 9$		& $9\times 9$ 			& $4\times 4$			& $9\times 9$		& $10\times 10$ \\
        Energy range    & 3 -- 10 keV 				& 3 -- 10 keV 		& 3 -- 10 keV 			& 3 -- 10 keV 			& 3 -- 10 keV		& 3 -- 10 keV \\
        Energy spectrum & $E^{-1.5}$ 				& $E^{-1.5}$		& $E^{-1.5}$			& $E^{-1.5}$			& $E^{-1.5}$		& $E^{-1.5}$ \\
        Background      & $2\times 10^{-2} \text{ s}^{-1}$	& None			& None				& None				& None			&  $8.73\times 10^{-2}$ s$^{-1}$\\
        CXB             & None  & None & None & None & None & $1.5\times 10^{-3}$  \\
        Exposure time   & $200 \text{ ks}$			& $200\text{ ks}$	& $200\text{ ks}$		& $200\text{ ks}$		& $200\text{ ks}$	& $200\text{ ks}$  \\
	Exposures	& Single				& Single		& $4\times2$ adjacent tiled	& Single			& Single		& $4\times 2$ tiled  \\
        $A$ 		& N/A 					& $1.5\times 10^{9}$ & $1.5\times 10^{12}$ & N/A & $8\times 10^{10}$ & $10^{12}$ \\
        $F_b$ 		& N/A 					& $1.2\times 10^{-8}$ & $3\times 10^{-9}$ & N/A & $6\times 10^{-9}$ & $6\times 10^{-9}$ \\
        $n_1, n_2$ 	& N/A					& $3,-2$            & $3,-2$ & N/A & $3,-2$ & $3,-2$ \\
        $N$         & N/A                   & N/A               & N/A    & $1.2\times 10^{4}$  & N/A & N/A \\
        $F$         & N/A                   & N/A               & N/A   & $3\times 10^{-10}$  & N/A & N/A
    \end{tabular}
    \caption{All scenario configuration options. Flux parameters are in units of [counts cm${}^{-2}$ keV${}^{-1}$ s${}^{-1}$]. $A$ parameters are in units of the inverse of the flux units. CXB flux is given in units of [photons s${}^{-1}$ cm${}^{-2}$ deg${}^{-2}$]. For scenarios with mosaiced exposures, the binning value gives the equivalent bin size for one observation. }
    \label{tbl:all_config}
\end{table}

\begin{table*}[b]
\begin{minipage}{0.495\textwidth}
    \centering
    \begin{tabular}{c|cc}
        Parameter & Lower limit & Upper limit \\ \hline \hline
        $A$ & $10^8\text{ cm}^2\text{ s}$ & $10^{14}\text{ cm}^2\text{ s}$ \\
        $F_{b(2)}$ & $10^{-15}\text{ cm}^{-2}\text{ s}^{-1}$ & $10^{-3}\text{ cm}^{-2}\text{ s}^{-1}$ \\
        $F_P$ & $10^{-15}\text{ cm}^{-2}\text{ s}^{-1}$ & $10^{8}\text{ cm}^{-2}\text{ s}^{-1}$ \\
        $n_1$ & $2$ & $5$ \\
        $n_2$ & $-3$ & $0$
    \end{tabular}
    \caption{Prior ranges for the demonstration using the standard differential source-count function parameterisation.}
    \label{tbl:priors_priors}
\end{minipage}
\begin{minipage}{0.495\textwidth}
    \centering
    \begin{tabular}{c|cc}
        Parameter & Lower limit & Upper limit \\ \hline \hline
        $N$ & 0.1 & $10^{20}$ \\
        $F_T$ & $10^{-15}\text{ cm}^{-2}\text{ s}^{-1}$ & $10^{-3}\text{ cm}^{-2}\text{ s}^{-1}$ \\
        $\psi_1$ & $\arctan{(2)}$ & $\arctan{(5)}$ \\
        $\psi_2$ & $\arctan{(-3)}$ & $\arctan{(0)}$ \\
        $\omega$ & $0$ & $1$ \\
    \end{tabular}
    \caption{Prior ranges for the demonstration using the natural coordinate system for specifying the priors.}
    \label{tbl:natural_priors}
\end{minipage}
\end{table*}

\begin{table*}[b]
\begin{minipage}{0.99\textwidth}
    \centering
    \begin{tabular}{c|cc}
        Parameter & Lower limit & Upper limit \\ \hline \hline
        $N$ & 0.1 & $10^{8}$ \\
        $F_T$ & $10^{-15}\text{ cm}^{-2}\text{ s}^{-1}$ & $10^{-3}\text{ cm}^{-2}\text{ s}^{-1}$ \\
        $\psi_1$ & $\arctan{(2)}$ & $\arctan{(10)}$ \\
        $\psi_2$ & $\arctan{(-5)}$ & $\arctan{(0)}$ \\
        $\omega$ & $0$ & $1$ \\
        $\lambda_{\rm bkg}$ & $6.984 \times 10^{-2}\text{ s}^{-1}$ & $10.476 \times 10^{-2}\text{ s}^{-1}$
    \end{tabular}
    \caption{Prior ranges for the realistic scenario.}
    \label{tbl:realistic_priors}
\end{minipage}
\end{table*}

\end{document}